\documentclass[11pt]{article}

\usepackage[margin=1in]{geometry}
\usepackage{amsmath,amssymb,amsthm,mathtools}
\usepackage{enumitem}
\usepackage{algorithm}
\usepackage{algpseudocode}
\usepackage[numbers,sort&compress]{natbib}
\usepackage{hyperref}
\usepackage{microtype}
\usepackage{graphicx}
\graphicspath{{shared/images/}}
\usepackage{pgfplots}
\pgfplotsset{compat=1.18}
\usetikzlibrary{arrows.meta,calc,decorations.pathreplacing,positioning}
\hypersetup{hidelinks}

\input{shared/definitions}

\title{Optimal Marginal Mixing for Transvection-Based Trapdoored Matrices over Large Fields}
\date{}
\author{Mark Braverman\\
  \small Princeton University
  \and Aayush Jain\\
  \small Carnegie Mellon University
  \and Stephen Newman\thanks{Corresponding author: \href{mailto:stephen.newman@princeton.edu}{\texttt{stephen.newman@princeton.edu}}.}\\
\small Princeton University}

\begin{document}
\maketitle

\begin{abstract}
  \emph{Trapdoored matrices} are pseudorandom matrices admitting a fast matrix-vector multiplication algorithm given the secret key. We study a candidate family of such matrices: matrices generated by short random \emph{transvection walks} on \(\mathrm{SL}_n(\mathbb{F}_q)\), whose steps are given by adding a random multiple of a random row of the current state to another.

  We show that the worst-case time for any nonempty set \(\mathcal{S}\subseteq [n]\times[n]\) of matrix entries to come within TV distance \(\varepsilon\leq \frac{1}{4}\) of its stationary marginal under this walk is \(\Theta(|\mathcal{S}|+n\log(n/\varepsilon))\), uniformly over all \(\mathcal{S}\) and \(q\geq Q_\mathrm{all}(n)\). This is asymptotically optimal, and resolves open questions about the mixing times of transvection walks. We derive concrete upper and lower bounds for \(n\geq 256\). Surprisingly, our bounds also hold conditioned on the row-pair structure of the walk, except on a fraction at most \(\varepsilon\) of such structures. This implies that these walks may be pseudorandom even after conditioning on the row-pair structure, and therefore that trapdoored matrices built with these walks may be universal.

  Assuming the conjectured structure-conditioned pseudorandomness of short walks, we exhibit highly efficient, highly parallelizable, and universal trapdoored matrices over \(\mathrm{SL}_n(\mathbb{F}_q)\), \(\mathrm{GL}_n(\mathbb{F}_q)\), and \(\mathbb{F}_q^{n\times n}\). The families over \(\mathrm{SL}_n(\mathbb{F}_q)\) and \(\mathrm{GL}_n(\mathbb{F}_q)\) admit equally efficient trapdoors for their inverses, making them the first to do so for finite-field trapdoored matrices. We discuss attacks, analyzing one in detail, and lay out directions for further investigation.
\end{abstract}

\clearpage
\section{Introduction}
Can a matrix look random while secretly admitting fast multiplication? This
question is captured in cryptography by \emph{trapdoored matrices}:
matrices that are computationally indistinguishable from random ones, yet
admit fast matrix--vector multiplication given a secret key. The standard algorithm for matrix-vector multiplication requires
\(\Theta(n^2)\) field operations, whereas the secret key should reduce this cost to
\(T(n)\ll n^2\)---ideally \(T(n)=c\cdot n\) for a small constant \(c\), or more
generally \(T(n)=\widetilde{O}(n)\). Although a weaker form of this primitive appears in the work of
Sotiraki~\cite{sotiraki2016authentication}, this natural notion was only recently
formalized and studied independently by Braverman and
Newman~\cite{braverman2025practical} and Vaikuntanathan and
Zamir~\cite{vaikuntanathan2025improving}. In the short time since, trapdoored
matrices have given rise to applications ranging from private
inference and secure delegated linear
algebra~\cite{braverman2025practical,benhamouda2025encrypted,merbaum2026maverick} and private
information retrieval~\cite{chen2026secret} to algorithmic
speedups~\cite{vaikuntanathan2025improving} and barriers in complexity
theory~\cite{raz2026natural}. The delegated matrix-matrix and matrix-vector multiplication protocols, and their consequences for private inference, are particularly important: unlike protocols involving heavier cryptographic machinery such as homomorphic encryption, these protocols are structure-preserving and carry small overheads, resulting in concrete delegator efficiency improvements for reasonable \(n\).

This motivates a search for even more efficient trapdoored matrix constructions. The LPN-based constructions of~\cite{braverman2025practical,benhamouda2025encrypted,vaikuntanathan2025improving} share an undesirable trait: their runtime grows as \(\Omega(n^{1+o(1)}\lambda)\) for security parameter \(\lambda\), causing significant overhead. \cite{benhamouda2025encrypted} proposed heuristic constructions from
linear-time encodable codes and accumulation, but these call for cryptanalysis, as the history of code-based cryptography shows that added structure often invites attacks~\cite{SidelnikovShestakov1992,Wieschebrink2006,Wieschebrink2010,FGOPT2013,CGGOT2014}.

\paragraph{The transvection-walk approach.}
In this work, we consider an elegant candidate proposed by Sotiraki~\cite{sotiraki2016authentication} and built from the simplest invertible linear
operations: adding a scalar multiple of one coordinate to another. For distinct
\(i,j\in[n]\) and \(a\in\F_q\) for \(q\) a prime power, the update \(i\mapsto j\) is left multiplication by
\[
  I_n+aE_{ji},\qquad E_{ji}=e_j e_i^{\top}.
\]

We form a trapdoored matrix by drawing, for each \(t\in [T]\), an independent and uniform pair of
distinct indices \((I_t, J_t)\sim ([n]\times [n])\setminus\left\{(i, i)\middle|i\in [n]\right\}\) and a coefficient \(a_t\sim \F_q\). This gives the matrix
\[
  X_T=(I_n+a_TE_{J_T,I_T})\cdots(I_n+a_1E_{J_1,I_1}).
\]
We refer to the sequence of pairs as the \emph{skeleton}, denoted \(\sigma=((I_1\mapsto J_1),\ldots,(I_T\mapsto J_T))\).
The public matrix is \(X_T\) (and possibly its skeleton); its secret description is the ordered list of transvections.
Given this list, one computes \(X_Tv\) by replaying the updates
\(v_{J_t}\leftarrow v_{J_t}+a_t v_{I_t}\), using at most \(2T\) field operations.
Reversing the list and negating the coefficients computes \(X_T^{-1}v\) at
the same cost.

The construction works over every finite field and requires
no special ring or code structure. \(T\) directly controls the
arithmetic cost, and increasing it provides a systematic way to increase
randomization. The central question for the transvection construction is therefore:

\begin{center}
  \emph{How long must the walk be to render its endpoint computationally\\
  indistinguishable from a random element of \(\mathrm{SL}_n(\F_q)\)?}
\end{center}

If the construction is indeed pseudorandom for small \(T\), its simplicity and efficiency
make it a compelling trapdoored matrix, for several reasons. First, it parallelizes: a length-\(T\) instance can be implemented by a circuit of expected depth \(\tilde{O}(T/n+1)\) with negligible overhead. Second, multiplication by
the matrix and by its inverse are equally fast, a property not shared by
the LPN- and code-based
candidates~\cite{braverman2025practical,vaikuntanathan2025improving,benhamouda2025encrypted}. Indeed, no finite-field construction yet proposed has this property, though Hhan and Lee~\cite{hhan2026trapdoored} give a construction achieving it up to a multiplicative dependence on the security parameter. Third, if the matrix remains pseudorandom even when the skeleton \(\sigma\) is public (Conjecture~\ref{conj:trapdooredness}), this construction satisfies the stronger notion of a \emph{universal} trapdoored matrix family --- a family whose trapdoor is implemented by a public circuit taking the secret as input \cite{benhamouda2025encrypted}.

A construction with similar intuition, and equally efficient inversion, for trapdoored matrices over the reals was proposed in~\cite{vaikuntanathan2025improving} and analyzed in~\cite{pillai2026kac,pillai2026pseudo}.

\subsection{Mixing and subset convergence}\label{sec:transvection-walk}
What kind of cryptanalytic evidence can we give for this construction? In
this work, we look at mixing. The construction corresponds to a random walk on
\(\mathrm{SL}_n(\mathbb{F}_q)\): each step applies a uniformly chosen
transvection to the current state, and the walk converges to \(\pi\), the
uniform distribution on \(\mathrm{SL}_n(\mathbb{F}_q)\). Full statistical convergence, however, is not the right target: it requires
\(\tilde{\Theta}(n^2)\) steps, which eliminates the desired saving in
multiplication. A useful trapdoored matrix must become computationally
pseudorandom long before it becomes statistically random. We therefore
study the convergence of marginals: how many steps are needed before the
joint law of any small set of entries is close
to its stationary marginal, both for the walk itself and conditioned on
its skeleton. Such a bound rules out every distinguisher that reads only
those entries, and, as we discuss below, it is the matrix analogue of the
\(k\)-wise independence that serves as standard evidence of cryptographic security for
block ciphers and local pseudorandom generators.

Simpler versions of this question have been studied before due to independent mathematical interest. Diaconis and Saloff-Coste~\cite{diaconis1996walks} introduced the \(\F_2\) full-matrix walk, and Ben-Hamou~\cite{ben2025mixing} showed that its half-lazy version (which corresponds exactly to our \(q=2\) case) has \(t_{\mathrm{mix}}(\varepsilon)\leq O(n^2\log n + n\log 1/\varepsilon)\). Pillai and Smith then showed that this walk restricted to \(k\) columns has \(t_{\mathrm{mix}}^{(k)}(\varepsilon)\leq O(nk[\log n+\log(1/\varepsilon)])\)~\cite[Theorem~2.7, Lemma~3.3, and Section~3.5]{PS}. By contrast, \cite[Theorem~1.2]{PS} shows the fixed-error lower bound \(t_{\mathrm{mix}}^{(k)}\geq\Omega\left(\max\left\{nk/\log n,n\log n\right\}\right)\). This leaves a gap, which is especially severe in the cryptographically relevant small-error regime.

To the best of our knowledge, no work has yet considered the skeleton-conditioned convergence of these walks.

\paragraph{Entry marginals and mixing.}
All logarithms are natural. For the mixing results, we allow the walk to start at any
\(X\in\mathrm{SL}_n(\F_q)\) and write \(X_0=X\).
Here \(M[i]\) denotes row \(i\), \(M[R]\) denotes the submatrix on row set \(R\), \(M[:, C]\) denotes the submatrix on column set \(C\), and \(M[R,C]\) denotes the submatrix on row set \(R\)
and column set \(C\).

We observe a fixed set \(\mc{S}\subseteq[n]\times[n]\) of entries. Put
\[
  \mc{S}_i=\{j:(i,j)\in\mc{S}\},\qquad
  \pi_{\mc{S}}=\pi|_{\mc{S}},
\]
and define
\begin{equation}\label{eq:entry-distance}
  d_{\mc{S}}(t)=
  \sup_{X\in\mathrm{SL}_n(\F_q)}
  \bigl\|\mathcal L_X(X_t|_{\mc{S}})-\pi_{\mc{S}}\bigr\|_{\TV}.
\end{equation}

For a length-\(T\) skeleton \(\sigma\), which records the row pairs
with repetitions but not the coefficients, write
\[
  K_\sigma(X,\cdot)=\mathcal L(X_T\mid \sigma,X_0=X),\qquad
  d_{\mc{S}}^\sigma(X)=
  \bigl\|K_{\sigma,\mc{S}}(X,\cdot)-\pi_{\mc{S}}\bigr\|_{\TV}.
\]
Here \(K_{\sigma,\mc{S}}\) is the marginal of \(K_\sigma\) on \(\mc{S}\).
Concatenation \(\sigma\tau\) runs \(\sigma\) first:
\(K_{\sigma\tau}=K_\sigma K_\tau\) and the matrix of \(\tau\) multiplies
that of \(\sigma\) on the left. Each full-state kernel preserves \(\pi\).

For \(C\subseteq[n]\) with \(|C|=k<n\), the columns \(X_t[:,C]\)
form a Markov chain on full-column-rank matrices, with uniform stationary
law.
By column symmetry, its mixing time is
\begin{equation}\label{eq:mixing-time-definition}
  t_{\mathrm{mix}}^{(k)}(\varepsilon)
  =\min\{t\ge0:d_{[n]\times[k]}(t)\le\varepsilon\}.
\end{equation}
At \(k=n\), this is the full-group mixing time.

\subsection{Our results}\label{sec:results}
\subsubsection{Mixing}
We prove optimal-order mixing bounds for arbitrary entry sets \(\mc{S}\)
over large fields. Surprisingly, we observe that such bounds hold even conditioned on the typical skeleton \(\sigma\):

\begin{theorem}[Mixing of entry marginals]\label{thm:main}
  Let \(n\ge256\) and \(q\ge Q_{\rm all}(n):=(5000n)^{6n^2}\).
  For every \(0<\varepsilon\leq \frac{1}{4}\) and
  \(\varnothing\ne\mc{S}\subseteq[n]\times[n]\),
  \begin{equation}\label{eq:main-conditional-mixing}
    d_{\mc{S}}(t)\le\varepsilon,\qquad
    \sup_X\Pr_{\sigma:|\sigma|=t}
    [d_{\mc{S}}^\sigma(X)>\varepsilon]\le\varepsilon
  \end{equation}
  at every integer time \(t\) greater than the minimum applicable bound below:
  \begin{equation}\label{eq:unified-main-time}
    \begin{array}{c|l}
      \text{hypothesis}&\text{sufficient time}\\ \hline
      |\mc{S}|\le n^2/10&
      21|\mc{S}|+89n\log n+66n\log(1/\varepsilon)\\
      \text{arbitrary }\mc{S}&
      37|\mc{S}|+135n\log n+100n\log(1/\varepsilon)\\
      \text{arbitrary }\mc{S}&
      14n^2+80n\log n+80n\log(1/\varepsilon)\\
      \text{arbitrary }\mc{S}&
      20n^2+24n\log(1/\varepsilon).
    \end{array}
  \end{equation}
  These results are tight up to constant factors.
\end{theorem}
For \(|\mc{S}|=o(n^2)\) and \(n\log(n/\varepsilon)=o(|\mc{S}|)\),
both guarantees hold by time \((12+o(1))|\mc{S}|\); see
\eqref{eq:entry-leading}.

As a corollary, we resolve the mixing-time gap of~\cite{ben2025mixing,PS} for large fields:

\begin{corollary}[Column mixing]\label{cor:column-mixing-intro}
  Let \(q\ge Q_{\rm all}(n)\), \(1\le k\le n\), and
  \(0<\varepsilon\le1/4\). As \(n\to\infty\),
  \[
    t_{\mathrm{mix}}^{(k)}(\varepsilon)
    =\Theta\!\left(nk+n\log\frac n\varepsilon\right),
  \]
  with absolute implied constants. Moreover, when \(n\ge256\), we have
  \[
    \begin{gathered}
      d_{[n]\times[k]}(t)\le\varepsilon,\qquad
      \sup_X\Pr_{\sigma:|\sigma|=t}
      [d_{[n]\times[k]}^\sigma(X)>\varepsilon]\le\varepsilon\\
      \forall\,t\ge
      \begin{cases}
        8nk+25n\log n+14n\log(1/\varepsilon),&k\le n/10,\\
        16nk+34n\log n+37n\log(1/\varepsilon),&n/10<k\le n/2,\\
        36nk+36n\log n+36n\log(1/\varepsilon),&n/2<k\le n.
      \end{cases}
    \end{gathered}
  \]
\end{corollary}

When \(\log(n/\varepsilon)=o(k)\), the column bounds improve to hold whenever
\begin{equation}\label{eq:unified-leading-constants}
  t\geq
  \begin{cases}
    (4+o(1))nk,&k=o(n),\\
    (6\log4+o(1))nk,&k\le n/2,\\
    (8\log2+o(1))n^2,&k=n.
  \end{cases}
\end{equation}

This also proves pre-cutoff for every sequence of nonempty entry patterns under
the field bound above (Corollary~\ref{cor:precutoff}), partially answering
the large-field analogue of a question of~\cite{ben2025mixing}.

\subsubsection{Trapdoored matrices}

Assuming the conjectured pseudorandomness of short transvection walks, we construct associated near-linear-time trapdoored matrix families over \(\mathrm{SL}_n(\F_q)\), \(\mathrm{GL}_n(\F_q)\), and \(\F_q^{n\times n}\). These families admit fast inversion (over the invertible distributions), have polylogarithmic expected circuit depth, and are universal trapdoored matrices.

We also provide cryptanalysis, considering the most natural attack on these constructions and analyzing its performance. In doing so, we observe an interesting behavior, consistent with the skeleton-based understanding of the walk: For \(T\ge Cn\log^2 n\) and \(q\geq 2^{\Omega(n)}\), except for a negligible fraction of bad skeletons, the skeleton-conditioned distributions of proper minors are exponentially close in total variation to uniform.

Lastly, we consider several open questions and directions for future work, including a conjecture that these families extend with uniform security to all field sizes; see Section~\ref{sec:conjectures} for details.

\paragraph{Marginal mixing as cryptanalytic evidence.} While \(|\mc{S}|\)-marginal mixing need not imply computational indistinguishability, it is evidence for it. In particular, small-set marginal mixing --- which is in our case statistically close to \(k\)-wise independence --- suffices to fool all \emph{local} attacks, namely, attacks that depend on a bounded number of fixed positions, columns, rows, or matrix--vector products chosen in advance. Evidence from restricted
attack models has long supported claims of pseudorandomness, including for Goldreich's local
pseudorandom generators~\cite{Goldreich00,MosselShpilkaTrevisan03,ODonnellWitmer2014,TCC:AppBogRos12,Applebaum2016survey,STOC:AppLov15}
and pseudorandom permutations from random reversible
circuits~\cite{Gowers1996,HMMR2005,BrodskyHoory2008,KaplanNaorReingold2009,GHKO2025,GHP2024}.
This viewpoint is also closely related to the \emph{low-degree method}
in average-case complexity, which uses lower bounds against low-degree
polynomial distinguishers as evidence of computational
hardness~\cite{Hopkins2018,BarakEtAl2016}. Recent cryptographic applications
include hardness evidence for the planted random subgraph problem
underlying private simultaneous messages~\cite{TCC:BJRZ24} and for
Goldreich-based public-key encryption~\cite{TCC:BogKotRos23}. We leave open finding other non-local attacks on our construction.

\subsection{A short overview of our mixing bound proofs}
\phantomsection\label{sec:proof-architecture}
We prove the marginal bounds conditioned on most skeletons.
First, we show that a rational section of the
coefficient-to-output map implies mixing. We then construct these
sections by choosing coefficients as rational functions of the desired
output, and show that the required transvections occur quickly.

\paragraph{Step 1: Rational sections imply mixing.}
For a fixed initial matrix and a length-\(T\) skeleton \(\sigma\), let
\(F_\sigma:\A^T\to Y\) map the transvection coefficients to the output.
For direct entry comparisons, we take \(Y=\A^{\mc{S}}\); for full-group
mixing, we take \(Y=\mathrm{SL}_n\). Both are viewed as varieties over \(\F_q\).

A \emph{rational section} \(s\) satisfies \(F_\sigma\circ s=\mathrm{id}\) on a
nonempty open subset of \(Y\). Differentiating this identity gives a full-rank Jacobian along the
section. Section~\ref{sec:generic-fibres} uses the resulting smooth
rational point on the generic fibre to prove that this fibre is
geometrically integral. After restricting both inputs to a dense open
set, the collision equation \(F_\sigma(a)=F_\sigma(a')\) also defines
a geometrically integral variety, of dimension \(2T-\dim Y\).
An effective Lang--Weil estimate bounds its number of points over
\(\F_q\) by \((1+o(1))q^{2T-\dim Y}\).

The connection to mixing is the second-moment identity: for independent
uniform coefficient vectors \(a,a'\in\F_q^T\),
\[
  \Pr[F_\sigma(a)=F_\sigma(a')]
  =\sum_{y\in Y(\F_q)}\Pr[F_\sigma(a)=y]^2.
\]
We apply this identity to the mass from the open set, bound the excluded
mass by Schwartz--Zippel, and use Cauchy--Schwarz to obtain a TV bound.
The error depends explicitly on the coordinate degrees and \(\dim Y\);
Section~\ref{sec:skeletons} bounds those degrees for a random skeleton.

This approach allows us to turn mixing-time questions into purely structural questions about the walk: in particular, whether the coefficient-to-endpoint map induced by the list of steps so far has the requisite algebraic property. This bypasses looser entropy bounds and allows us to obtain asymptotically tight results. While such algebraic techniques may appear out of place in a mixing proof, similar arguments have been used before. In particular, \cite{LarsenShalevTiep2019} use the section/Lang--Weil estimate combination to prove asymptotic uniformity for certain word maps as the field size grows when group rank is bounded, and \cite{GowersViola2019} combined Lang--Weil counts of trace fibres with a collision-probability reduction to prove near-uniformity of a distribution. Related ideas also appear in \cite{GlazerHendel2024,Green2025}.

\paragraph{Step 2: Rational sections for bounded row demands.}
Suppose \(|\mc{S}_i|\le k\le n/2\). In Section~\ref{sec:rectangular},
we partition the requested rows into a bounded number of target blocks,
each with enough rows outside it to supply sources. Assign a separate
transvection interval to each block. With high probability, each row
\(i\) receives updates from \(|\mc{S}_i|\) distinct sources outside
its block during its interval. If the source rows restricted to columns \(\mc{S}_i\) form
a nonsingular matrix, a square linear system determines the selected
coefficients from the desired entries in row \(i\), and we set all others to zero.

Process the blocks in order. Earlier blocks have known recovered rows;
later blocks retain their initial rows. Thus the same linear solve
applies to each block, and completed blocks are never changed again.
For a fixed initial matrix, the burn-in in Section~\ref{sec:routing-burnin}
routes linearly independent initial rows to each required source set. This makes
the source minors nonzero polynomials in the burn-in coefficients,
including when different rows request different column sets.
At the endpoint obtained by setting all selected coefficients to zero,
every recovery formula uses these nonsingular minors. This common
baseline makes the formulas simultaneously valid on a nonempty open
set. Choosing the auxiliary burn-in coefficients as rational functions
of the endpoint gives a section of the full walk.
The burn-in and collection take \(O(nk+n\log(n/\varepsilon))\) steps.

\paragraph{Step 3: Rational sections for low-density entry patterns.}
For patterns below a fixed density, Section~\ref{sec:entry-mixing} groups rows into high-, medium-, and low-density classes according to their number of requested entries.

We first set the high-density rows by applying Step 2 after transposition.
For each medium-density row, we build a disjoint tree of helper rows whose
later transvections carry contributions toward that row. Updates into any
node of its tree can then supply the requested entries at the root. We
recover these contributions by treating each tree as a target block in the
ordered recovery of Step 2. Finally, we apply Step 2 directly to the low-density
rows. For \(|\mc{S}|\) below a constant
fraction of \(n^2\), this gives the scale
\(O(|\mc{S}|+n\log(n/\varepsilon))\).

\paragraph{Step 4: Rational sections for the full matrix.}
For dense patterns, \(|\mc{S}|=\Omega(n^2)\), mixing the full matrix has
the desired scale. Section~\ref{sec:four-segment-alternative} factors a
generic matrix in \(\mathrm{SL}_n\) rationally into four nearly balanced
block shears. Each shear adds linear combinations of one block of rows
to a disjoint block. The transvection construction produces a
\emph{quasi-shear} of the form
\[
  \begin{pmatrix}I&U\\0&M
  \end{pmatrix},
\]
where \(U\) is arbitrary and the invertible source block \(M\) is fixed.
The source blocks can be absorbed into the four-shear factorization,
so it remains to construct each quasi-shear.

To build a quasi-shear, we use three transvection intervals. For each target row, the first and
third intervals supply most source rows directly. Match each missing
source to one appearing in both outer intervals, using an update from
the missing source to its partner in the middle interval. The partner's
two outer occurrences can then supply both source directions. This
matching proves that the required source minor is a nonzero polynomial
in the middle coefficients. Generic middle coefficients therefore let
us solve for every target row, giving rational recovery of \(U\).
Four such constructions give a section on \(\mathrm{SL}_n\) in
\(O(n^2+n\log(1/\varepsilon))\) steps.

\paragraph{Step 5: Every error level over one field.}
For very small errors, the \(\log \frac{1}{\varepsilon}\) term dominates, so
it suffices to improve the full-matrix bound. Section~\ref{sec:mixing-assembly}
repeats independent attempts at a strengthened version of Step 4. Their TV
contraction factors multiply, giving all error levels under one field threshold.

\subsection{Next steps, conjectures, and open questions}
\label{sec:conjectures}

\paragraph{Mixing.}
While we show asymptotically tight mixing, our constants are still loose. We conjecture that for large fields, the correct leading constant for the \(|\mc{S}|\) term in the mixing time is 1 when \(|\mc{S}|\gg n\log n\) (by the information-theoretic intuition), and the correct constant for the \(n\log n\) term is 1 when \(\mc{S}\) intersects \(\ll \log n\) columns\footnote{More broadly, we expect this term to correspond to the time to achieve density across all columns that \(\mc{S}\) intersects.}. We conjecture further that sequences of walks on either \(k_n=\omega(\log n)\) or \(k_n=o(\log n)\) columns experience the cutoff phenomenon.

For small \(q\), our skeleton-unconditioned lower bound decreases with the ratio of element entropy to transvection entropy. We conjecture that both this and our conditioned lower bound are asymptotically tight for every \(q\):

\begin{conjecture}[Entry-marginal mixing at the lower-bound scale]\label{conj:mixing}
  There is an absolute constant \(C>0\) such that, for every prime power \(q\),
  integer \(n\ge2\), nonempty \(\mc{S}\subseteq[n]\times[n]\), and
  \(\varepsilon\in(0,1)\), the following bounds hold.
  Unconditionally, \(d_{\mc{S}}(t)\le\varepsilon\) at every integer time
  \[
    t\ge C\left(\frac{\log q}{\log(qn)}|\mc{S}|
    +n\log\frac n\varepsilon\right).
  \]
  For any start, the skeleton-conditional bound
  \[
    \sup_{X\in\mathrm{SL}_n(\F_q)}
    \Pr_{\sigma:|\sigma|=t}
    [d_{\mc{S}}^\sigma(X)>\varepsilon]\le\varepsilon
  \]
  holds at every integer time
  \[
    t\ge C\left(|\mc{S}|+n\log\frac n\varepsilon\right).
  \]
\end{conjecture}

\paragraph{Pseudorandomness and attacks.} We make the following all-\(q\) conjecture:
\begin{conjecture}[Pseudorandomness with a public skeleton]\label{conj:trapdooredness}
  There is a function \(f(n)=O(n\log^2 n)\) such that the joint distribution over uniformly chosen length-\(f(n)\) skeletons \(\sigma_n\) and their randomly sampled endpoints \(X^{(n)}_{f(n)}\) satisfies
  \[
    (\sigma_n,X^{(n)}_{f(n)})\approx_c(\sigma_n,U_n),
    \qquad U_n\sim\Unif(\mathrm{SL}_n(\F_{q_n}))\textrm{ independently},
  \]
  uniformly over every sequence of prime powers \(2\le q_n\le2^{n^{O(1)}}\) and starting matrices \(X^{(n)}_0\in \mathrm{SL}_n(\F_{q_n})\). Here, \(\approx_c\)
  denotes computational indistinguishability with negligible advantage in \(n\).
\end{conjecture}

As an intermediate to the above, the behavior described by Proposition~\ref{prop:sampled-dependence-upper} merits further investigation. For \(q\ge2^{\Omega(n)}\), apart from a negligible fraction of bad skeletons, each proper minor is exponentially close in total variation to its stationary distribution after \(Cn\log^2 n\) steps. The sampled rank-deficiency attack therefore has negligible advantage even with runtime \(2^{o(n)}\). We conjecture that this minor convergence extends to all \(q\). We also suspect that this phenomenon extends to many other properties of the walk matrix, and ask whether efficient nonlocal attacks can exploit properties to which it does not extend.

Finally, we note that the bad skeleton events described in this work are typically easy to avoid by deterministically choosing a small subset of the source-target pairs in the skeleton (e.g. to obtain the graph property used in Proposition~\ref{prop:sampled-dependence-upper}), while leaving the rest random. We ask which such choices are likely to aid in defeating broad classes of attacks.

\section{Basic results}
\label{sec:preliminaries} 

Proofs of results in this section appear in Appendix~\ref{app:standard-estimates}.

\begin{lemma}[extension of uniform bounds]\label{lem:fixed-time-extension}
  Let \(P\) be a Markov kernel on a finite state space \(\mathcal X\)
  with stationary distribution \(\pi\).
  \begin{enumerate}[label={(\roman*)}]
    \item Let \(R\) be auxiliary randomness independent of the chain and
      \(\Phi(X,R)\ge0\). Then
      \[
        \sup_{x\in\mathcal X}\E_x[\Phi(X_t,R)]
      \]
      is nonincreasing in \(t\).
    \item The quantity
      \(\sup_{x\in\mathcal X}\|P^t(x,\cdot)-\pi\|_{\TV}\)
      is nonincreasing in \(t\).
  \end{enumerate}
\end{lemma}

\begin{lemma}[geometric waiting-time tail]\label{lem:geometric-waiting}
  A process completes \(m\ge1\) stages in order, running trials
  until the first success at each stage. Conditional on the past, every
  trial in stage \(i\) succeeds with probability at least
  \(p_i\in(0,1]\). Let \(T\) be the total number of trials, and put
  \(\mu=\sum_{i=1}^m p_i^{-1}\) and \(p_*=\min_i p_i\).
  Then, for every \(\varepsilon_0\in(0,1)\), 
  \[
    \Pr\left[
      T>\mu+\sqrt{\frac{2\mu}{p_*}\log\frac1{\varepsilon_0}}
      +\frac1{p_*}\log\frac1{\varepsilon_0}
    \right]\le\varepsilon_0.
  \]
\end{lemma}

\begin{lemma}[full-rank probability]\label{lem:full-rank-uniform}
  Let \(X\) be uniform on \(\F_q^{n\times k}\), where
  \(1\le k<n\). Then
  \[
    \Pr[\rank X<k]
    \le \frac{q^k-1}{q-1}q^{-n}
    \le q^{k-n}\le\frac12.
  \]
  Consequently,
  \[
    \|\Unif(\F_q^{n\times k})-\pi_{[n]\times[k]}\|_{\TV}
    =\Pr[\rank X<k]
    \le q^{k-n}.
  \]
\end{lemma}

\section{Mixing from rational sections}\label{sec:rational-sections}

After bounding coordinate degrees (Section~\ref{sec:skeletons}), we turn
a rational section of the coefficient map into a TV bound in three steps:
\begin{enumerate}
  \item Restrict to a dense open set of coefficient vectors and show that
    the closure of its collision locus satisfies the hypotheses of Lang--Weil
    (Section~\ref{sec:generic-fibres}).
  \item Apply Lang--Weil to bound the collision count, hence the
    \(\ell_2\) norm of the pushforward mass from this open set
    (Section~\ref{sec:LW}).
  \item Account for the excluded coefficients and use Cauchy--Schwarz to
    bound the TV distance of the full pushforward from uniform
    (Section~\ref{sec:LW}).
\end{enumerate}

\subsection{Coefficient maps and polynomial degree}\label{sec:skeletons}

Fix an initial matrix \(X\in\mathrm{SL}_n(\F_q)\). For each application,
choose a target variety \(Y\) over \(\F_q\) and an output map
\(\operatorname{pr}_Y:\mathrm{SL}_n\to Y\) as follows:
\begin{itemize}
  \item To compare a nonempty entry pattern
    \(\mc{S}\subseteq[n]\times[n]\) with the product-uniform law,
    take \(Y=\A^{\mc{S}}\) and
    \(\operatorname{pr}_Y(M)=M|_{\mc{S}}\).
  \item For the full group, take \(Y=\mathrm{SL}_n\) and
    \(\operatorname{pr}_Y(M)=M\).
\end{itemize}
These targets are smooth and geometrically integral, of dimensions
\(|\mc{S}|\) and \(n^2-1\), respectively.
Write \(\pi_Y=\Unif(Y(\F_q))\). This is the product-uniform law in
the first case and \(\pi\) in the second.

For a length-\(T\) skeleton
\[
  \sigma=((I_1\mapsto J_1),\ldots,(I_T\mapsto J_T)),\qquad I_t\ne J_t
\]
the coefficient map is
\[
  F_\sigma:\A^T\longrightarrow Y,\qquad
  F_\sigma(a_1,\ldots,a_T)
  =\operatorname{pr}_Y\!\left((I_n+a_TE_{J_T,I_T})\cdots(I_n+a_1E_{J_1,I_1})X\right).
\]
A transmission sequence is a sublist
\((I_{t_1}\mapsto J_{t_1}),\ldots,(I_{t_h}\mapsto J_{t_h})\), with
\(t_1<\cdots<t_h\), in which each target is the next source:
\(J_{t_r}=I_{t_{r+1}}\) for \(1\le r<h\).

\begin{lemma}[Degree of the coefficient map]\label{lem:transmission-degree}
  Let \(T\ge1\), let \(0<\eta<1\), and let \(\sigma\) be a random
  skeleton of length \(T\). Except with probability \(\eta/16\),
  every coordinate of \(F_\sigma\) has total degree at most
  \begin{equation}\label{eq:common-degree-cutoff}
    H(T,\eta)=\min\left\{T,
    \ceil{\frac{eT}{n}+\log\frac{16n}{\eta}}\right\}.
  \end{equation}
\end{lemma}

\begin{proof}
  For any skeleton, each coordinate degree is at most the length of its
  longest transmission sequence. Indeed, a degree-\(h\) term in the
  product of elementary transvections contains
  \[
    E_{J_{t_h},I_{t_h}}\cdots E_{J_{t_1},I_{t_1}}.
  \]
  This product can be nonzero only when
  \(I_{t_{r+1}}=J_{t_r}\) for every \(r<h\). For any fixed increasing
  sequence of times, this has probability \(n^{-(h-1)}\). For
  \(1\le h\le T\), a union bound gives
  \[
    \Pr[\text{a transmission sequence has length at least }h]
    \le
    \binom Th n^{-(h-1)}
    \le n\left(\frac{eT}{nh}\right)^h
    \le n\exp\left(\frac{eT}{n}-h\right),
  \]
  where the last step follows from \(\log z\le z-1\).
  If \(H(T,\eta)<T\), take \(h=H(T,\eta)+1\); otherwise the degree
  bound is deterministic.
\end{proof}

\subsection{Rational sections and smooth collisions}\label{sec:generic-fibres}

Let \(F:\A^T\to Y\) be a polynomial map over \(\F_q\), with every matrix
entry of total degree at most \(H\). A \emph{rational section} of \(F\)
is a map \(s\), defined over \(\F_q\) by rational functions on a nonempty
open subset of \(Y\), such that \(F\circ s=\mathrm{id}\). In other words, it is a function recovering coefficients to obtain a generic desired endpoint.
Larsen, Shalev, and Tiep~\cite[Corollary~2.3]{LarsenShalevTiep2019}
use sections similarly to prove equidistribution. Here a section makes
the collision variety geometrically integral after we exclude the zeros
of one Jacobian minor.

For \(Y=\mathrm{SL}_n\), we use all entries except \((n,n)\) as
coordinates on the open set where \(\operatorname{cof}_{nn}\) is nonzero.
Write \(\operatorname{cof}_{ij}(Z)\) for the signed \((i,j)\)-cofactor
of a matrix \(Z\). Expanding \(\det Z=1\) along the last row gives
\[
  Z_{nn}
  =\frac{1-\sum_{j=1}^{n-1}Z_{nj}\operatorname{cof}_{nj}(Z)}
  {\operatorname{cof}_{nn}(Z)}
  \qquad\text{when }\operatorname{cof}_{nn}(Z)\ne0.
\]
Every cofactor in this formula is independent of the last row, so the
retained entries are coordinates on this open subset of \(\mathrm{SL}_n\).
This open set contains \(I_n\), hence also the generic point.

Take \(F_1,\ldots,F_{\dim Y}\) to be all output entries, omitting
entry \((n,n)\) when \(Y=\mathrm{SL}_n\). The corresponding entries are
local coordinates at the generic point of \(Y\).

The identity \(F\circ s=\mathrm{id}\) makes \(F\) dominant and its
differential surjective along the section. We can therefore choose input
variables \(c_1,\ldots,c_{\dim Y}\) such that
\begin{equation}\label{eq:J-def}
  \mathcal J(\mathbf a)
  =\det\left(
    \frac{\partial(F_1,\ldots,F_{\dim Y})}
    {\partial(c_1,\ldots,c_{\dim Y})}
  \right)
\end{equation}
does not vanish identically along the section. Also,
\[
  \deg\mathcal J\le(\dim Y)(H-1).
\]
Put \(X^\circ=D(\mathcal J)\subseteq\A^T\) where \(D(\mathcal J)=\{a:\mathcal J(a)\ne0\}\).

In the full-group case, \(F(X^\circ)\) lies in this coordinate chart:
otherwise, differentiating \(\det F=1\) at a point where
\(\operatorname{cof}_{nn}(F)=0\) would
give a linear relation among the retained Jacobian rows. Since
\(\mathcal J\ne0\), all remaining cofactors would vanish, contradicting
\(\det F=1\). Thus on \(X^\circ\), equality of the retained entries
implies equality of the full endpoints.

\paragraph{The collision variety.}
For each target, the Jacobian criterion now shows that
\(F^\circ:X^\circ\to Y\) is
smooth~\cite[\href{https://stacks.math.columbia.edu/tag/01V9}{Tag~01V9}]{stacks-project}
of relative dimension
\(T-\dim Y\)~\cite[\href{https://stacks.math.columbia.edu/tag/02G1}{Tag~02G1}]{stacks-project}.

We count pairs of coefficient vectors in \(X^\circ\) with the same
endpoint. These are the points of
\[
  Z^\circ=X^\circ\times_Y X^\circ.
\]

\begin{lemma}\label{lem:collision-integral}
  The generic fibre of \(F\) is geometrically integral of dimension
  \(T-\dim Y\). Moreover, \(Z^\circ\) is smooth and geometrically integral
  of dimension \(2T-\dim Y\).
\end{lemma}
\begin{proof}
  Dominance of \(F\) and integrality of \(\A^T\) make the generic fibre
  integral over \(\F_q(Y)\). The section supplies a smooth rational point over this field,
  so this fibre is geometrically integral
  \cite[\href{https://stacks.math.columbia.edu/tag/0CDW}{Tag~0CDW}]{stacks-project}.
  Its nonempty open part in \(X^\circ\) has dimension \(T-\dim Y\).

  The projection \(Z^\circ\to X^\circ\) is the base change of \(F^\circ\);
  hence it is smooth of relative dimension \(T-\dim Y\)
  \cite[\href{https://stacks.math.columbia.edu/tag/01VB}{Tag~01VB}]{stacks-project}.
  Thus \(Z^\circ\) is smooth of dimension \(2T-\dim Y\).
  Over \(\overline{\F}_q\), \(X^\circ\) is integral.
  The projection's generic fibre is obtained by extending scalars
  on a nonempty open subset of the geometrically integral generic fibre
  of \(F\), so it is integral as well.
  After base change, the projection remains smooth, hence open
  \cite[\href{https://stacks.math.columbia.edu/tag/056G}{Tag~056G}]{stacks-project}.
  Thus every nonempty open subset of \(Z^\circ_{\overline{\F}_q}\)
  meets this integral generic fibre. Any two such subsets therefore
  intersect, so \(Z^\circ_{\overline{\F}_q}\) is irreducible.
  Smoothness makes it reduced~\cite[\href{https://stacks.math.columbia.edu/tag/056T}{Tag~056T}]{stacks-project}, hence integral.
\end{proof}

All endpoint polynomials and \(\mathcal J\) are defined over \(\F_q\).
Thus \(Z^\circ\) is the intersection of the open set where
\(\mathcal J(\mathbf a)\ne0\) and \(\mathcal J(\mathbf a')\ne0\) with the locus
\[
  F_j(\mathbf a)=F_j(\mathbf a'),
  \qquad 1\le j\le\dim Y.
\]
By Lemma~\ref{lem:collision-integral}, the closure \(\overline{Z^\circ}\)
in \(\A^{2T}_{\F_q}\) is an \(\F_q\)-defined geometrically integral
component of that locus, of dimension
\(2T-\dim Y\).

\subsection{Explicit Lang--Weil counting and conditional mixing}\label{sec:LW}
We convert collision counts to TV distance by a standard second-moment
bound. We estimate these counts using a Lang--Weil bound, as in~\cite{GlazerHendel2024,Green2025}; related ideas appear in~\cite{GowersViola2019}.

\paragraph{From collisions to total variation.}

Schwartz--Zippel and the degree bound on \(\mathcal J\) give
\begin{equation}\label{eq:Xgood-count}
  q^T-(\dim Y)(H-1)q^{T-1}
  \le |X^\circ(\F_q)|\le q^T.
\end{equation}
Let \(\nu\) be the image under \(F\) of the uniform law on \(\A^T(\F_q)\).
For each target,
\[
  |Y(\F_q)|\le q^{\dim Y}.
\]
For \(\mathrm{SL}_n(\F_q)\), this follows from
\(|\mathrm{SL}_n(\F_q)|=q^{n^2-1}\prod_{j=2}^n(1-q^{-j})\).

\begin{proposition}[collision count implies mixing]\label{prop:collision-to-mixing}
  If \(\mathcal E\ge0\) and
  \begin{equation}\label{eq:abstract-collision-count}
    |Z^\circ(\F_q)|\le(1+\mathcal E)q^{2T-\dim Y},
  \end{equation}
  then
  \begin{equation}\label{eq:conditional-TV}
    \|\nu-\pi_Y\|_{\TV}
    \le
    \frac{(\dim Y)(H-1)}{2q}
    +\frac12\sqrt{\mathcal E+2(\dim Y)(H-1)q^{-1}}.
  \end{equation}
\end{proposition}
\begin{proof}
  Put \(x=(\dim Y)(H-1)/q\). For \(y\in Y(\F_q)\), define the mass
  contributed by coefficients in \(X^\circ\):
  \[
    g_y=q^{-T}\#\{\mathbf a\in X^\circ(\F_q):F(\mathbf a)=y\},
    \qquad \sum_y g_y\ge1-x.
  \]
  Then \(\nu(y)\ge g_y\) and
  \begin{equation}\label{eq:collision-second-moment}
    |Y(\F_q)|\sum_y g_y^2
    =|Y(\F_q)|q^{-2T}|Z^\circ(\F_q)|
    \le1+\mathcal E.
  \end{equation}
  The triangle inequality and Cauchy--Schwarz give
  \begin{align*}
    2\|\nu-\pi_Y\|_{\TV}
    &\le x+\sum_y|g_y-|Y(\F_q)|^{-1}|\\
    &\le x+\sqrt{|Y(\F_q)|\sum_y g_y^2-2\sum_y g_y+1}\\
    &\le x+\sqrt{\mathcal E+2x}.
  \end{align*}
\end{proof}

\paragraph{Bounding the collision count.}

We use the effective Lang--Weil bound of Cafure and
Matera~\cite[Theorem~7.1]{CM}, with the field-size restriction removed
as in~\cite[Theorem~4.6 and Appendix~A]{GVJZ}. An \(\F_q\)-defined
absolutely irreducible affine variety
\(V\subseteq\A^N_{\overline{\F}_q}\) of dimension \(r>0\) satisfies
\begin{equation}\label{eq:CM}
  \left|\,|V(\F_q)|-q^r\right|
  \le
  (\deg V-1)(\deg V-2)q^{r-1/2}
  +5(\deg V)^{13/3}q^{r-1}.
\end{equation}

The locus containing \(\overline{Z^\circ}\) is cut out by \(\dim Y\) equations of
degree at most \(H\). Set \(B=H^{\dim Y}\).
The componentwise affine B\'ezout
inequality~\cite[Theorem~1]{Heintz}, also stated in~\cite[Section~2, (7)]{CM},
bounds the sum of the degrees of its
irreducible components, and gives \(\deg\overline{Z^\circ}\le B\).

Apply~\eqref{eq:CM} to \(V=(\overline{Z^\circ})_{\overline{\F}_q}\), which is affine,
\(\F_q\)-defined, and absolutely irreducible of dimension
\(r=2T-\dim Y\ge\dim Y>0\). Here \(T\ge\dim Y\) follows from dominance.
Since \(|Z^\circ(\F_q)|\le|V(\F_q)|\), this gives
\eqref{eq:abstract-collision-count} with
\[
  \mathcal E
  =(B-1)(B-2)q^{-1/2}
  +5B^{13/3}q^{-1}.
\]

For \(H\ge2\), an integer \(m\ge1\), and \(0<\varepsilon\le1/4\), define
\begin{equation}\label{eq:appendix-geometric-threshold}
  Q_{\mathrm{geom}}(H,m;\varepsilon)
  =256\varepsilon^{-4}H^{13m/3}.
\end{equation}

\begin{corollary}[conditional mixing from a rational section]
  \label{cor:section-mixing}
  Let \(Y=\A^m\) with \(m\ge1\), or \(Y=\mathrm{SL}_n\), and let
  \(F:\A^T\to Y\) be a polynomial map over \(\F_q\) with a rational
  section over \(\F_q\) and coordinate degrees at most \(H\ge2\).
  For \(0<\varepsilon\le1/4\), if
  \begin{equation}\label{eq:local-Q}
    q\ge Q_{\mathrm{geom}}(H,\dim Y;\varepsilon),
  \end{equation}
  then \(F\) sends uniform coefficients to a law within \(\varepsilon/8\)
  of the uniform law \(\pi_Y\) on \(Y(\F_q)\) in total variation.
\end{corollary}
\begin{proof}
  Since \(H\ge2\), we have \(B\ge2\) and
  \(B^{-1/6}<9/10\). Since \((\dim Y)(H-1)\le B\),
  the field bound gives
  \begin{align*}
    \frac{(\dim Y)(H-1)}q
    &\le\frac{\varepsilon^4}{256}\le\frac{\varepsilon}{256},\\
    \mathcal E+\frac{2(\dim Y)(H-1)}q
    &\le\frac{\varepsilon^2 B^{-1/6}}{16}
    +\frac{7\varepsilon^4}{256}\\
    &\le\frac{9\varepsilon^2}{160}+\frac{7\varepsilon^4}{256}
    \le\left(\frac{63\varepsilon}{256}\right)^2.
  \end{align*}
  Proposition~\ref{prop:collision-to-mixing} bounds the distance by
  \(\tfrac12(\varepsilon/256+63\varepsilon/256)=\varepsilon/8\).
\end{proof}

\begin{lemma}[Returning to the stationary marginal]\label{lem:marginal-comparison}
  Let \(\nu\) be a law on the requested entries. Suppose that, for every
  initial state \(X\), a random length-\(T\) skeleton satisfies
  \[
    \Pr_\sigma[\|K_{\sigma,\mc{S}}(X,\cdot)-\nu\|_{\TV}>\delta]
    \le\rho.
  \]
  Then \(\|\pi_{\mc{S}}-\nu\|_{\TV}\le\rho+\delta\),
  \(d_{\mc{S}}(T)\le2(\rho+\delta)\), and
  \[
    \sup_X\Pr_\sigma[d_{\mc{S}}^\sigma(X)>\rho+2\delta]\le\rho.
  \]
\end{lemma}
\begin{proof}
  Averaging first over the skeleton gives error at most \(\rho+\delta\)
  from \(\nu\). Average also over a stationary initial state to obtain
  the bound for \(\pi_{\mc{S}}\), then use the triangle inequality.
\end{proof}

\section{Rank burn-in}\label{sec:routing-burnin}

Over sufficiently large fields, a fixed set of at least \(k\) rows
of a full-column-rank \(n\times k\) walk is typically full-rank after
\(O(n\log n)\) steps, provided \(k/n\) stays bounded away from \(1\).
We prove this by routing basis rows to the target set along disjoint
paths (Lemma~\ref{lem:routing-rank}). Appendix~\ref{app:small-field-burnin}
combines this with a small-field bound to prove the field-uniform
Theorem~\ref{thm:field-uniform-subset-burnin}.
Thus extending the mixing proof to all fields requires a field-uniform
replacement for the post-burn-in argument in
Sections~\ref{sec:rational-sections} and~\ref{sec:assembly}.

The \emph{time-expanded graph} of a skeleton
\[
  \sigma=((I_1\mapsto J_1),\ldots,(I_T\mapsto J_T)),
  \qquad I_t\ne J_t,
\]
is the directed graph with vertices \([n]\times\{0,\ldots,T\}\) and edges
\begin{align*}
  \underbrace{\bigl\{(r,t-1)\to(r,t):r\in[n],\ t\in[T]\bigr\}}_{\text{persistence edges}}
  \cup
  \underbrace{\bigl\{(I_t,t-1)\to(J_t,t):t\in[T]\bigr\}}_{\text{transmission edges}}.
\end{align*}

\begin{lemma}[disjoint-path routing]\label{lem:routing}
  Let \(1\le k<n\) and \(\mathcal B,\mathcal T\subseteq[n]\) satisfy
  \[
    |\mathcal B|=|\mathcal T|=k.
  \]
  For every \(\zeta\in(0,1/4]\) and every integer 
  \[
    T\ge8\frac{n^2}{n-k}\log\frac{n}{\zeta}
  \]
  the time-expanded graph of a random skeleton of length \(T\) contains, with
  probability at least \(1-\zeta\), \(k\) vertex-disjoint directed paths from
  \(\{(b,0):b\in\mathcal B\}\) to
  \(\mathcal T\times\{T\}\). Moreover, if
  \(\zeta\le(16n)^{-1}\), one may take
  \begin{equation}\label{eq:routing-explicit-length}
    T\ge
    \frac92\frac{n^2}{n-k}\log\frac n\zeta.
  \end{equation}
\end{lemma}

\begin{figure}[htbp]
  \centering
  \begin{tikzpicture}[
      x=1cm,
      y=1cm,
      >={Latex[length=1.8mm]},
      region one/.style={draw=blue!32,fill=blue!5,rounded corners=7pt},
      region two/.style={draw=orange!45!black,fill=orange!7,rounded corners=7pt,
      dashed},
      region three/.style={draw=teal!42!black,fill=teal!6,rounded corners=7pt,
      densely dotted},
      tree one/.style={draw=blue!72!black,line width=0.9pt,
      -{Latex[length=1.2mm]}},
      tree two/.style={draw=orange!88!black,line width=0.9pt,dashed,
      -{Latex[length=1.2mm]}},
      tree three/.style={draw=teal!70!black,line width=0.95pt,densely dotted,
      -{Latex[length=1.2mm]}},
      chosen one/.style={tree one,line width=1.75pt,-{Latex[length=1.8mm]}},
      chosen two/.style={tree two,line width=1.75pt,-{Latex[length=1.8mm]}},
      chosen three/.style={tree three,line width=1.75pt,-{Latex[length=1.8mm]}},
      candidate/.style={draw=black!30,line width=0.55pt,-{Latex[length=1.4mm]}},
      vertex/.style={circle,draw=black!55,fill=white,minimum size=4.2pt,
      inner sep=0pt},
      root/.style={vertex,minimum size=6.2pt,line width=0.8pt},
      target/.style={rectangle,rounded corners=0.6pt,draw=black!65,fill=white,
      minimum width=6.5pt,minimum height=6.5pt,inner sep=0pt},
      note/.style={font=\footnotesize,align=center}
    ]
    \path[region one] (0.25,4.35) rectangle (5.15,6.45);
    \path[region two] (0.25,2.00) rectangle (5.15,4.10);
    \path[region three] (0.25,-0.10) rectangle (5.15,1.75);

    \draw[black!35,dashed] (5.55,-0.25)--(5.55,6.60);
    \draw[decorate,decoration={brace,amplitude=3pt}]
    (0.25,6.65)--(5.45,6.65)
    node[midway,above=4pt,note] {grow disjoint parent-pointer trees};
    \draw[decorate,decoration={brace,amplitude=3pt}]
    (5.65,6.65)--(10.85,6.65)
    node[midway,above=4pt,note] {join color classes to the target family};

    \coordinate (r1) at (0.55,5.35);
    \coordinate (r2) at (0.55,3.05);
    \coordinate (r3) at (0.55,0.75);

    \draw[tree one]
    (r1)--(1.15,5.82)
    (r1)--(1.85,4.78)
    (1.15,5.82)--(2.05,6.12)
    (1.15,5.82)--(2.75,5.52)
    (2.05,6.12)--(3.55,6.22)
    (3.55,6.22)--(4.65,5.95)
    (2.75,5.52)--(4.55,5.42)
    (1.85,4.78)--(2.65,4.55)
    (2.65,4.55)--(3.45,4.85)
    (3.45,4.85)--(4.75,4.62);
    \draw[tree two]
    (r2)--(1.10,3.52)
    (1.10,3.52)--(2.00,3.78)
    (2.00,3.78)--(3.50,3.92)
    (2.00,3.78)--(4.25,3.57)
    (r2)--(2.05,2.55)
    (2.05,2.55)--(2.80,2.25)
    (2.80,2.25)--(4.45,2.18)
    (2.05,2.55)--(4.00,2.72)
    (1.10,3.52)--(2.90,3.18)
    (2.90,3.18)--(4.60,3.32);
    \draw[tree three]
    (r3)--(1.05,1.18)
    (1.05,1.18)--(1.75,1.48)
    (1.75,1.48)--(3.15,1.58)
    (3.15,1.58)--(4.60,1.42)
    (r3)--(1.75,0.32)
    (1.75,0.32)--(2.55,0.12)
    (2.55,0.12)--(4.25,0.28)
    (1.75,0.32)--(3.35,0.60)
    (1.05,1.18)--(2.45,0.92)
    (2.45,0.92)--(4.55,0.88);

    \foreach \x/\y in {0.55/5.35,1.15/5.82,1.85/4.78,2.05/6.12,
      2.75/5.52,3.55/6.22,4.65/5.95,4.55/5.42,2.65/4.55,
    3.45/4.85,4.75/4.62}
    \node[vertex] at (\x,\y) {};
    \foreach \x/\y in {0.55/3.05,1.10/3.52,2.00/3.78,3.50/3.92,
      4.25/3.57,2.05/2.55,2.80/2.25,4.45/2.18,4.00/2.72,
    2.90/3.18,4.60/3.32}
    \node[vertex] at (\x,\y) {};
    \foreach \x/\y in {0.55/0.75,1.05/1.18,1.75/1.48,3.15/1.58,
      4.60/1.42,1.75/0.32,2.55/0.12,4.25/0.28,3.35/0.60,
    2.45/0.92,4.55/0.88}
    \node[vertex] at (\x,\y) {};
    \node[root] at (r1) {};
    \node[root] at (r2) {};
    \node[root] at (r3) {};

    \node[note,anchor=east] at (0.10,5.35) {\(b_1\)};
    \node[note,anchor=east] at (0.10,3.05) {\(b_2\)};
    \node[note,anchor=east] at (0.10,0.75) {\(b_3\)};

    \node[target] (s1) at (10.55,5.45) {};
    \node[target] (s2) at (10.55,3.10) {};
    \node[target] (s3) at (10.55,0.75) {};
    \draw[candidate] (4.65,5.95)--(s2);
    \draw[candidate] (4.75,4.62)--(s3);
    \draw[candidate] (3.50,3.92)--(s1);
    \draw[candidate] (4.45,2.18)--(s3);
    \draw[candidate] (4.60,1.42)--(s1);
    \draw[candidate] (4.25,0.28)--(s2);

    \draw[chosen one]
    (r1)--(1.15,5.82)--(2.75,5.52)--(4.55,5.42)--(s1);
    \draw[chosen two]
    (r2)--(1.10,3.52)--(2.90,3.18)--(4.60,3.32)--(s2);
    \draw[chosen three]
    (r3)--(1.05,1.18)--(2.45,0.92)--(4.55,0.88)--(s3);

    \node[note,anchor=west] at (10.78,5.45) {\(s_1\)};
    \node[note,anchor=west] at (10.78,3.10) {\(s_2\)};
    \node[note,anchor=west] at (10.78,0.75) {\(s_3\)};
    \draw[decorate,decoration={brace,amplitude=3pt,mirror}]
    (11.35,5.85)--(11.35,0.35)
    node[midway,right=5pt,note]
    {reserved targets\\\(\mc{R}=\mathcal T\setminus\mathcal B\)};

    \draw[-{Latex[length=2mm]},black!55]
    (0.40,-0.70)--(10.90,-0.70) node[right,note] {time};
    \draw[black!45] (0.55,-0.78)--(0.55,-0.62)
    node[below=3pt,note] {\(0\)};
    \draw[black!45] (5.55,-0.78)--(5.55,-0.62)
    node[below=3pt,note] {growth cutoff};
    \draw[black!45] (10.55,-0.78)--(10.55,-0.62)
    node[below=3pt,note] {\(T\)};
  \end{tikzpicture}
  \caption{Routing in the time-expanded graph. The bold edges show the perfect
    matching and the resulting root-to-target paths. Persistence edges, unused
    transmission edges, and the paths through \(\mathcal B\cap\mathcal T\) are
  omitted.}
  \label{fig:routing-schematic}
\end{figure}
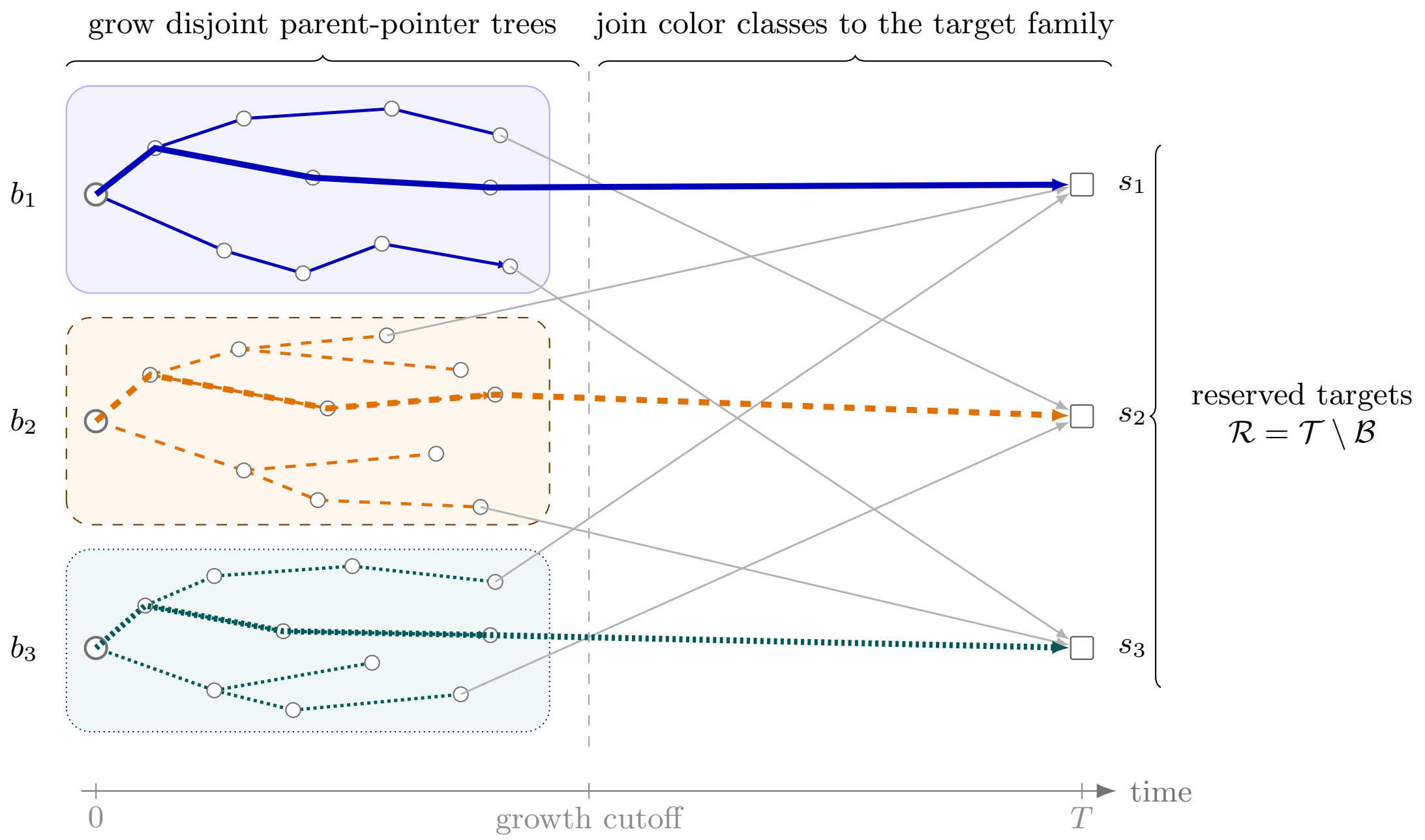

\begin{proof}
  For \(n=2\), the failure probability is at most \(2^{-T}\), which gives
  both assertions. Assume henceforth that \(n\ge3\).
  Note that \(\log(n/\zeta)>2\). When
  \(\zeta\le(16n)^{-1}\), assume only the shorter bound on \(T\).

  Use persistence paths for \(\mathcal B\cap\mathcal T\), and let
  \(\mc{R}=\mathcal T\setminus\mathcal B\). If \(|\mc{R}|=0\),
  these paths suffice. Otherwise, we construct disjoint paths through a two-stage process: we grow a disjoint tree of reachable nodes from each source, then match the trees to targets by connecting them with edges.

  \paragraph{Growing trees.}
  Assign distinct colors to the rows in \(\mathcal B\setminus\mathcal T\). The
  workspace \([n]\setminus\mathcal T\) has size \(n-k\). Cap each color
  class at
  \[
    m=\max\left\{1,\floor{\frac{n-k}{2|\mc{R}|}}\right\}.
  \]
  Go sequentially through the transvections. When a transvection \(i\mapsto j\) has \(i\) in a
  color class with fewer than \(m\)
  elements and \(j\notin \mathcal T\) is uncolored, assign \(j\) the color of \(i\) and
  record the corresponding transmission edge as its parent
  edge. Each color class therefore forms a tree rooted at its row in
  \(\mathcal B\setminus\mathcal T\).

  When \(m=1\), set \(P_1=0\). Otherwise, \((n-k)/3<m|\mc{R}|\le(n-k)/2\), and
  a class of size \(z<m\) grows with conditional probability at least \(zp\), where
  \[
    p=\frac{n-k-m|\mc{R}|}{n(n-1)},\qquad P_1=\ceil{\frac{\log(n/\zeta)}p}.
  \]
  Its growth time is stochastically dominated by the sum of independent geometric
  waiting times with probabilities \(p,2p,\ldots,(m-1)p\).
  Reversing the order of the waiting times gives the time to collect \(m-1\) coupons,
  each drawn with probability \(p\) per step. A union bound over coupons
  and classes bounds the chance that any color class fails to reach size \(m\) over the first \(P_1\) steps by
  \[
    |\mc{R}|(m-1)(1-p)^{P_1}\le\frac{|\mc{R}|(m-1)\zeta}{n}<\frac\zeta2.
  \]

  \paragraph{An intermediate bound.}
  Let the remaining number of steps be \(P_2=T-P_1\) and let \(a=P_2m|\mc{R}|/(n(n-1))\). The next phase will require two bounds on \(e^{-a}\), so we prove them here. We first show that
  \[
    a\ge\max\{1,5|\mc{R}|/n\}
    \begin{cases}
      \log(n/\zeta),&\zeta\le(16n)^{-1},\\
      2\log(n/\zeta),&\text{otherwise}.
    \end{cases}
  \]
  If \(m=1\), then \(P_1=0\) and \(a=T|\mc{R}|/(n(n-1))\).
  Since \(n-k<4|\mc{R}|\) and \(n-k\le n-|\mc{R}|\),
  \(4|\mc{R}|/(n-k)\ge\max\{1,5|\mc{R}|/n\}\), so the claim follows
  from the stated bounds on \(T\).
  If \(m\ge2\), then \(5|\mc{R}|\le n\), so the maximum is \(1\). The inequalities
  \[
    \frac1{n-k-m|\mc{R}|}+\frac1{m|\mc{R}|}\le\frac9{2(n-k)},\qquad
    \frac1{n-k-m|\mc{R}|}+\frac2{m|\mc{R}|}\le\frac{15}{2(n-k)}
  \]
  bound \(P_1\) after multiplication by \(n(n-1)\log(n/\zeta)\) and hence lower-bound \(P_2=T-P_1\) as required. The slack from
  \(n(n-1)<n^2\) absorbs the ceiling in \(P_1\).
  We next apply \(x\log(n/\zeta)\ge\log(n/\zeta)+2\log x\), valid for
  \(x\ge1\), with \(x=\max\{1,5|\mc{R}|/n\}\), and exponentiate to obtain:
  \[
    e^{-a}\le\frac1{\max\{n,5|\mc{R}|\}^2}
    \begin{cases}
      \zeta n,&\zeta\le(16n)^{-1},\\
      \zeta^2,&\text{otherwise}
    \end{cases}
    \le\min\left\{\frac{\zeta}{5|\mc{R}|},\frac1{400|\mc{R}|^2}\right\}.
  \]

  \paragraph{Matching trees to targets.}

  Condition on all color classes reaching size \(m\), and freeze them. During the
  remaining \(P_2\) steps, add an edge from a color class to a reserved target
  \(j\) whenever a transvection \(i\mapsto j\) occurs with \(i\) in that
  class.

  We show that a matching of color classes to targets occurs with high probability. If Hall's
  condition fails, some \(i\) colors have edges into only \(i-1\) reserved
  targets. A union bound over these two sets gives
  \begin{align*}
    \Pr[\text{Hall fails}]
    &\le
    \sum_{i=1}^{|\mc{R}|}
    \binom{|\mc{R}|}{i}\binom{|\mc{R}|}{i-1}
    \left(1-\frac{(mi)(|\mc{R}|-i+1)}{n(n-1)}\right)^{P_2}\\
    &\le
    \sum_{i=1}^{|\mc{R}|}
    \binom{|\mc{R}|}{i}\binom{|\mc{R}|}{i-1}
    e^{-a i(|\mc{R}|-i+1)/|\mc{R}|}.
  \end{align*}
  For \(|\mc{R}|\ge3\), group terms by \(j=\min\{i,|\mc{R}|-i+1\}\).
  Using \(\binom{x}{y}\le x^{y(x-y)/(x-1)}\) bounds each summand by
  \begin{align*}
    &|\mc{R}|e^{-a}
    \left(|\mc{R}|e^{-a(|\mc{R}|-1)/(2|\mc{R}|)}\right)^{2(j-1)(|\mc{R}|-j)/(|\mc{R}|-1)}\\
    &\quad\le |\mc{R}|e^{-a}
    \left(\frac{(20|\mc{R}|)^{1/|\mc{R}|}}{20}\right)^{2(j-1)(|\mc{R}|-j)/(|\mc{R}|-1)}\\
    &\quad\le |\mc{R}|e^{-a}5^{-2(j-1)(|\mc{R}|-j)/(|\mc{R}|-1)}
    \le |\mc{R}|e^{-a}5^{1-j}.
  \end{align*}
  Summing the geometric series therefore gives
  \[
    \Pr[\text{Hall fails}]
    \le2|\mc{R}|e^{-a}
    \sum_{j=1}^{\floor{(|\mc{R}|+1)/2}}5^{1-j}
    \le\frac52|\mc{R}|e^{-a}
    \le \frac\zeta2
  \]
  as desired. For \(|\mc{R}|\le2\), the original Hall sum is at most
  \(2|\mc{R}|e^{-a}\le2\zeta/5\)
  by the same estimate.

  \paragraph{Concluding the proof.} Conditional on successful growth, a perfect matching therefore exists
  except with probability \(\zeta/2\). For each matched class, follow
  its tree to the matching edge, then use persistence edges to reach time
  \(T\). These paths and those through \(\mathcal B\cap\mathcal T\)
  are disjoint. Including growth failure gives total failure probability
  at most \(\zeta\).
\end{proof}

\begin{lemma}[routed paths give a nonzero minor]\label{lem:routing-determinant}
  Fix a full-column-rank \(X_0\in\F_q^{n\times k}\) and a skeleton \(\sigma\) of length \(T\). Let \(\mathcal B\) index \(k\) linearly independent rows of \(X_0\), and suppose the time-expanded graph contains
  vertex-disjoint paths from \(\{(b,0):b\in\mathcal B\}\) to distinct final
  target rows \(s_1,\ldots,s_k\).  Then
  \[
    \det X_T[\{s_1,\ldots,s_k\}]
  \]
  is a nonzero multiaffine polynomial of degree at most \(T\).
\end{lemma}

\begin{proof}
  Order each row set increasingly. For a \(k\)-set \(S\) and the
  transvection \(I_t\mapsto J_t\), multilinearity gives
  \begin{equation}\label{eq:routing-minor-recurrence}
    \det X_t[S]=
    \begin{cases}
      \det X_{t-1}[S]+\epsilon_{S,t}a_t
      \det X_{t-1}[(S\setminus\{J_t\})\cup\{I_t\}],
      &J_t\in S,\ I_t\notin S,\\
      \det X_{t-1}[S],&\text{otherwise},
    \end{cases}
  \end{equation}
  where \(\epsilon_{S,t}\in\{1,-1\}\) comes from reordering the rows. Every
  \(\det X_t[S]\) is multiaffine in \(a_1,\ldots,a_t\), with degree at most
  \(t\).

  Fixing the transmission times determines a unique path backward through the recurrence of~\eqref{eq:routing-minor-recurrence}, which gives that the coefficient of
  \[
    \prod_{\substack{1\le t\le T:\\
    \text{the path family uses }I_t\mapsto J_t\text{ at time }t}}a_t
  \]
  in \(\det X_T[\{s_1,\ldots,s_k\}]\) is \(\pm\det X_0[\mathcal B]\ne0\).
\end{proof}

\begin{lemma}[routing implies a fixed-subset rank bound]\label{lem:routing-rank}
  Let \(\zeta\in(0,1/4]\), \(1\leq k<n\), and let
  \[
    T=\left\lceil8\frac{n^2}{n-k}\log\frac n\zeta\right\rceil.
  \]
  For every fixed \(S\subseteq[n]\) with \(|S|\ge k\), prime power \(q\), and
  integer \(t\ge T\),
  \begin{equation}\label{eq:routing-rank-bound}
    \sup_{X_0\in \F_q^{n\times k}: \rank X_0=k}
    \Pr_{X_0}[\rank X_t[S]<k]\le\zeta+\frac Tq.
  \end{equation}
  The same conclusion holds with the shorter length
  \eqref{eq:routing-explicit-length} when \(\zeta\le(16n)^{-1}\).
\end{lemma}

\begin{proof}
  Fix a \(k\)-set \(\mathcal T\subseteq S\) and a basis-row set
  \(\mathcal B\) of \(X_0\). Lemma~\ref{lem:routing} supplies paths from
  \(\mathcal B\) to \(\mathcal T\) by time \(T\) with probability at least
  \(1-\zeta\). Conditional on such a skeleton,
  Lemma~\ref{lem:routing-determinant} and Schwartz--Zippel~\cite{Schwartz1980,Zippel1979}
  bound the probability of a singular \(\mathcal T\)-minor by \(T/q\).
  This proves~\eqref{eq:routing-rank-bound} at time \(T\); Lemma~\ref{lem:fixed-time-extension} extends it to later times.
\end{proof}

\section{Constructing sections and proving mixing}\label{sec:assembly}

We first construct sections for entry patterns with a bounded number of entries
per row, then extend the
construction to arbitrary entry patterns. Dense patterns use a separate
section for the full group. We combine their time estimates to prove
Theorem~\ref{thm:main}.

From this point on, we typically measure two types of error: \(\eta\), the probability of a bad event in the skeleton, and \(u\), the TV error of the skeleton-conditional distribution given that no bad event occurs in the skeleton.

\subsection{Bounded row demands: ordered target blocks}\label{sec:rectangular}

Suppose \(|\mc{S}_i|\le k<n\). We prescribe the requested entries
one block of rows at a time. Rows outside the current block supply sources;
once a block is completed, later blocks never change it.

Choose disjoint nonempty target blocks containing every row with
\(\mc{S}_i\ne\varnothing\):
\begin{equation}\label{eq:phase-order}
  \{i:\mc{S}_i\ne\varnothing\}=\bigsqcup_{v=0}^{p-1}R_v,
  \qquad p\ge1,\qquad
  |R_v|+\max_{i\in R_v}|\mc{S}_i|\le n.
\end{equation}
In phase \(v\), select, for each requested row \(i\in R_v\),
\(|\mc{S}_i|\) distinct sources outside \(R_v\), and one update from
each source to \(i\). Set all other coefficients to zero.
The selected updates in a phase commute and fix their source rows.

Write \(R_{<v}=\bigsqcup_{u<v}R_u\). At phase \(v\), rows in \(R_{<v}\)
are known from earlier recovery; all other rows retain their initial values.

\begin{figure}[!ht]
  \centering
  \begin{tikzpicture}[
      x=1cm,
      y=1cm,
      >={Latex[length=1.7mm]},
      source/.style={draw=black!55,line width=0.65pt,rounded corners=3pt,
      fill=black!4},
      past/.style={fill=teal!11},
      future/.style={fill=black!4},
      current/.style={draw=blue!65!black,line width=0.9pt,rounded corners=3pt,
      fill=blue!10},
      source dot/.style={circle,draw=black!60,fill=white,minimum size=4.2pt,
      inner sep=0pt},
      target dot/.style={rectangle,rounded corners=0.5pt,draw=blue!70!black,
      fill=white,minimum width=5.5pt,minimum height=5.5pt,inner sep=0pt},
      update/.style={-{Latex[length=1.6mm]},draw=black!57,line width=0.65pt},
      phase/.style={font=\small},
      annotation/.style={font=\scriptsize,align=center}
    ]
    \node[phase] at (2.35,4.85) {phase \(0\)};
    \path[source] (0.15,1.05) rectangle (2.75,4.25);
    \path[current] (3.15,1.05) rectangle (4.55,4.25);
    \node[annotation] at (1.45,3.95) {\([n]\setminus R_0\)};
    \node[annotation] at (3.85,3.95) {\(R_0\)};
    \node[source dot] (g1) at (0.60,3.15) {};
    \node[source dot] (g2) at (1.30,2.80) {};
    \node[source dot] (g3) at (2.05,3.25) {};
    \node[source dot] (g4) at (0.75,2.05) {};
    \node[source dot] (g5) at (1.55,1.90) {};
    \node[source dot] (g6) at (2.25,2.25) {};
    \node[target dot] (d01) at (3.55,3.10) {};
    \node[target dot] (d02) at (4.12,2.58) {};
    \node[target dot] (d03) at (3.55,1.95) {};
    \draw[update] (g2)--(d01);
    \draw[update] (g3)--(d01);
    \draw[update] (g1)--(d02);
    \draw[update] (g6)--(d02);
    \draw[update] (g4)--(d03);
    \draw[update] (g5)--(d03);

    \node[phase] at (7.58,4.85) {phase \(v\), \(1\le v<p-1\)};
    \path[past] (5.35,2.65) rectangle (8.00,4.25);
    \path[future] (5.35,1.05) rectangle (8.00,2.65);
    \path[draw=black!55,line width=0.65pt,rounded corners=3pt]
    (5.35,1.05) rectangle (8.00,4.25);
    \draw[black!35] (5.35,2.65)--(8.00,2.65);
    \path[current] (8.40,1.05) rectangle (9.80,4.25);
    \node[annotation] at (6.675,3.95) {\(R_{<v}\)};
    \node[annotation] at (6.675,1.35) {\(R_{>v}\)};
    \node[annotation] at (9.10,3.95) {\(R_v\)};
    \node[source dot] (p1) at (5.72,3.55) {};
    \node[source dot] (p2) at (6.48,2.92) {};
    \node[source dot] (p3) at (7.46,3.45) {};
    \node[source dot] (f1) at (5.65,1.55) {};
    \node[source dot] (f2) at (6.72,2.28) {};
    \node[source dot] (f3) at (7.35,1.72) {};
    \node[target dot] (dv1) at (9.20,3.35) {};
    \node[target dot] (dv2) at (8.72,2.55) {};
    \node[target dot] (dv3) at (9.30,1.75) {};
    \draw[update] (p1)--(dv1);
    \draw[update] (f2)--(dv1);
    \draw[update] (p3)--(dv2);
    \draw[update] (f1)--(dv2);
    \draw[update] (p2)--(dv3);
    \draw[update] (f3)--(dv3);

    \node[phase] at (12.83,4.85) {phase \(p-1\)};
    \path[past,draw=black!55,line width=0.65pt,rounded corners=3pt]
    (10.60,1.05) rectangle (13.25,4.25);
    \path[current] (13.65,1.05) rectangle (15.05,4.25);
    \node[annotation] at (11.925,3.95) {\(R_{<p-1}\)};
    \node[annotation] at (14.35,3.95) {\(R_{p-1}\)};
    \node[source dot] (z1) at (10.98,3.60) {};
    \node[source dot] (z2) at (11.95,3.30) {};
    \node[source dot] (z3) at (12.65,2.75) {};
    \node[source dot] (z4) at (11.18,1.55) {};
    \node[source dot] (z5) at (12.10,1.90) {};
    \node[source dot] (z6) at (12.82,1.30) {};
    \node[target dot] (last1) at (14.55,3.45) {};
    \node[target dot] (last2) at (13.95,2.45) {};
    \node[target dot] (last3) at (14.55,1.55) {};
    \draw[update] (z1)--(last1);
    \draw[update] (z5)--(last1);
    \draw[update] (z2)--(last2);
    \draw[update] (z4)--(last2);
    \draw[update] (z3)--(last3);
    \draw[update] (z6)--(last3);

    \draw[-{Latex[length=2mm]},black!55]
    (0.10,0.30)--(15.35,0.30) node[right,annotation] {time};
    \node[annotation,below=3pt] at (2.35,0.30) {\(\sigma_0\)};
    \node[annotation] at (4.95,2.65) {\(\cdots\)};
    \node[annotation,below=3pt] at (7.58,0.30) {\(\sigma_v\)};
    \node[annotation] at (10.20,2.65) {\(\cdots\)};
    \node[annotation,below=3pt] at (12.83,0.30) {\(\sigma_{p-1}\)};
  \end{tikzpicture}
  \caption{Ordered target blocks. Each row \(i\in R_v\) selects
    \(|\mc{S}_i|\) sources outside \(R_v\). Earlier blocks have known
    recovered rows; later blocks retain their initial rows. Solving on
    columns \(\mc{S}_i\) sets the requested entries, and later phases
    never change a completed block. The schematic shows a complete
  partition; unused rows also remain available as sources.}
  \label{fig:multiphase-schematic}
\end{figure}
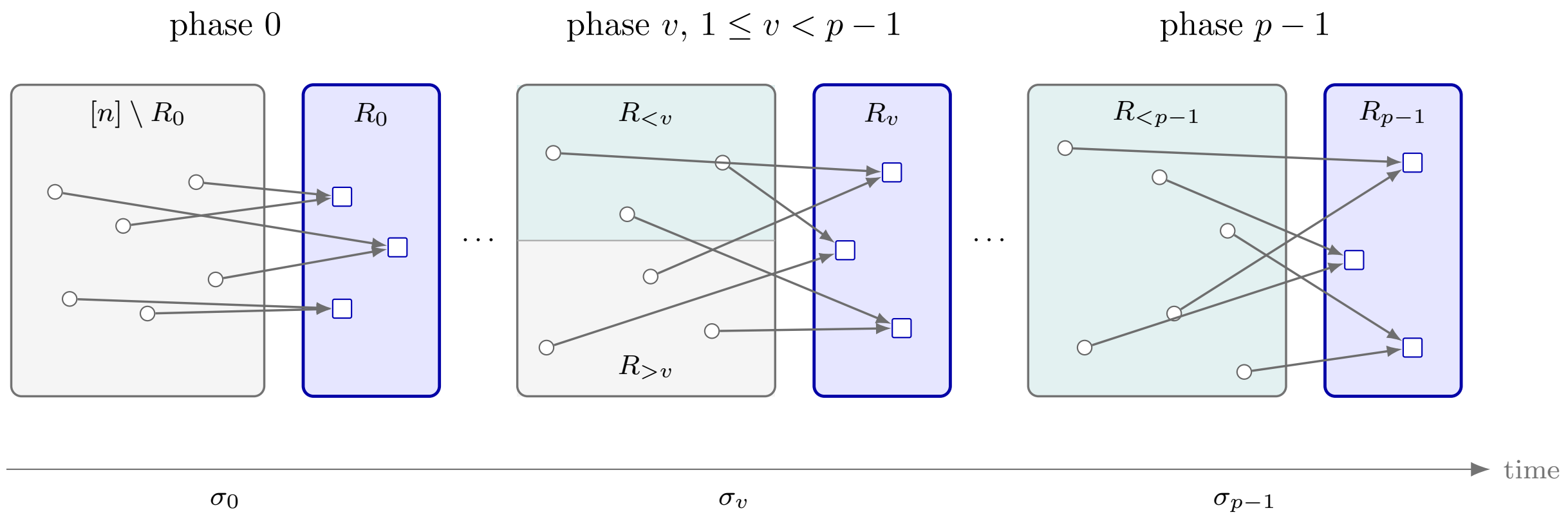

\subsubsection{Recovering the selected coefficients}
\label{sec:coefficient-slice} 

\begin{proposition}[Recovering requested entries]\label{prop:birational}
  Let \(V\) be an initial matrix over a field \(K\). For each requested
  row \(i\), let \(S_i\) be its selected source set in the preceding
  construction. If every \(V[S_i,\mc{S}_i]\) is nonsingular, the map
  from the \(|\mc{S}|\) selected coefficients to the requested endpoint
  entries is birational over \(K\).
\end{proposition}
\begin{proof}
  Recover the selected coefficients block by block. Just before row \(i\)
  is updated, its current row and all its source rows are known. Let \(V_i\) denote
  this current matrix and put
  \begin{equation}\label{eq:hybrid-source-matrix}
    M_i=V_i[S_i,\mc{S}_i].
  \end{equation}
  For a requested endpoint \(y\) with rows \((y_i)_{i\in [n]}\), choose the coefficient vector
  \begin{equation}\label{eq:coefficient-inverse}
    (y_i-V_i[i,\mc{S}_i])M_i^{-1}.
  \end{equation}
  This sets the requested entries and determines the entire row for
  later phases.

  At \(y=V|_{\mc{S}}\), these formulas return zero coefficients in
  every phase and \(M_i=V[S_i,\mc{S}_i]\). Their denominators are
  therefore nonzero rational functions of \(y\). The formulas give the
  unique selected coefficients on a nonempty open set.
\end{proof}

The same recovery can be applied after earlier transvections, using their coefficients to determine the current matrix. In our applications, these earlier coefficients are sometimes themselves left indeterminate. To check that the recovery denominators are not identically zero, we exhibit one target at which all coefficients being recovered are zero and all source minors are nonsingular. We then choose the indeterminate coefficients in the infinite field \(K(y)\) so that every denominator remains nonzero. All coefficients are now rational functions of \(y\) over \(K\), giving a section over \(K\).

\subsubsection{Collecting sources and bounding time}\label{app:time-constant}

\begin{corollary}[Collecting distinct sources]\label{cor:target-source-collection}
  Let \(S,R\subseteq[n]\) be disjoint nonempty source and target sets, let
  \(1\le s\le |S|\) be an integer, and let \(L>0\). A random skeleton
  of length at least \(\ceil{\mu+\sqrt{2\mu vL}+vL}\) contains updates
  from at least \(s\) distinct rows of \(S\) into \(R\), except with
  probability at most \(e^{-L}\), where
  \[
    \mu=\frac{n(n-1)}{|R|}
    \log\frac{|S|+1/2}{|S|-s+1/2},\qquad
    v=\frac{n(n-1)}{|R|(|S|-s+1)}.
  \]
  The receiving rows in \(R\) need not be distinct.

  More generally, fix \(p\) consecutive phases, each with one or more
  collection requirements \((S,R,s)\) as above, and let \(N\) be the
  total number of requirements. Different requirements may use overlapping
  row sets. Let \(\mu_a\) be the largest \(\mu\) among phase \(a\)'s
  requirements, and let \(v_a\) be their largest \(v\). Run that phase for
  \[
    \ceil{\mu_a+\sqrt{2\mu_av_aL}+v_aL}
  \]
  independent uniform transvections. All \(N\) requirements are then
  met within their assigned phases except with probability at most
  \(Ne^{-L}\), and the total number of steps is at most
  \[
    \sum_{a=1}^p\mu_a
    +\sqrt{2L\left(\sum_{a=1}^p\mu_a\right)
    \left(\sum_{a=1}^p v_a\right)}
    +L\sum_{a=1}^p v_a+p.
  \]
\end{corollary}
\begin{proof}
  After \(j<s\) distinct sources have appeared, the next new source arrives
  with probability \(|R|(|S|-j)/(n(n-1))\).
  Convexity of \(1/t\) gives
  \[
    \sum_{j=|S|-s+1}^{|S|}\frac1j
    \le\int_{|S|-s+1/2}^{|S|+1/2}\frac{dt}{t}
    =\log\frac{|S|+1/2}{|S|-s+1/2}.
  \]
  Thus \(\mu\) bounds the mean collection time, and
  Lemma~\ref{lem:geometric-waiting}, with \(p_*^{-1}=v\), gives
  the single-collection bound.
  For several phases, union-bound over all \(N\) requirements and use
  Cauchy--Schwarz to sum the square-root terms. Rounding adds at most one
  step per phase.
\end{proof}

For a nonempty target block \(R\), let \(S=[n]\setminus R\), put
\(k_R=\max_{i\in R}|\mc{S}_i|\) and define
\begin{align*}
  s_R&=\log\frac{|S|+1/2}{|S|-k_R+1/2},\\
  N_R(\eta)&=\ceil{n(n-1)\left(
      s_R+\sqrt{\frac{2s_R\log(4n/\eta)}{|S|-k_R+1}}
  +\frac{\log(4n/\eta)}{|S|-k_R+1}\right)}.
\end{align*}
Set \(N_R(\eta)=0\) when no entry is requested in \(R\).
Otherwise Corollary~\ref{cor:target-source-collection}, with source set \(S\)
and target set \(\{i\}\), gives every row \(i\in R\) enough sources.
Bounding \(k_R\) by \(k\) gives the uniform
estimates below. For a balanced partition of all rows into \(p\)
blocks, call \(p\) admissible if \(2\le p\le n\) and \(n/p\le n-k\).
Put
\[
  T_{\rm burn}(\eta)=
  \ceil{\frac{9n^2}{2(n-k)}\log\frac{4n^2}{\eta}},
  \qquad
  T_p(\eta)=\sum_{v=0}^{p-1}N_{R_v}(\eta).
\]
For partial patterns, partition only rows with requested entries and
minimize the sum of the \(N_R(\eta)\)'s. One block suffices when its
complement has enough sources; unused blocks cost no phase.
The balanced complete partitions below give uniform upper bounds.

\begin{proposition}[Mixing from a target schedule]\label{prop:exact-density-constant}
  Let \(0<\eta,u\le1/4\), and take any target partition from
  Section~\ref{sec:rectangular}, including a single block when it fits.
  Put \(T=T_{\rm burn}(\eta)+\sum_v N_{R_v}(\eta)\). If
  \(q\ge Q_{\rm geom}(H(T,\eta),|\mc{S}|;u)\), then, for every fixed
  \(X\in\mathrm{SL}_n(\F_q)\),
  \begin{equation}\label{eq:rectangular-conditional-mixing}
    \Pr_{\sigma:|\sigma|=T}\left[
      \|K_{\sigma,\mc{S}}(X,\cdot)-\Unif(\F_q^{\mc{S}})\|_{\TV}>u/8
    \right]\le9\eta/16.
  \end{equation}
  If \(\mc{S}=[n]\times C\), the same estimate holds with
  \(\pi_{\mc{S}}\) in place of the product law. For columns,
  \(u=\eta\), and any admissible complete partition,
  \begin{equation}\label{eq:exact-density-time}
    t_{\mathrm{mix}}^{(k)}(\eta)\le T_{\rm burn}(\eta)+T_p(\eta).
  \end{equation}
  For \(n\ge256\), the schedule lengths satisfy
  \begin{align}
    T_{\rm burn}(\eta)+T_2(\eta)
    &\le8nk+25n\log n+
    \left(\frac9{2(1-k/n)}+\frac{20}{3(1-2k/n)}\right)
    n\log\frac1{4\eta},
    &&k\le n/10,\label{eq:two-block-affine}\\
    T_{\rm burn}(\eta)+T_4(\eta)
    &\le16nk+34n\log n+
    \frac{37n^2}{2(n-k)}\log\frac1{4\eta},
    &&k\le n/2.\label{eq:four-block-affine}
  \end{align}
\end{proposition}
\begin{proof}
  The source collections succeed except with probability at most \(\eta/4\).
  Conditional on the selected source sets \(S_i\), apply
  Lemmas~\ref{lem:routing} and~\ref{lem:routing-determinant}
  to the burn-in skeleton of length \(T_{\rm burn}(\eta)\) acting on
  \(X[:,\mc{S}_i]\), with
  \(\zeta=\eta/(4n)\) for each nonempty \(\mc{S}_i\).
  There are at most \(n\) tests, all of size at most \(k\), and the
  source sets are independent of the burn-in skeleton. Except with probability
  \(\eta/4\), every \((AX)[S_i,\mc{S}_i]\) is nonsingular over
  the field of burn-in coefficients, where \(A\) is the burn-in product.
  Proposition~\ref{prop:birational} and specialization of those
  coefficients give the required section.

  The degree estimate fails with probability at most \(\eta/16\).
  Corollary~\ref{cor:section-mixing} now proves
  \eqref{eq:rectangular-conditional-mixing}.

  For \(\mc{S}=[n]\times C\), let
  \(\mathcal R\subseteq\F_q^{n\times|C|}\) be the full-rank matrices.
  Since \(\pi_{\mc{S}}=\Unif(\mathcal R)\), every law \(\mu\)
  supported on \(\mathcal R\) satisfies
  \begin{align*}
    \|\mu-\pi_{\mc{S}}\|_{\TV}
    &=\sum_{M\in\mathcal R}\bigl(\mu(M)-|\mathcal R|^{-1}\bigr)_+\\
    &\le\sum_{M\in\mathcal R}\bigl(\mu(M)-q^{-|\mc{S}|}\bigr)_+
    =\|\mu-\Unif(\F_q^{\mc{S}})\|_{\TV},
  \end{align*}
  where \(z_+=\max\{z,0\}\). Each conditional column law is supported
  on \(\mathcal R\), so the same estimate holds with
  \(\pi_{\mc{S}}\) in place of the product law. Averaging with
  \(u=\eta\) gives distance at most \(11\eta/16\).

  We now bound \(T_{\mathrm{burn}}(\eta)+T_p(\eta)\) for the two complete
  partitions, assuming \(n\ge256\).

  \emph{Two blocks when \(k\le n/10\).}
  Since \(|S|+1/2\ge n/2\), convexity gives
  \[
    s_R\le\log\frac n{n-2k}\le10\log(5/4)\,\frac{k}{n}.
  \]
  Also \(|S|-k+1\ge(n-2k)/2\). Apply
  \(u+\sqrt{2uv}+v\le(7/4)u+(5/3)v\) in the phase-length formula
  and sum over the two blocks:
  \begin{equation}\label{eq:two-block-phase-bound}
    T_2(\eta)\le35\log(5/4)\,nk+
    \frac{20n^2}{3(n-2k)}\log\frac{4n}{\eta}+2.
  \end{equation}
  Add burn-in and use \(\log16\le\tfrac12\log n\). The coefficient
  of \(n\log n\) is bounded by
  \[
    \frac{45}{4(1-k/n)}+\frac{10}{1-2k/n}\le25.
  \]
  Since \(k\ge1\) and
  \[
    35\log(5/4)+\frac3n<8
    \qquad(n\ge256),
  \]
  this proves~\eqref{eq:two-block-affine}.

  \emph{Four blocks when \(k\le n/2\).}
  Each block has
  \(|R|\le\ceil{n/4}\), so \(|S|-k+1/2\ge(n-k)/2\).
  Hence, by convexity,
  \begin{equation}\label{eq:midpoint-harmonic-bound}
    s_R\le\log\frac{n+k}{n-k}\le2\log3\,\frac{k}{n}.
  \end{equation}
  Apply \(u+\sqrt{2uv}+v\le(5/3)u+(7/4)v\) in the formula for
  \(N_R\). Since \(\log16\le\tfrac12\log n\), summing the four
  phase lengths and burn-in gives
  \[
    T_{\mathrm{burn}}(\eta)+T_4(\eta)
    \le\frac{40}{3}\log3\,nk+
    \frac{n^2}{n-k}\left(\frac{129}{4}\log n+\frac{37}{2}\log\frac1{4\eta}\right)+5.
  \]
  Subtract the right side of~\eqref{eq:four-block-affine}. The resulting
  expression is convex in \(k\), so it suffices to check \(k=0\) and
  \(k=n/2\). The first endpoint is immediate; the second follows from
  \[
    \frac{40}{3}\log3+\frac{61\log n}{n}+\frac{10}{n^2}<16
    \qquad(n\ge256).
  \]
\end{proof}

\subsection{Arbitrary entry patterns}\label{sec:entry-mixing}

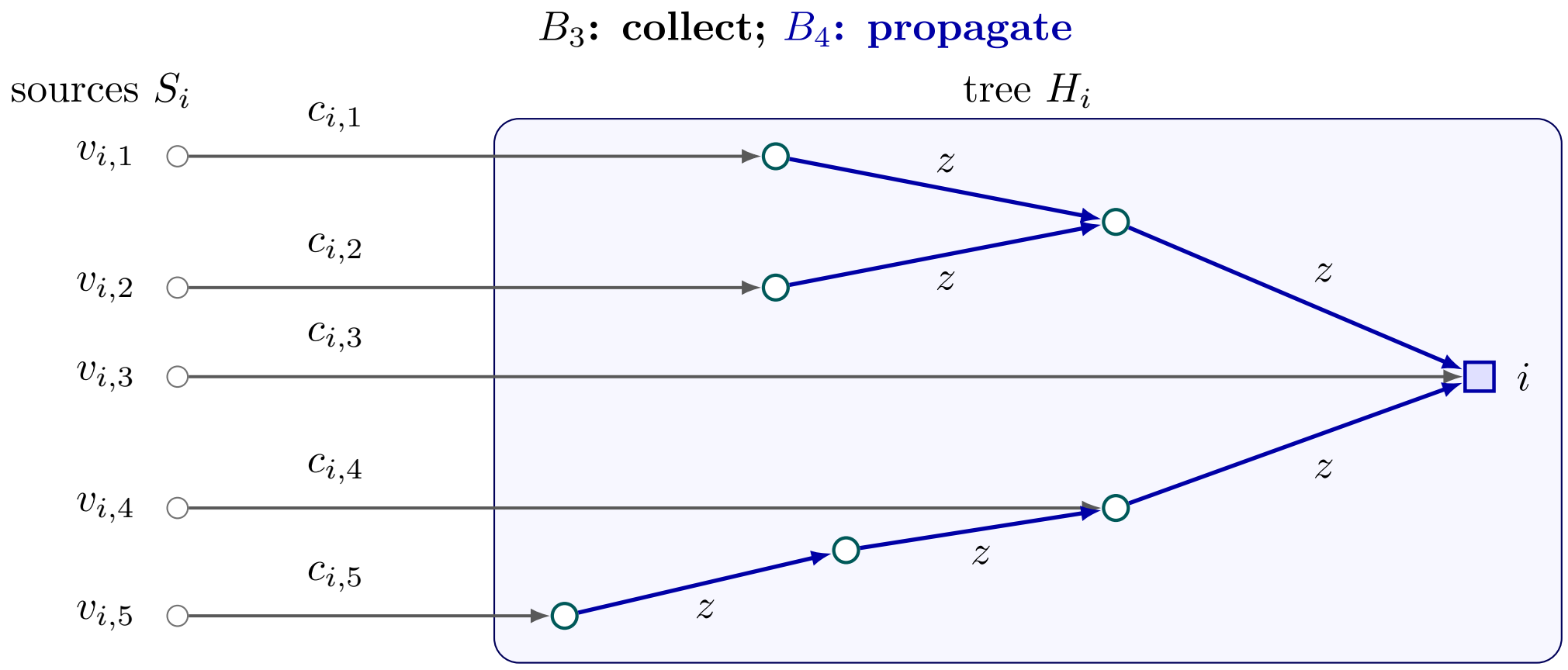
\begin{figure}[!htbp]
  \centering
  \begin{tikzpicture}[
      x=1cm,y=0.8cm,
      note/.style={font=\small,align=center},
      source/.style={circle,draw=black!55,fill=white,
      minimum size=5pt,inner sep=0pt},
      root/.style={rectangle,draw=blue!65!black,fill=blue!12,
      minimum size=7pt,inner sep=0pt,line width=0.85pt},
      helper/.style={circle,draw=teal!70!black,fill=white,
      minimum size=6pt,inner sep=0pt,line width=0.8pt},
      collect/.style={draw=black!65,line width=0.75pt,
      -{Latex[length=1.7mm]}},
      flow/.style={draw=blue!65!black,line width=1pt,
      -{Latex[length=1.8mm]}}
    ]
    \node[note] at (7.55,7.10)
    {\textbf{\(B_3\): collect; \textcolor{blue!65!black}{\(B_4\): propagate}}};
    \node[note] at (1.55,6.45) {sources \(S_i\)};
    \node[note] at (9.45,6.45) {tree \(H_i\)};
    \path[draw=blue!35!black,fill=blue!3,rounded corners=6pt]
    (4.90,0.35) rectangle (14.00,6.15);

    \node[root] (etroot) at (13.30,3.40) {};
    \node[helper] (etupper) at (10.20,5.05) {};
    \node[helper] (etlower) at (10.20,2.00) {};
    \node[helper] (etleafone) at (7.30,5.75) {};
    \node[helper] (etleaftwo) at (7.30,4.35) {};
    \node[helper] (etmiddle) at (7.90,1.55) {};
    \node[helper] (etdeep) at (5.50,0.85) {};
    \node[note,anchor=west] at (13.48,3.40) {\(i\)};

    \foreach \ell/\height/\receiver in {
      1/5.75/etleafone,2/4.35/etleaftwo,3/3.40/etroot,
    4/2.00/etlower,5/0.85/etdeep} {
      \node[source] (es\ell) at (2.20,\height) {};
      \node[note,anchor=east] at (1.98,\height) {\(v_{i,\ell}\)};
      \draw[collect] (es\ell)--(\receiver);
      \node[note,above=2pt] at (3.55,\height) {\(c_{i,\ell}\)};
    }
    \draw[flow] (etleafone)--(etupper) node[midway,above,note] {\(z\)};
    \draw[flow] (etleaftwo)--(etupper) node[midway,below,note] {\(z\)};
    \draw[flow] (etupper)--(etroot) node[midway,above right,note] {\(z\)};
    \draw[flow] (etdeep)--(etmiddle) node[midway,below,note] {\(z\)};
    \draw[flow] (etmiddle)--(etlower) node[midway,below,note] {\(z\)};
    \draw[flow] (etlower)--(etroot) node[midway,below right,note] {\(z\)};
  \end{tikzpicture}
  \caption{Collection in \(B_3\) and propagation in \(B_4\). Five updates are shown.}
  \label{fig:entry-tree-collection}
\end{figure}

We handle rows according to their demands. Transposition lets us set the
high-demand rows first, helper trees collect contributions for medium-demand
rows next, and the ordered blocks of Section~\ref{sec:rectangular} finish the
low-demand rows. Figure~\ref{fig:entry-tree-collection} illustrates
collection into a helper tree.

\subsubsection{The construction}
Fix \(X\in\mathrm{SL}_n(\F_q)\), and let \(y_i\) be variables for the
requested entries in row \(i\). We choose the transvection coefficients
rationally in \(y\).
Take \(0<\overline d<n\) and \(\underline d>0\), and put
\[
  \mathcal H=\{i:|\mc{S}_i|>\overline d\},\quad
  \mathcal M=\{i\notin\mathcal H:|\mc{S}_i|>\underline d\},\quad
  \mathcal L=[n]\setminus(\mathcal H\cup\mathcal M).
\]

Choose the cutoffs and target partitions before exposing the skeleton.
Every collection pool below must contain enough distinct sources for
its demand.
Divide the transvections into five independent intervals of deterministic
lengths, with products \(B_1,\ldots,B_5\). We will make
\((B_5B_4B_3B_2B_1X)|_{\mc{S}}=y\).
We use interval 1 to fix the rows in \(\mc{H}\) after interval 2 is applied; interval 2 for burn-in for the rest
of the construction; interval 3 to supply sources to the trees of rows in \(\mc{M}\),
which are grown in interval 4; and interval 5 to supply sources to the rows in \(\mc{L}\).
Set all unselected coefficients outside \(B_2\) to zero.

\paragraph{Burn-in.}
Leave the coefficients \(\mathbf b_2\) of \(B_2\) indeterminate.
Conditional on the source selections below, Lemmas~\ref{lem:routing}
and~\ref{lem:routing-determinant} make the minors
\(B_2[\mathcal H,Q_j]\) and \((B_2X)[S_i,\mc{S}_i]\)
nonsingular over \(\F_q(\mathbf b_2)\), with failure probability estimated
in the next section. Here \(Q_j\) is the source set for transposed target \(j\) in the
\(\mathcal H\) step, and \(S_i\) is the source set for row \(i\) in the
\(\mathcal M\) or \(\mathcal L\) step. These sets are independent of
the skeleton of \(B_2\).

\paragraph{Fixing \(\mathcal H\).}
Form the \(|\mathcal H|\times n\) matrix \(W\) from
\((B_2X)[\mathcal H,:]\) by replacing its requested entries with \(y_i\).
Use the procedure of Section~\ref{sec:rectangular} on the reversed,
transposed skeleton of \(B_1\); let \(Q_j\) be its \(|\mathcal H|\)
selected sources for target \(j\). Apply
Proposition~\ref{prop:birational} over \(\F_q(\mathbf b_2)\) to the initial matrix
\(B_2[\mathcal H,:]^{\top}\), choosing the coefficients of \(B_1\) so that
\[
  B_1^{\top}B_2[\mathcal H,:]^{\top}
  =(X^{-1})^{\top}W^{\top}.
\]
Then \((B_2B_1X)[\mathcal H,:]=W\). Nonzero transvections in
\(B_3,B_4,B_5\) never target these rows.

\paragraph{Fixing \(\mathcal M\).}
Initialize a helper tree \(H_i=\{i\}\) for every
\(i\in\mathcal M\). Read the transvections of \(B_4\) backwards. When
\(u\mapsto v\) has \(u\notin\mathcal H\cup\bigcup_jH_j\) and
\(v\in H_i\) for an unfinished tree, attach \(u\) to \(H_i\) and record
this edge. A tree is finished when
\[
  |H_i|=\floor{|\mc{S}_i|/\underline d}\ge1.
\]
Give every recorded edge the same coefficient \(z\), an indeterminate, and let \(B_4(z)\) be the associated operator. In forward time, a contribution at depth \(e\) reaches
the root multiplied by \(z^e\).

We now use \(B_3\) to assign \(|\mc{S}_i|\) contributions to tree \(H_i\).
Order the trees by nondecreasing \(|\mc{S}_i|/|H_i|\), breaking
ties by row index. Write \(j\succeq i\) when \(j=i\) or tree \(j\)
comes later in this order. In interval 3, select \(|\mc{S}_i|\) updates
from distinct sources
\(S_i\subseteq[n]\setminus\bigcup_{j\succeq i}H_j\)
into nodes of \(H_i\); receiving nodes may repeat.

Recover the trees in this order. For selected update \(\ell\) into
\(H_i\), let \(v_{i,\ell}\) be its source row at that time and
\(e_{i,\ell}\) the receiving node's depth. The source lies outside
every unrecovered tree, so its value is determined by previously recovered coefficients. Choose the coefficients
\(c_{i,\ell}\) by the linear solve~\eqref{eq:coefficient-inverse}, with
the factors \(z^{e_{i,\ell}}\):
\[
  (B_4(z)B_2B_1X)[i,\mc{S}_i]
  +\sum_{\ell=1}^{|\mc{S}_i|}
  c_{i,\ell}z^{e_{i,\ell}}v_{i,\ell}[\mc{S}_i]=y_i.
\]
Then \(B_4B_3B_2B_1X\) matches the target on rows in \(\mc{M}\) as well as \(\mc{H}\).

\paragraph{Fixing \(\mathcal L\).}
In interval 5, partition the low rows with requested entries into blocks \(R\) with
\(|R|+\max_{i\in R}|\mc{S}_i|\le n\). Apply
Proposition~\ref{prop:birational} to \(B_4B_3B_2B_1X\) with targets
\(y_i\), and let \(S_i\) be the selected source set for row \(i\).
These updates preserve the completed rows, so the final matrix agrees
with \(y\) on all of \(\mc{S}\).

\paragraph{Checking recovery.}
To show that all recovery formulas are defined on a common nonempty open
set, evaluate their denominators at the baseline
\[
  y^0=(B_4(z)B_2X)|_{\mc{S}}.
\]
Since \(B_4\) fixes \(\mathcal H\), the high-row target is then
\(W=(B_2X)[\mathcal H,:]\), so Proposition~\ref{prop:birational}
returns zero coefficients in \(B_1\), hence \(B_1=I\).
Inductively, earlier tree coefficients also vanish. The coefficient
matrix for tree \(H_i\) has determinant
\[
  \pm z^{\sum_\ell e_{i,\ell}}
  \det\bigl((B_2X)[S_i,\mc{S}_i]\bigr)\ne0
\]
over \(\F_q(\mathbf b_2,z)\), so its recovered coefficients are zero.
Thus \(B_3=I\) at the baseline.
For the low rows, the source minors are
\((B_4(z)B_2X)[S_i,\mc{S}_i]\). Their determinants have nonzero constant
term in \(z\), namely \(\det((B_2X)[S_i,\mc{S}_i])\).
The low-row coefficients therefore vanish successively as well.
Writing the recovered coefficients as ratios of polynomials, the product
of their denominators and \(z\) is therefore nonzero. For nonempty
\(\mc{S}\), the argument at the end of
Section~\ref{sec:coefficient-slice} lets us fix
\(\mathbf b_2,z\in\F_q(y)\) so that this product remains nonzero.
The resulting coefficients give the required rational section over \(\F_q\).

\subsubsection{Choosing the cutoffs and bounding time}\label{app:entry-estimates}
Assume \(n\ge256\) and \(0<|\mc{S}|<n^2/3\), and let \(0<\varepsilon<1\).
Set
\[
  \begin{gathered}
    u=\frac13+\frac{5|\mc{S}|}{6n^2},\qquad
    c=\frac4{3(1-u)},\qquad s_{\mathrm{rem}}=\sum_{i\notin\mathcal H}|\mc{S}_i|,
  \end{gathered}
\]
and \(\mc{H}/\mc{M}/\mc{L}\) cutoffs
\[\overline d=un, \qquad \underline d=\frac{c s_{\mathrm{rem}}}{n}.\]
Put \(L=\log(7n/\varepsilon)\). If \(s_{\mathrm{rem}}=0\), take
\(\underline d=\overline d\) and omit intervals 3--5.

\paragraph{Interval 2: source minors.}
Let \(k\) be the largest required minor order. The high-row minors have
order \(|\mathcal H|\le|\mc{S}|/(un)<un\), since \(u^2>|\mc{S}|/n^2\).
Every remaining required minor has order at most \(un\).
Thus \(k\le un\), so give interval 2 length
\[
  \ceil{\frac{9n^2}{2(n-k)}\log\frac{32n^2}{\varepsilon}}\leq \frac{9nL}{1-u}-4.
\]
Conditional on the selected sources, Lemmas~\ref{lem:routing}
and~\ref{lem:routing-determinant} with \(\zeta=\varepsilon/(32n)\)
make the at most \(2n\) required source minors nonsingular over
\(\F_q(\mathbf b_2)\) except with probability at most \(\varepsilon/16\).

\paragraph{Interval 4: helper trees.}
Omit this interval if every prescribed tree has size one; otherwise use length
\[
  \ceil{\frac{n(n-1)L}{n-|\mathcal H|-\sum_{i\in\mathcal M}|H_i|}},
\]
provided \(|\mathcal H|+\sum_{i\in\mathcal M}|H_i|<n\).
The growth argument in the proof of Lemma~\ref{lem:routing}
bounds the probability that any tree
is unfinished by \(e^{-L}\sum_{i\in\mathcal M}|H_i|\).
For the chosen cutoffs, there are at most \(n/c\le n/2\) tree nodes,
leaving at least \(n(1-|\mc{S}|/(un^2)-1/c)>0\) unused rows.
Since \(L>5\), the interval length is at most
\[
  \frac{nL}{1-|\mc{S}|/(un^2)-1/c}-4.
\]
The failure probability is at most
\((n/2)e^{-L}=\varepsilon/14\).

\paragraph{Intervals 1, 3, 5: source collections.}
Use three balanced blocks in interval 1 and a
single phase for all trees in interval 3.
In interval 5, order the rows in \(\mc{L}\) with
positive demand by decreasing demand and use blocks of \(\ceil{n/4}\) rows.
Use one phase per nonempty block.
The collection requirements for each target are
\[
  \begin{array}{c|ccc}
    \text{task}&\text{source count}&\text{demand}&\text{receivers}\\\hline
    \text{interval 1: transposed block }R&n-|R|&|\mathcal H|&1\\
    \text{interval 3: tree }H_i&n-\sum_{j\succeq i}|H_j|&|\mc{S}_i|&|H_i|\\
    \text{interval 5: block }R\subseteq\mc{L}&n-|R|&\max_{i\in R}|\mc{S}_i|&1
  \end{array}
\]
For a bound uniform over initial matrices, count all \(n\) transposed
targets when \(\mc{H}\ne\varnothing\), unless a smaller set suffices
uniformly.
For \(j\in\{1,3,5\}\), let \(\mu_j\) and \(v_j\) sum the largest mean
and inverse-probability bounds, respectively, taken in each phase of
interval \(j\). Use the phase lengths in
Corollary~\ref{cor:target-source-collection}.
Rounding adds one step per nonempty phase, so interval \(j\) has
length at most
\[
  \mu_j+\sqrt{2\mu_jv_jL}+v_jL+
  \begin{cases}
    3,&j=1,\\
    1,&j=3,\\
    4,&j=5.
  \end{cases}
\]
Conditional on the completed trees from interval 4, a union bound over at
most \(2n\) collections gives failure probability at most
\(2ne^{-L}=2\varepsilon/7\). Put
\(\mu=\mu_1+\mu_3+\mu_5\) and \(v=v_1+v_3+v_5\).

\emph{Mean bounds.}
For \(p=3,4\), retaining \(n(n-1)\) absorbs block rounding:
\[
  (n-1)\log\frac{n-\ceil{n/p}+1/2}{n-\ceil{n/p}+1/2-d}
  \le n\log\frac{1-1/p}{1-1/p-d/n},
\]
for \(0\le d/n\le6/11\) when \(p=3\), and \(0\le d/n\le11/18\) when \(p=4\).
Differentiating the right side minus the left shows that its minimum is
at an endpoint; both endpoint values are nonnegative for \(n\ge256\).
Thus Corollary~\ref{cor:target-source-collection} gives
\[
  \mu_1\le3n^2\log\frac{2/3}{2/3-|\mathcal H|/n}
  \le16(|\mc{S}|-s_{\mathrm{rem}}).
\]
When \(\mathcal H\ne\varnothing\), use \(|\mathcal H|/n\le6/11\) and
\(u\ge2/(6-5|\mathcal H|/n)\), then check the endpoints of the resulting
convex bound on \(\mu_1/(|\mc{S}|-s_{\mathrm{rem}})\) to obtain the
second inequality.

If \(s_{\mathrm{rem}}=0\), then \(\mu=\mu_1\le16|\mc{S}|\).
For \(s_{\mathrm{rem}}>0\), put
\(\beta=s_{\mathrm{rem}}/n^2\) and
\(\theta=\frac{1}{s_{\mathrm{rem}}}\sum_{i\in\mathcal M}|\mc{S}_i|\).
For \(\mc{L}\), each block maximum after the first is at most the
preceding block's average; the block maxima therefore sum to at most
\((c+4(1-\theta))s_{\mathrm{rem}}/n\).

If \(c\beta\ge u\), there are no trees and all remaining demands are at
most \(un\). Corollary~\ref{cor:target-source-collection} and convexity
in the demand give
\[
  \frac{\mu_5}{s_{\mathrm{rem}}}
  \le\frac{c+4}{u}\log\frac{3/4}{3/4-u}.
\]
The right-hand side increases with \(u\) and is below \(16\) at
\(u=5/12\) and below \(26\) at \(u=11/18\).

For \(c\beta<u\), fix a tree \(H_i\) and put
\(r=\frac{n|\mc{S}_i|}{s_{\mathrm{rem}}|H_i|}\).
The tree order and size give
\begin{equation}\label{eq:ordered-source-pool}
  n-\sum_{j\succeq i}|H_j|\ge n(1-\theta/r),\qquad
  c\le r<2c,\qquad r\le u/\beta.
\end{equation}
The source bound~\eqref{eq:ordered-source-pool},
Corollary~\ref{cor:target-source-collection}, and convexity of the logarithms give
\begin{align*}
  \frac{\mu_3}{n^2}
  &\le\max_{c\le r\le\min(2c,u/\beta)}
  \frac{\beta r}{u}\log\frac{1-\theta/r}{1-u-\theta/r},\\
  \frac{\mu_5}{n^2}
  &\le\left(1+\frac{4(1-\theta)}c\right)\log\frac{3/4}{3/4-c\beta}.
\end{align*}
To bound \((\mu_3+\mu_5)/s_{\mathrm{rem}}\), divide by \(\beta\).
The bound for \(\mc{L}\) then increases with \(\beta\), so for each
\(r\) use \(\beta\le u/r\). After substituting \(\beta=u/r\), the
bound is convex separately in \(r,\theta\). It suffices to check
\(r=c,2c\) and \(\theta=0,1\). These four values increase with \(u\);
they are below \(16\) at \(u=5/12\) and below \(26\) at \(u=11/18\).

Adding the \(\mc{H}\) contribution gives \(\mu\le16|\mc{S}|\) when \(u\le5/12\), and
\(\mu\le26|\mc{S}|\) throughout.

\emph{Tail bounds.}
If \(s_{\mathrm{rem}}=0\), Corollary~\ref{cor:target-source-collection} gives
\(v=v_1\le3n/(2/3-|\mc{S}|/(un^2))\).
This is below \(18n\) when \(u\le5/12\), and below \(32n\) throughout.
Now assume \(s_{\mathrm{rem}}>0\) and put \(d=\min(c\beta,u)\).
For \(\mc{H}\) and \(\mc{L}\), Corollary~\ref{cor:target-source-collection} gives
\begin{align*}
  v_1&\le\frac{3n(n-1)}{n-\ceil{n/3}-|\mc{H}|+1}
  \le\frac{3n}{2/3-(|\mc{S}|/n^2-\beta)/u},\\
  v_5&\le n(n-1)\sum_R
  \frac1{n-\ceil{n/4}-\max_{i\in R}|\mc{S}_i|+1}
  \le n\left(\frac{16}3+
  \frac{d+4(1-\theta)\beta}{(3/4)(3/4-d)}\right),
\end{align*}
using for \(v_5\) that the normalized block maxima sum to at most \(d+4(1-\theta)\beta\).

If \(c\beta\ge u\), there are no trees, so \(v_3=0\), \(\theta=0\), and \(d=u\).
For \(c\beta<u\), when trees are present,
\(|\mc{S}_i|/(n|H_i|)<2c\beta\), so their contribution satisfies
\[
  \frac{v_3}{n}\le\max_i\frac1{|H_i|(1-\theta/c-|\mc{S}_i|/n)}
  \le\min\left\{\frac1{1-\theta/c-u},\,
  \frac1{1-\theta/c}+\frac{2c\beta}{(1-\theta/c)(1-\theta/c-u)}\right\}.
\]
The two bounds agree at \(\beta=u/(2c)\).
The resulting bound on \(v/n\) is convex in \(\theta\) and in
\(\beta\) on each interval cut out by
\(0,u/(2c),u/c,|\mc{S}|/n^2\).
Check \(\theta=0,1\) and those endpoints in
\(0\le\beta\le|\mc{S}|/n^2\); beyond \(u/c\), omit the tree term and
take \(\theta=0\). These endpoint bounds give \(v\le18n\) when
\(u\le5/12\), and \(v\le32n\) throughout.

\paragraph{Total time.}
Summing the interval bounds as in
Corollary~\ref{cor:target-source-collection} gives
\begin{equation}\label{eq:entry-schedule-time}
  \begin{aligned}
    T&\le\sum_{i\in\{1,3,5\}}
    \left(\mu_i+\sqrt{2\mu_iv_iL}+v_iL\right)
    +8+\frac{9nL}{1-u}-4+\frac{nL}{1-|\mc{S}|/(un^2)-1/c}-4\\
    &\le\mu+\sqrt{2\mu vL}+vL+
    \begin{cases}
      19nL,&u\le5/12,\\
      \tfrac{59}{2}nL,&u\le11/18.
    \end{cases}
  \end{aligned}
\end{equation}
Substitute the bounds for \(\mu,v\),
using \(\sqrt{576|\mc{S}|nL}\le5|\mc{S}|+(144/5)nL\) when \(u\le5/12\),
and \(\sqrt{1664|\mc{S}|nL}\le11|\mc{S}|+38nL\) throughout. This gives
\begin{align}
  T&\le21|\mc{S}|+89n\log n+66n\log(1/\varepsilon),
  &&|\mc{S}|\le n^2/10,\label{eq:entry-sparse-time}\\
  T&\le37|\mc{S}|+135n\log n+100n\log(1/\varepsilon),
  &&|\mc{S}|<n^2/3.\label{eq:entry-dense-time}
\end{align}

Combining the collection and growth failures gives probability at most
\(5\varepsilon/14\).
Burn-in and the degree bound of Lemma~\ref{lem:transmission-degree}
add at most \(\varepsilon/8\), for total failure probability at most
\(27\varepsilon/56\). The resulting section yields a conditional TV bound
under~\eqref{eq:local-Q}.

\subsection{The full group: four shears}\label{sec:four-segment-alternative}

Assume \(n\ge256\). We factor a generic matrix into four shears and use
this factorization to construct a section along the skeleton. All recovered
factors and coefficients are rational functions of the endpoint.

Let \(B_0=\{1,\ldots,\floor{n/2}\}\) and
\(B_1=\{\floor{n/2}+1,\ldots,n\}\). Write
\[
  L(A)=
  \begin{pmatrix}I_{\floor{n/2}}&0\\A&I_{\ceil{n/2}}
  \end{pmatrix},
  \qquad
  U(B)=
  \begin{pmatrix}I_{\floor{n/2}}&B\\0&I_{\ceil{n/2}}
  \end{pmatrix}.
\]
Let \(\widetilde U\) be a shear from \(B_1\setminus\{n\}\) to
\(B_0\cup\{n\}\). In the coordinate order
\((B_0,B_1\setminus\{n\},\{n\})\), it has the form
\[
  \widetilde U=
  \begin{pmatrix}
    I_{\floor{n/2}}&*&0\\
    0&I_{\ceil{n/2}-1}&0\\
    0&*&1
  \end{pmatrix},
\]
with arbitrary starred blocks.

\subsubsection{Factoring a generic matrix}

\begin{lemma}[Four shears]\label{lem:four-segment-factor}
  Over any field, a generic \(W\in\mathrm{SL}_n\) factors as
  \[
    W=\widetilde U L(A)U(B)L(C),
  \]
  with every factor rational in \(W\) over that field.
\end{lemma}
\begin{proof}
  We give linear systems for recovering the factors and exhibit a point
  where they are uniquely solvable. A dimension count then shows that
  recovery works for a generic matrix in \(\mathrm{SL}_n\).

  For \(W\) already written as a product of the prescribed shears, write
  \(\widetilde U=I_n+N\). The structure of \(\widetilde U\) gives
  \(N^2=0\), hence \((I_n+N)^{-1}=I_n-N\). Define
  \[
    M\coloneqq(I_n-N)W=L(A)U(B)L(C).
  \]
  Multiplying the three shears and their inverses gives \(M_{12}=B\)
  and \((M^{-1})_{12}=-B\), where subscripts denote blocks for the balanced
  split. Thus \(M_{12}+(M^{-1})_{12}=0\). Substituting
  \(M=W-NW\) and \(M^{-1}=W^{-1}+W^{-1}N\) gives
  \begin{equation}\label{eq:four-segment-linear-system}
    (NW-W^{-1}N)_{12}=(W+W^{-1})_{12}.
  \end{equation}
  Solve this linear system for the allowed entries of \(N\), then recover
  \[
    M=(I_n-N)W,\qquad B=M_{12},\qquad
    AB=M_{22}-I_{\ceil{n/2}},\qquad C=M_{21}-AM_{11}.
  \]
  These formulas recover all factors whenever the linear system
  in~\eqref{eq:four-segment-linear-system} uniquely determines \(N\)
  and \(B\) has rank \(\floor{n/2}\).

  We now verify that recovery holds at one point. Let \(J_s\) have ones just
  above the diagonal and zeros elsewhere. Let \(B_*\) consist of the last
  \(\floor{n/2}\) rows of \(I_{\ceil{n/2}}\), and put
  \(A_*=J_{\ceil{n/2}}B_*^{\top}\). Then
  \[
    A_*B_*=J_{\ceil{n/2}},\qquad B_*A_*=J_{\floor{n/2}},\qquad
    B_*e_{\ceil{n/2}}=e_{\floor{n/2}}.
  \]
  Take \(W_0=L(A_*)U(B_*)\), whose \(B_*\) block has rank \(\floor{n/2}\).
  An allowed \(N\) has the form
  \[
    N=
    \begin{pmatrix}0&D\\0&e_{\ceil{n/2}}u^{\top}
    \end{pmatrix},
    \qquad De_{\ceil{n/2}}=0,\quad u^{\top}e_{\ceil{n/2}}=0.
  \]
  At \(W_0\), the homogeneous system becomes
  \[
    DJ_{\ceil{n/2}}-J_{\floor{n/2}}D=-e_{\floor{n/2}}u^{\top}.
  \]
  Solving row by row gives the rows of \(D\) as
  \(d^{\top},d^{\top}J_{\ceil{n/2}},\ldots,
  d^{\top}J_{\ceil{n/2}}^{\floor{n/2}-1}\),
  and \(u^{\top}=-d^{\top}J_{\ceil{n/2}}^{\floor{n/2}}\).
  The last column of \(N\) is zero, so
  \(d^{\top}J_{\ceil{n/2}}^je_{\ceil{n/2}}=0\) for \(0\le j\le\floor{n/2}\).
  Since \(\ceil{n/2}\le\floor{n/2}+1\), these equations force \(d=0\).
  Thus \(D=u=0\), proving injectivity of the linear map at \(W_0\)
  over every field.

  We now generalize to a dense set. In each recovery system, retain a square
  subsystem invertible at
  \(W_0\). Solving these subsystems gives rational formulas in \(W\).
  On the nonempty open set of factor choices where these denominators are
  nonzero, uniqueness makes multiplication and recovery inverse. This set
  contains the factorization of \(W_0\), and its image has dimension
  \[
    3\floor{n/2}\ceil{n/2}+(\ceil{n/2}-1)(\floor{n/2}+1)
    =n^2-1=\dim\mathrm{SL}_n.
  \]
  Since \(\mathrm{SL}_n\) is irreducible, the image is dense and the
  factorization holds for generic \(W\).
\end{proof}

\subsubsection{Realizing a factorization along a skeleton}

Split the skeleton into four consecutive segments
\(\sigma=\sigma_1\sigma_2\sigma_3\sigma_4\). Updates multiply on the left,
so their chronological order reverses the four-shear factorization:
\[
  \begin{array}{c|c|c|c}
    \text{segment} & \text{source set} & \text{target set}
    & \text{shear factor}\\ \hline
    \sigma_1 & B_0 & B_1 & L(C)\\
    \sigma_2 & B_1 & B_0 & U(B)\\
    \sigma_3 & B_0 & B_1 & L(A)\\
    \sigma_4 & B_1\setminus\{n\} & B_0\cup\{n\} & \widetilde U
  \end{array}
\]
Split each segment into three consecutive sublists. The retained middle
transvections transform the source rows. On a nonempty open set of their
coefficients, the outer coefficients can prescribe any source-to-target
block without changing this transformation.
We call the resulting matrix a quasi-shear.
The first lemma constructs one quasi-shear; the second combines four
of them to reach a generic endpoint.

\begin{lemma}[Building one quasi-shear]\label{lem:single-quasi-shear}
  Let \(S\sqcup R=[n]\) be nonempty source and target sets, and let
  \(\tau=\tau_1\tau_2\tau_3\) be a fixed skeleton split into three
  consecutive sublists. For each target \(i\in R\), define
  \[
    S_{a,i}=\{s\in S:(s\mapsto i)\in\tau_a\},\qquad a\in\{1,3\}.
  \]
  Let \(P_i\) be the sources that update \(i\) in neither outer sublist,
  and \(Q_i\) those that update \(i\) in both:
  \[
    P_i=S\setminus(S_{1,i}\cup S_{3,i}),\qquad
    Q_i=S_{1,i}\cap S_{3,i}.
  \]
  Suppose that, for every \(i\in R\), the middle sublist contains
  updates matching all of \(P_i\) to distinct rows of \(Q_i\).

  Retain every \(S\)-to-\(R\) update in the outer sublists and every
  \(S\)-to-\(S\) update in the middle sublist; set all other coefficients
  to zero. Let \(M(z)\) be the retained middle product on \(S\), with
  coefficients \(z\). There is one open set of \(z\), defined over
  \(\F_q\) and nonempty over \(\overline{\F}_q\), with the following
  property. For every field extension \(K/\F_q\), every \(K\)-valued
  point \(z\) of this open set, and every \(H\in K^{|R|\times|S|}\),
  the retained outer coefficients can realize
  \[
    \begin{pmatrix}I_{|R|}&H\\0&M(z)
    \end{pmatrix}
  \]
  in coordinate order \((R,S)\). The coefficients can be chosen linearly
  in \(H\) and rationally in \(z\), by formulas over \(\F_q\).
\end{lemma}
\begin{proof}
  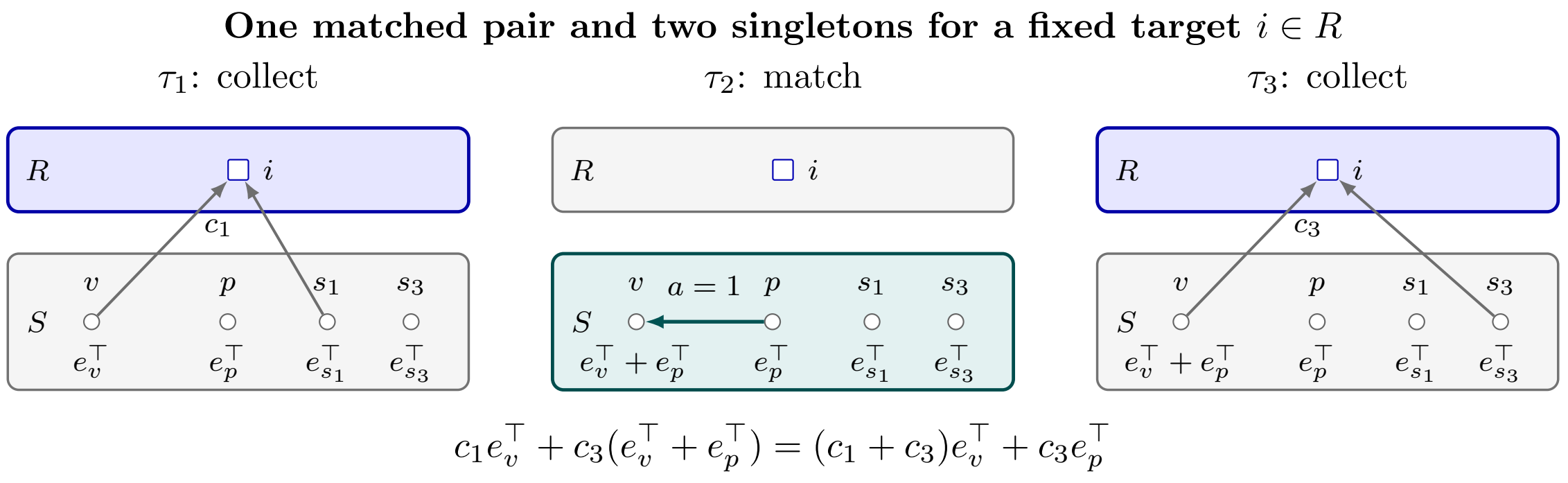
\begin{figure}[!htp]
    \centering
    \begin{tikzpicture}[
        x=1cm,y=1cm,
        source/.style={draw=black!55,line width=0.65pt,rounded corners=3pt,
        fill=black!4},
        target/.style={draw=blue!65!black,line width=0.9pt,rounded corners=3pt,
        fill=blue!10},
        matching/.style={draw=teal!60!black,line width=0.9pt,rounded corners=3pt,
        fill=teal!11},
        source dot/.style={circle,draw=black!60,fill=white,minimum size=4.2pt,
        inner sep=0pt},
        target dot/.style={rectangle,rounded corners=0.5pt,draw=blue!70!black,
        fill=white,minimum width=5.5pt,minimum height=5.5pt,inner sep=0pt},
        update/.style={-{Latex[length=1.7mm]},draw=black!57,line width=0.75pt},
        match update/.style={-{Latex[length=1.8mm]},draw=teal!65!black,
        line width=1pt},
        annotation/.style={font=\scriptsize,align=center},
        aligned label/.style={annotation,text height=1.5ex,text depth=0.5ex}
      ]
      \node[font=\small\bfseries] at (7.55,15.95)
      {One matched pair and two singletons for a fixed target \(i\in R\)};
      \foreach \x/\title in {
        2.35/{\(\tau_1\): collect},
        7.55/{\(\tau_2\): match},
      12.75/{\(\tau_3\): collect}} {
        \node[font=\small] at (\x,15.48) {\title};
      }
      \foreach \x in {0.15,10.55} {
        \path[target] (\x,14.20) rectangle ++(4.40,0.80);
        \path[source] (\x,12.50) rectangle ++(4.40,1.30);
      }
      \path[source] (5.35,14.20) rectangle (9.75,15.00);
      \path[matching] (5.35,12.50) rectangle (9.75,13.80);
      \foreach \x in {0.43,5.63,10.83} {
        \node[annotation] at (\x,14.60) {\(R\)};
        \node[annotation] at (\x,13.15) {\(S\)};
      }
      \foreach \x/\j in {2.35/1,7.55/2,12.75/3} {
        \node[target dot,label={[annotation]right:\(i\)}]
        (quasi-i\j) at (\x,14.60) {};
      }
      \foreach \x/\j/\row in {
      0.95/1/{e_v^{\top}},6.15/2/{e_v^{\top}+e_p^{\top}},11.35/3/{e_v^{\top}+e_p^{\top}}} {
        \node[source dot,label={[aligned label]above:\(v\)},
        label={[aligned label]below:\(\row\)}]
        (quasi-v\j) at (\x,13.15) {};
      }
      \foreach \x/\j in {2.25/1,7.45/2,12.65/3} {
        \node[source dot,label={[aligned label]above:\(p\)},
        label={[aligned label]below:\(e_p^{\top}\)}]
        (quasi-p\j) at (\x,13.15) {};
      }
      \foreach \x/\j in {3.20/1,8.40/2,13.60/3} {
        \node[source dot,label={[aligned label]above:\(s_1\)},
        label={[aligned label]below:\(e_{s_1}^{\top}\)}]
        (quasi-single-first\j) at (\x,13.15) {};
      }
      \foreach \x/\j in {4.00/1,9.20/2,14.40/3} {
        \node[source dot,label={[aligned label]above:\(s_3\)},
        label={[aligned label]below:\(e_{s_3}^{\top}\)}]
        (quasi-single-third\j) at (\x,13.15) {};
      }
      \draw[update] (quasi-v1)--node[annotation,pos=0.65,right=2pt] {\(c_1\)}(quasi-i1);
      \draw[match update] (quasi-p2)--node[annotation,above=3pt]
      {\(a=1\)}(quasi-v2);
      \draw[update] (quasi-v3)--node[annotation,pos=0.65,right=2pt] {\(c_3\)}(quasi-i3);
      \draw[update] (quasi-single-first1)--(quasi-i1);
      \draw[update] (quasi-single-third3)--(quasi-i3);
      \node[font=\small] at (7.55,11.95) {
        \(\displaystyle
          c_1e_v^{\top}+c_3(e_v^{\top}+e_p^{\top})
          =(c_1+c_3)e_v^{\top}+c_3e_p^{\top}
      \)};
    \end{tikzpicture}
    \caption{Building a quasi-shear from source set \(S\) to target set \(R\) for a particular row \(i\).}
    \label{fig:quasi-shear-schematic}
  \end{figure}
  Fix a target \(i\) (Figure~\ref{fig:quasi-shear-schematic}).
  An occurrence of \(s\mapsto i\) in \(\tau_1\)
  supplies \(e_s^{\top}\), and one in \(\tau_3\) supplies row \(s\) of
  \(M(z)\). Thus the outer coefficients map linearly to \(H[i]\)
  through a matrix polynomial in \(z\). Choose a matching
  \(\phi_i:P_i\to Q_i\). Give one occurrence of each
  \(p\mapsto\phi_i(p)\) coefficient \(1\), and every other middle
  coefficient \(0\). Since \(P_i,Q_i\) are disjoint
  and \(\phi_i\) is injective, only rows \(\phi_i(p)\) change, to
  \(e_{\phi_i(p)}^{\top}+e_p^{\top}\). Subtracting the two outer contributions
  from each \(\phi_i(p)\in Q_i\) gives \(e_p^{\top}\). For \(s\notin P_i\),
  use \(e_s^{\top}\) from \(\tau_1\) if \(s\in S_{1,i}\), and otherwise
  from its unchanged row in \(\tau_3\), since then \(s\notin Q_i\).
  Thus every row \(H[i]\) can be
  produced at this assignment.

  To recover all rows on one open set, select for each target \(i\)
  exactly \(|S|\) outer occurrences whose supplied vectors are
  linearly independent at its matching assignment. Let \(V_i(z)\)
  be the matrix with these supplied vectors as rows. Each \(\det V_i(z)\)
  is a nonzero polynomial, so
  \[
    \left\{z:\prod_{i\in R}\det V_i(z)\ne0\right\}
  \]
  is an open set independent of \(H\), defined over \(\F_q\) and
  nonempty over \(\overline{\F}_q\).

  On this set, use \(H[i]V_i(z)^{-1}\) as the selected coefficient vector
  for target \(i\), and set the remaining outer coefficients to zero.
  The targets use disjoint outer coefficients, so these choices produce
  all of \(H\) without changing \(M(z)\).
\end{proof}

Fix the four middle matrices \(M_j\). Moving them to the right
rewrites the product as four shears followed by a fixed matrix \(Z\).
To reach a generic endpoint \(W\), factor \(WZ^{-1}\).

\begin{lemma}[Combining four quasi-shears]
  \label{lem:four-segment-construction}
  Split a skeleton as \(\sigma=\sigma_1\sigma_2\sigma_3\sigma_4\),
  using the source and target sets in the preceding table. If each
  segment splits into three sublists satisfying the matching condition
  of Lemma~\ref{lem:single-quasi-shear}, then the coefficient-to-endpoint
  map from \(I_n\) to \(\mathrm{SL}_n\) has a rational section over
  \(\F_q\).
\end{lemma}
\begin{proof}
  Fix the middle matrices \(M_j\), each of determinant one. Let \(D_j\)
  act by \(M_j\) on its segment's source set and identically on its
  target set. Each quasi-shear \(G_j\) is an arbitrary shear of its
  prescribed type followed by \(D_j\): in target--source coordinates,
  \[
    \begin{pmatrix}I&H\\0&M
    \end{pmatrix}
    =
    \begin{pmatrix}I&HM^{-1}\\0&I
    \end{pmatrix}
    \begin{pmatrix}I&0\\0&M
    \end{pmatrix}.
  \]
  Every \(D_j\) is block diagonal for \(B_0\sqcup B_1\). For invertible
  \(R=\operatorname{diag}(R_0,R_1)\),
  \[
    R L(A)=L(R_1AR_0^{-1})R,\qquad
    R U(B)=U(R_0BR_1^{-1})R.
  \]
  Moving the \(D_j\) to the right therefore gives, after invertible
  linear changes of the ordinary shear blocks,
  \[
    G_4G_3G_2G_1=\widetilde U L(A)U(B)L(C)Z,
    \qquad Z=D_4D_3D_2D_1.
  \]
  The modified shear stays leftmost and is not conjugated.
  Since \(Z\in\mathrm{SL}_n\), factor \(WZ^{-1}=\widetilde U L(A)U(B)L(C)\) for generic \(W\)
  using Lemma~\ref{lem:four-segment-factor}, then multiply by \(Z\) and undo these changes
  to recover the free quasi-shear blocks rationally in \(W\) and the \(M_j\).

  Leave the middle coefficients \(z\) indeterminate. For each segment,
  Lemma~\ref{lem:single-quasi-shear} supplies \(M_j(z)\) and recovers
  the outer coefficients by linear formulas. Substituting these into
  the factorization gives a section over \(\F_q(z)\).
  Choose \(z\) in the infinite field \(\F_q(\mathrm{SL}_n)\) so that
  every recovery denominator is nonzero. This gives a section over \(\F_q\).
\end{proof}

\subsubsection{Bounding mixing time}

We bound the probability that a random skeleton fails to supply the
matchings in Lemma~\ref{lem:four-segment-construction}.

\begin{lemma}[Quasi-shear matching failure]\label{lem:quasi-shear-hall-tail}
  Let \(S,R\) be nonempty disjoint source and target sets, and let
  \(\tau_1,\tau_2,\tau_3\) be independent random skeletons, each of
  length \(N\).
  Define \(P_i,Q_i\) as in Lemma~\ref{lem:single-quasi-shear}.
  The probability that, for some \(i\in R\), \(\tau_2\) fails to match
  all of \(P_i\) to distinct rows of \(Q_i\) is at most
  \begin{equation}\label{eq:quasi-shear-hall-failure}
    |R|5^{|S|}\exp\left(-\frac{(|S|+1)N}{n(n-1)}\right).
  \end{equation}
\end{lemma}
\begin{proof}
  Fix a target \(i\). By Hall's theorem, failure gives a nonempty
  \(A\subseteq P_i\) whose middle-sublist neighbors in \(Q_i\) form
  a set \(B\) with \(|B|<|A|\). Put \(C=Q_i\setminus B\).
  Every source in \(A\) misses \(i\) in both outer sublists;
  every source outside \(A\cup B\cup C\) misses \(i\) in at least one.
  Choose one such missing sublist for each of these remaining sources.
  The middle sublist contains no update from \(A\) to \(C\).

  Each source has one of five roles: in \(A\), \(B\), or \(C\), or
  assigned to miss \(i\) in the first or third sublist. Each assignment
  forbids the following number of pairs across the three sublists:
  \[
    \left(2|A|\right)+\left(|S|-|A|-|B|-|C|\right)+\left(|A||C|\right)
    =|S|+|A|-|B|+(|A|-1)|C|\ge|S|+1.
  \]
  Each sublist has \(N\) independent steps, so the probability
  of avoiding these pairs is at most
  \(\exp(-(|S|+1)N/(n(n-1)))\).
  Union-bound over at most \(5^{|S|}\) assignments and \(|R|\) targets.
\end{proof}

For \(0<\eta\le1/4\), set
\begin{equation}\label{eq:four-shear-parameters}
  \lambda=\min\left\{1,\sqrt{\frac9n\log\frac{3n}{\eta}}\right\}.
\end{equation}
\(\lambda\) interpolates between schedule lengths effective for small and large \(n\). Give each of the four segments three sublists of lengths
\(N^{\mathrm{out}},N^{\mathrm{mid}},N^{\mathrm{out}}\), where
\begin{equation}\label{eq:four-shear-intervals}
  \begin{aligned}
    N^{\mathrm{out}}
    &=\ceil{n(n-1)\bigl((1-\lambda)\log2+\lambda\log5\bigr)
    +2(n-1)\log\frac{3n}{\eta}},\\
    N^{\mathrm{mid}}
    &=\ceil{n(n-1)\lambda\log5+2(n-1)\log\frac{3n}{\eta}}.
  \end{aligned}
\end{equation}
Set
\begin{equation}\label{eq:four-segment-time}
  T_{\mathrm{shears}}(\eta)=4(2N^{\mathrm{out}}+N^{\mathrm{mid}}).
\end{equation}

\begin{proposition}[Mixing from four shears]\label{prop:four-segment-mixing}
  Let \(0<\eta,u\le1/4\), and
  \[
    q\ge Q_{\mathrm{geom}}\bigl(H(T_{\mathrm{shears}}(\eta),\eta),n^2-1;u\bigr).
  \]
  For every fixed \(X\in\mathrm{SL}_n(\F_q)\), a random skeleton \(\sigma\) of
  length \(T_{\mathrm{shears}}(\eta)\) satisfies
  \begin{equation}\label{eq:shears-conditional-mixing}
    \Pr_\sigma\left[
      \|K_\sigma(X,\cdot)-\pi\|_{\TV}>u/8
    \right]\le\eta/4.
  \end{equation}
  Taking \(u=\eta\) gives \(t_{\mathrm{mix}}^{(n)}(\eta)\le T_{\mathrm{shears}}(\eta)\).
  The schedule length satisfies each applicable bound below:
  \begin{equation}\label{eq:four-segment-bounds}
    T_{\mathrm{shears}}(\eta)\le
    \begin{cases}
      20n^2+24n\log\frac1{4\eta},&\text{always},\\[1mm]
      14n^2+80n\log\frac n\eta,&\text{always},\\[1mm]
      (8\log2+o(1))n^2,&\log(n/\eta)=o(n).
    \end{cases}
  \end{equation}
\end{proposition}
\begin{proof}
  \emph{Matching failure.}
  The four target-set sizes sum to \(2n+1\).
  If \(\lambda=1\), all sublists have the same length, and
  Lemma~\ref{lem:quasi-shear-hall-tail} bounds total matching failure by
  \((2n+1)\eta/(15n)<3\eta/16\). Then suppose \(\lambda<1\), so \(\lambda^2=(9/n)\log(3n/\eta)\).
  Fix a target \(i\), and let \(W\) be the sum of its missed-source
  counts in the two outer sublists. Thus \(|Q_i|-|P_i|=|S|-W\).
  Conditional on the outer sublists, scan the middle updates, greedily
  matching an unmatched source in \(P_i\) to an unused row in \(Q_i\)
  whenever an update joins them.
  If \(W<|S|\), each still-unmatched source has at least \(|S|-W\)
  available partners at every step. Since
  \(N^{\mathrm{mid}}\ge\tfrac43\lambda n(n-1)\), a union bound over sources gives
  conditional failure probability at most \(|S|e^{-4\lambda(|S|-W)/3}\).
  This also holds for \(W\ge|S|\), when the bound is at least one.

  To average this bound over the outer sublists, we estimate
  \(\E[e^{4\lambda W/3}]\). Since \(\lambda\in[0,1]\), the outer length gives
  \[
    e^{-N^{\mathrm{out}}/(n(n-1))}
    \le\tfrac12e^{-\lambda\log(5/2)-2\lambda^2/9}
    \le\tfrac12-\tfrac\lambda3.
  \]
  A fixed set of \(j\) sources is therefore missed in a given outer
  sublist with probability at most \((1/2-\lambda/3)^j\).
  For \(a\in\{1,3\}\) and \(s\in S\), let \(I_{a,s}\) indicate that
  source \(s\) is missed in outer sublist \(a\). By independence of the sublists,
  any fixed \(j\) of these \(2|S|\) occurrences are all missed with probability
  at most \((1/2-\lambda/3)^j\). Then
  \[
    \begin{aligned}
      \E[e^{4\lambda W/3}]
      &=\E\left[\prod_{a\in\{1,3\}}\prod_{s\in S}
      \bigl[1+(e^{4\lambda/3}-1)I_{a,s}\bigr]\right]\\
      &\le\sum_{j=0}^{2|S|}\binom{2|S|}{j}
      \left(\left(\tfrac12-\tfrac\lambda3\right)(e^{4\lambda/3}-1)\right)^j\\
      &=\left(1+\left(\tfrac12-\tfrac\lambda3\right)(e^{4\lambda/3}-1)\right)^{2|S|}
      \le\exp\left(\frac43|S|\lambda-\frac49|S|\lambda^2\right).
    \end{aligned}
  \]
  The last inequality uses \(\log(1-p+pe^t)\le pt+t^2/8\) for
  \(0\le p\le1\). Averaging the conditional failure bound multiplies this
  by \(|S|e^{-4\lambda|S|/3}\), giving
  \[
    \Pr[\text{greedy matching fails for }i]
    \le |S|e^{-4|S|\lambda^2/9}
    \le 2|S|\frac{\eta^2}{9n^2},
  \]
  since \(|S|\ge n/2-1\) and \(4\log(3n/\eta)/n<4/9<\log2\).
  A union bound over \(2n+1\le3n\) targets gives total matching failure
  probability at most \(2\eta^2/3\le\eta/6\).

  \emph{Mixing.}
  In either case, all required matchings exist except with probability at most
  \(3\eta/16\), so Lemma~\ref{lem:four-segment-construction} gives a
  rational section on \(\mathrm{SL}_n\).
  The degree bound fails with probability at most \(\eta/16\), by
  Lemma~\ref{lem:transmission-degree}. When both conditions hold,
  Corollary~\ref{cor:section-mixing} gives conditional distance at most
  \(u/8\) from uniform. Right multiplication by \(X\) preserves \(\pi\),
  proving \eqref{eq:shears-conditional-mixing} for every \(X\).
  Taking \(u=\eta\) and averaging gives ordinary distance at most
  \(3\eta/8<\eta\).

  \emph{Schedule length.}
  The twelve ceilings contribute at most \(12\). For \(0\le a\le6\),
  \[
    8(1-\lambda)\log2+12\lambda\log5
    \le12\log5-a(1-\lambda^2),
  \]
  since \(12\log5-8\log2>12\ge a(1+\lambda)\).
  Using \(\lambda^2\le(9/n)\log(3n/\eta)\) in the interval formulas gives
  \begin{equation}\label{eq:four-shear-affine-bound}
    T_{\mathrm{shears}}(\eta)
    \le(12\log5-a)n(n-1)+(24+9a)(n-1)\log\frac{3n}{\eta}+12
    \qquad(0\le a\le6).
  \end{equation}
  Taking \(a=0,28/5\) gives the first two bounds in
  \eqref{eq:four-segment-bounds}, using \(n\ge256\).
  Completing the square gives
  \(8(1-\lambda)\log2+12\lambda\log5\le37/3+7\lambda^2\), so
  \begin{equation}\label{eq:four-shear-dense-bound}
    T_{\mathrm{shears}}(\eta)
    \le\frac{37}{3}n(n-1)+87(n-1)\log\frac{3n}{\eta}+12.
  \end{equation}
  If \(\log(n/\eta)=o(n)\), then \(\lambda=o(1)\), so the interval
  formulas give \(T_{\mathrm{shears}}(\eta)=(8\log2+o(1))n^2\).
\end{proof}

\subsection{One field threshold for every error level}
\label{sec:mixing-assembly}\label{app:thresholds}

We now extend to the very-small-error regime. Here, it suffices to
amplify the full-group construction.

\begin{proposition}[Full-group amplification]\label{prop:full-group-amplification}
  For \(n\ge256\) and \(q\ge Q_{\rm all}(n)\), both guarantees in
  \eqref{eq:main-conditional-mixing} hold for \(\mc{S}=[n]\times[n]\)
  at every integer time at least
  \[
    \ceil{187n^2+13n\log(1/\varepsilon)},\qquad 0<\varepsilon<1.
  \]
\end{proposition}
\begin{proof}
  Split a \(187n^2\)-step block into eleven blocks of \(17n^2\) steps. We try to form the desired four shears within these eleven smaller blocks by sequentially attempting to form the next unfinished shear in each one.
  Divide a smaller block into three intervals of length
  \[
    N=\floor{17n^2/3}\ge n(n-1)\log5+2(n-1)(2n+\log n)
    \qquad(n\ge256),
  \]
  so that \eqref{eq:quasi-shear-hall-failure} bounds failure for the next
  desired shear by \(e^{-2n}\), conditional on the earlier blocks.

  The probability of having fewer than four block successes is then at most
  \[
    \binom{11}{8}e^{-16n}\le\frac{e^{-15n}}{16}.
  \]
  On success, Lemma~\ref{lem:four-segment-construction} gives a section
  for the whole block. Set unused coefficients to zero.

  Put \(\eta=e^{-15n}\). By Lemma~\ref{lem:transmission-degree},
  every coordinate degree is at most \(H(187n^2,\eta)<525n\),
  except with probability \(\eta/16\).
  The field hypothesis suffices for Corollary~\ref{cor:section-mixing}
  with \(u=\eta^{40}\), since
  \[
    256\eta^{-160}(525n)^{13(n^2-1)/3}\le Q_{\rm all}(n).
  \]
  Indeed, \(\log256+2400n\le n^2\log(5000n)\).
  Thus one event of probability at least \(1-\eta/8\) gives
  \[
    \sup_X\|K_\sigma(X,\cdot)-\pi\|_{\TV}\le\eta^{40}/8.
  \]
  The event is common to all initial states because right multiplication
  by \(X\) preserves the uniform law on \(\mathrm{SL}_n(\F_q)\).

  The resulting \(187n^2\)-step kernel preserves \(\pi\). Its TV contraction factor is at most
  \(\eta^{40}\) on this event, by the triangle inequality, and at most one always.
  Independent blocks therefore give
  \[
    \E_\sigma\left[
      \left(\sup_X d_{[n]\times[n]}^\sigma(X)\right)^{1/40}
    \right]\le(2\eta)^r
  \]
  after \(r\) blocks. Take
  \[
    r=\ceil{\frac{41\log(1/\varepsilon)}{40(15n-\log2)}},
  \]
  giving \((2\eta)^r\le \varepsilon^{41/40}\); Markov's
  inequality and averaging then give the conditional and ordinary guarantees.
  Finally,
  \[
    187rn^2\le187n^2+13n\log(1/\varepsilon).
  \]
  Additional steps preserve \(\pi\) and contract TV, so both guarantees
  hold at every later time.
\end{proof}

\begin{proof}[Proof of Theorem~\ref{thm:main}]
  If \(\log(1/\varepsilon)\ge200n\), the preceding proposition gives
  both conclusions by time \(14n\log(1/\varepsilon)\), which fits all
  stated bounds. It also covers every time
  \(t\ge187n^2+13n\log(1/\varepsilon)\).
  For the remaining times, \(\log(1/\varepsilon)<200n\) and \(t<2800n^2\), so
  \(H(t,\varepsilon/14)<8000n\). For every target dimension \(m\le n^2\),
  \[
    Q_{\rm geom}(H(t,\varepsilon/14),m;\varepsilon/14)
    \le256(14/\varepsilon)^4(8000n)^{13n^2/3}
    \le Q_{\rm all}(n).
  \]
  Thus the common field threshold covers all constructions below.
  To use a construction beyond its scheduled length, place its section on a
  suffix and set earlier coefficients to zero. The degree bound still
  applies to the full coefficient map.

  For \(0<|\mc{S}|<n^2/3\), Section~\ref{app:entry-estimates}
  bounds total collection, growth, burn-in, and degree failure by
  \(27\varepsilon/56\).
  Take \(u=\varepsilon/14\) in Corollary~\ref{cor:section-mixing}.
  On success, the distance to \(\Unif(\F_q^{\mc{S}})\) is at most
  \(\varepsilon/112\). Lemma~\ref{lem:marginal-comparison}
  gives ordinary error at most \(55\varepsilon/56\), and conditional
  error at most \(\varepsilon/2\) outside a set of probability at most
  \(27\varepsilon/56\).
  For \(|\mc{S}|\ge n^2/3\), use the full-group construction with
  \(\eta=u=\varepsilon/2\), and project.
  Bound~\eqref{eq:four-shear-dense-bound} gives the time bound in
  \eqref{eq:entry-dense-time}, using
  \(\log(3n/\eta)\le\log(12n/\varepsilon)\), \(n\ge256\), and
  \((37/3)n^2\le37|\mc{S}|\).

  For the last two bounds in~\eqref{eq:unified-main-time}, take
  \(\eta=u=\varepsilon\) and project the full-group construction.
  The first two bounds in~\eqref{eq:four-segment-bounds} suffice.
  One may minimize over
  the admissible entry-pattern schedules, these full-group bounds, and the
  projected time in Proposition~\ref{prop:full-group-amplification}.
  For columns, each marginal bound also bounds the mixing time.
\end{proof}

\begin{proof}[Proof of Corollary~\ref{cor:column-mixing-intro}]
  The lower bound follows from Corollary~\ref{cor:all-error-lower}
  under the field hypothesis. For \(k\le n/2\), take
  \(\eta=u=\varepsilon/2\) in Proposition~\ref{prop:exact-density-constant}
  for the pattern \([n]\times[k]\).
  Bounds~\eqref{eq:two-block-affine} and~\eqref{eq:four-block-affine}
  give the first two cases in the corollary.
  For \(k>n/2\), put \(\eta=u=\varepsilon\) and take \(a=4/3\) in
  \eqref{eq:four-shear-affine-bound}. Since \(n\ge256\) and
  \(k\ge(n+1)/2\), this gives
  \[
    \begin{aligned}
      T_{\mathrm{shears}}(\varepsilon)
      &\le\left(12\log5-\frac43\right)n(n-1)
      +36(n-1)\log\frac{3n}{\varepsilon}+12\\
      &\le36n\bigl(k+\log(n/\varepsilon)\bigr).
    \end{aligned}
  \]
  The field-threshold and amplification argument in the proof of
  Theorem~\ref{thm:main} applies to these schedules as well, giving both
  guarantees at every integer time at least the corresponding bound.
\end{proof}

\subsection{Sharper asymptotic constants}
\label{sec:leading-constants}

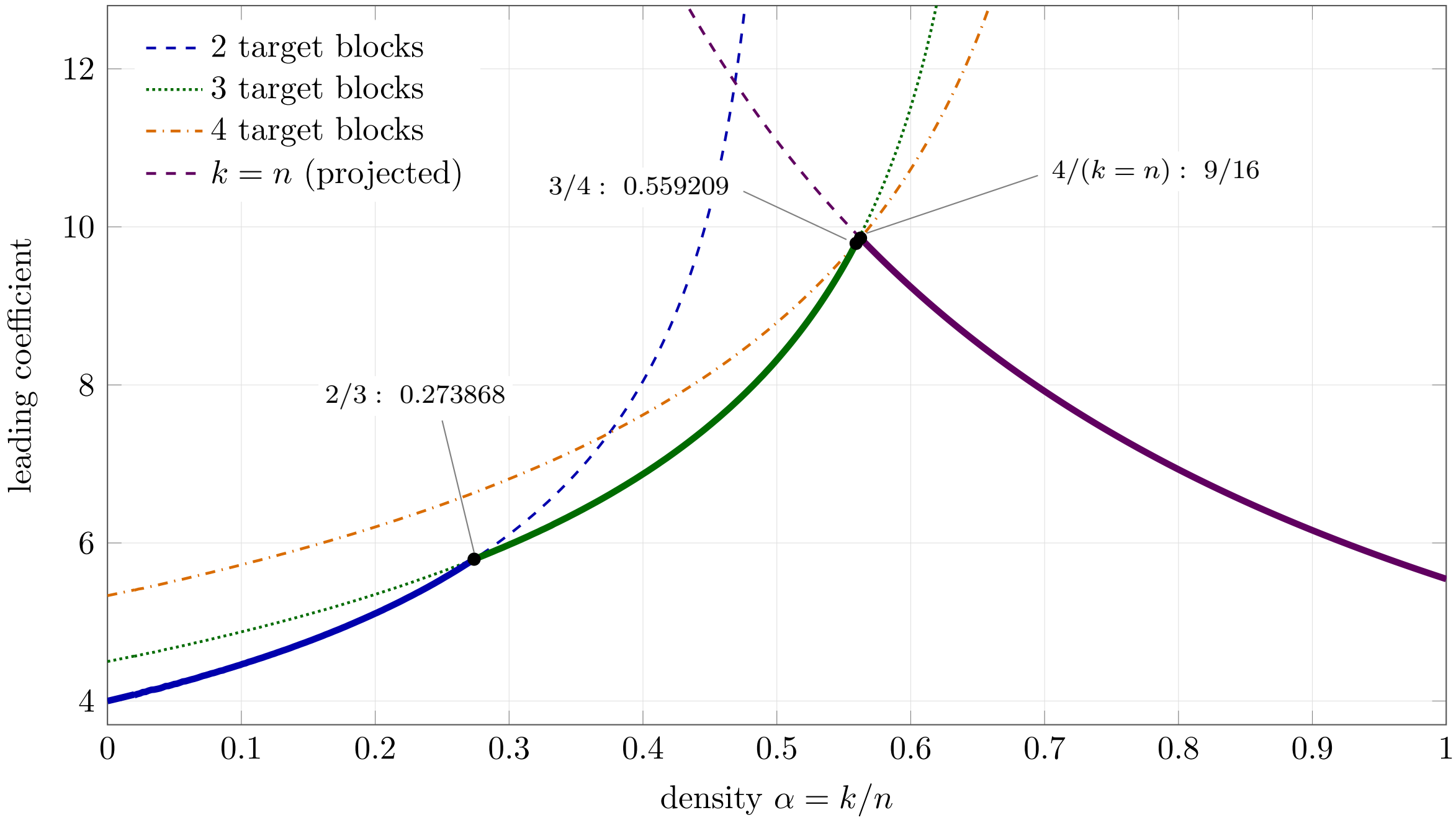
\begin{figure}[!ht]
  \centering
  \begin{tikzpicture}
    \begin{axis}[
        width=0.96\textwidth,height=0.56\textwidth,
        xmin=0,xmax=1,ymin=3.7,ymax=12.8,
        xlabel={density \(\alpha=k/n\)},ylabel={leading coefficient},
        xtick={0,0.1,0.2,0.3,0.4,0.5,0.6,0.7,0.8,0.9,1},
        ytick={4,6,8,10,12},
        tick label style={font=\small},label style={font=\small},
        grid=major,grid style={draw=black!12,line width=0.2pt},
        axis line style={draw=black!65},
        legend style={draw=none,fill=white,font=\small,
        at={(0.02,0.98)},anchor=north west},
        legend columns=1,legend cell align={left},unbounded coords=discard
      ]
      \addplot[blue!68!black,dashed,line width=0.8pt,
      domain=0.02:0.498,samples=220]
      {2/x*ln(1/(1-2*x))};
      \addlegendentry{2 target blocks}
      \addplot[green!42!black,densely dotted,line width=0.85pt,
      domain=0.02:0.665,samples=220]
      {3/x*ln(2/(2-3*x))};
      \addlegendentry{3 target blocks}
      \addplot[orange!85!black,dashdotted,line width=0.85pt,
      domain=0.02:0.748,samples=220]
      {4/x*ln(3/(3-4*x))};
      \addlegendentry{4 target blocks}
      \addplot[violet!75!black,dashed,line width=0.85pt,
      domain=0.4:1,samples=160]
      {8*ln(2)/x};
      \addlegendentry{\(k=n\) (projected)}
      \addplot[blue!68!black,line width=2pt,forget plot,
      domain=0.02:0.273868471279,samples=120]
      {2/x*ln(1/(1-2*x))};
      \addplot[green!42!black,line width=2pt,forget plot,
      domain=0.273868471279:0.559208838308,samples=120]
      {3/x*ln(2/(2-3*x))};
      \addplot[orange!85!black,line width=2pt,forget plot,
      domain=0.559208838308:0.5625,samples=70]
      {4/x*ln(3/(3-4*x))};
      \addplot[violet!75!black,line width=2pt,forget plot,
      domain=0.5625:1,samples=140]
      {8*ln(2)/x};

      \addplot[blue!68!black,line width=2pt,forget plot]
      coordinates {(0,4.000000000000) (0.002,4.008021397539) (0.004,4.016085848632) (0.006,4.024193744756) (0.008,4.032345482471) (0.01,4.040541463504) (0.015,4.061227664628) (0.02,4.082199452026)};
      \addplot[green!42!black,densely dotted,line width=0.85pt,forget plot]
      coordinates {(0,4.500000000000) (0.002,4.506763530448) (0.004,4.513554244172) (0.006,4.520372326075) (0.008,4.527217962851) (0.01,4.534091343014) (0.015,4.551397424523) (0.02,4.568881122706)};
      \addplot[orange!85!black,dashdotted,line width=0.85pt,forget plot]
      coordinates {(0,5.333333333333) (0.002,5.340457111758) (0.004,5.347606326595) (0.006,5.354781131510) (0.008,5.361981681488) (0.01,5.369208132856) (0.015,5.387388618005) (0.02,5.405734477584)};
      \addplot[only marks,mark=*,mark size=1.8pt,black,forget plot]
      coordinates {(0.273868471279,5.794688809859)
        (0.559208838308,9.791646177114)
      (0.5625,9.858093234630)};
      \node[font=\scriptsize,anchor=south,fill=white,inner sep=2pt]
      at (axis cs:0.23,7.6) {\(2/3:\ 0.273868\)};
      \draw[black!50,thin]
      (axis cs:0.25,7.55) -- (axis cs:0.273868471279,5.88);
      \node[font=\scriptsize,anchor=east,fill=white,inner sep=2pt]
      at (axis cs:0.47,10.5) {\(3/4:\ 0.559209\)};
      \draw[black!50,thin]
      (axis cs:0.475,10.45) -- (axis cs:0.552,9.84);
      \node[font=\scriptsize,anchor=west,fill=white,inner sep=2pt]
      at (axis cs:0.70,10.7) {\(4/(k=n):\ 9/16\)};
      \draw[black!50,thin]
      (axis cs:0.695,10.65) -- (axis cs:0.567,9.92);
    \end{axis}
  \end{tikzpicture}
  \caption{Asymptotic leading-coefficient bounds across all densities.}
  \label{fig:kappa-density}
\end{figure}

When \(k=\omega(\log(n/\varepsilon))\), the exact schedules give smaller
leading constants.

\begin{proposition}[Sharper asymptotic constants]
  \label{prop:optimized-density-constant}
  Let \(n\to\infty\), \(1\le k_n\le n\), \(0<\varepsilon_n\le1/4\),
  \(q_n\ge Q_{\rm all}(n)\), and \(k_n=\omega(\log(n/\varepsilon_n))\).
  Suppose \(k_n/n\to\alpha\in[0,1]\). Interpreting \(8\log2/0=+\infty\),
  \begin{equation}\label{eq:optimized-density-mixing}
    \limsup_{n\to\infty}
    \frac{t_{\mathrm{mix}}^{(k_n)}(\varepsilon_n)}{nk_n}
    \le\min\left\{\min_{p\in \mathbb{Z}_{\geq 2}}\kappa_p(\alpha),
    \frac{8\log2}{\alpha}\right\},
  \end{equation}
  where
  \begin{equation}\label{eq:kappa-candidates}
    \kappa_p(\alpha)=
    \begin{cases}
      \displaystyle\frac{p^2}{p-1},&\alpha=0,\\[1mm]
      \displaystyle\frac p\alpha
      \log\frac{p-1}{p-1-p\alpha},&0<\alpha<1-1/p,\\[1mm]
      +\infty,&1-1/p\le\alpha\le1.
    \end{cases}
  \end{equation}
  The same right-hand side bounds \(\limsup_{n\to\infty}t_n/(nk_n)\),
  where \(t_n\) is the least integer time at which the conditional
  guarantee holds for any initial state:
  \[
    \sup_{X\in\mathrm{SL}_n(\F_{q_n})}
    \Pr_{\sigma:|\sigma|=t_n}
    \left[d_{[n]\times[k_n]}^\sigma(X)>\varepsilon_n\right]\le\varepsilon_n.
  \]
\end{proposition}

Figure~\ref{fig:kappa-density} shows the stated bounds and their
intersections.

\begin{proof}
  For any fixed \(p\ge2\) with \(\alpha<1-1/p\), blocks have size
  \(n/p+O(1)\), so the collection means satisfy
  \(n(n-1)s_R/(nk_n)\to\kappa_p(\alpha)/p\), including at \(\alpha=0\).
  The square-root and logarithmic terms in \(N_R\), and the burn-in,
  are \(o(nk_n)\) by assumption. Thus
  \[
    T_{\mathrm{burn}}(\varepsilon_n)+T_p(\varepsilon_n)
    =(\kappa_p(\alpha)+o(1))nk_n.
  \]

  We also have a mixing bound from the shear schedule by projection, with
  \begin{align*}
    T_{\mathrm{shears}}(\varepsilon_n)&=(8\log2+o(1))n^2\\
    &=\left(\frac{8\log2}{\alpha}+o(1)\right) nk_n\qquad(\alpha>0).
  \end{align*}
  The constants in~\eqref{eq:unified-leading-constants} follow from
  \(\kappa_2(0)=4\), \(\kappa_3(\alpha)\le6\log4\) for
  \(\alpha\le1/2\), and the full-group bound.
\end{proof}

Equation~\eqref{eq:unified-leading-constants} follows by taking a subsequence
attaining the limsup of the normalized sufficient times, then a further
subsequence along which \(k_n/n\) converges.

\paragraph{A sharper leading constant for sparse entry patterns.}
Suppose \(q\ge Q_{\rm all}(n)\), \(0<\varepsilon<1\),
\(|\mc{S}|=o(n^2)\), and \(n\log(n/\varepsilon)=o(|\mc{S}|)\).
The construction in Section~\ref{sec:entry-mixing} gives both guarantees
in~\eqref{eq:main-conditional-mixing} at a sufficient time
\begin{equation}\label{eq:entry-leading}
  (12+o(1))|\mc{S}|.
\end{equation}
Take \(\overline d=n/3\) and, for \(s_{\mathrm{rem}}>0\),
\(\underline d=9s_{\mathrm{rem}}/(5n)\), with two low blocks.
If \(s_{\mathrm{rem}}=0\), take \(\underline d=\overline d\) and omit intervals 3--5.
Now \(|\mathcal H|=o(n)\), \(s_{\mathrm{rem}}=o(n^2)\), and
\(\sum_{i\in\mathcal M}|H_i|\le5n/9\), eventually leaving \(n/3\) unused rows.
Use \(\log(14n/\varepsilon)\) in the collection and growth bounds to
preserve the failure allowances. Burn-in and growth take
\(O(n\log(n/\varepsilon))\) steps. The source margin
\(1-5/9-1/3=1/9\) gives \(v=O(n)\), so the tail terms are
\(O(\sqrt{|\mc{S}|n\log(n/\varepsilon)}+n\log(n/\varepsilon))
=o(|\mc{S}|)\).

The two transposed blocks for \(\mathcal H\) cost at most
\((12+o(1))(|\mc{S}|-s_{\mathrm{rem}})\).
For \(s_{\mathrm{rem}}>0\), put
\(\theta=s_{\mathrm{rem}}^{-1}\sum_{i\in\mathcal M}|\mc{S}_i|\).
The two low blocks cost at most
\((18/5+4(1-\theta)+o(1))s_{\mathrm{rem}}\).
For the trees, the ordered-pool bound~\eqref{eq:ordered-source-pool} and
Corollary~\ref{cor:target-source-collection} give
\[
  \frac{\mu_3}{s_{\mathrm{rem}}}\le
  \max_{9/5\le r\le18/5}
  3r\log\frac{1-\theta/r}{2/3-\theta/r}.
\]
The function inside the maximum equals
\(3\int_0^{1/3}r^2/((1-t)r-\theta)\,dt\), so it is convex in
\(r\) and \(\theta\).
It therefore suffices to check \(r\in\{9/5,18/5\}\) and
\(\theta\in\{0,1\}\). Including the low blocks, the two
\(\theta\)-endpoint bounds are
\[
  \frac{38}{5}+\frac{54}{5}\log\frac32<12,
  \qquad \frac{18}{5}+\frac{27}{5}\log4<12.
\]
Adding the \(\mathcal H\) contribution proves~\eqref{eq:entry-leading}; the field and error
arguments in Section~\ref{sec:mixing-assembly} apply unchanged.

\section{Lower bounds for entry marginals}\label{sec:lower-bounds}

Two obstructions match the upper-bound scale: a short walk has too few
possible outputs, and information from one initial entry needs time to
reach another row. Pillai and Smith~\cite[Theorem~1.2]{PS} give the
corresponding support and propagation obstructions for binary column
projections. We use the same principles over every finite field.

\subsection{The number of requested entries}
\begin{proposition}\label{prop:reachability-lower}
  For every nonempty \(\mc{S}\subseteq[n]\times[n]\), initial state \(X\),
  and integer \(t\ge0\),
  \begin{equation}\label{eq:reachability-lower-general}
    \|\mathcal L_X(X_t|_{\mc{S}})-\pi_{\mc{S}}\|_{\TV}
    \ge1-2q^{1-|\mc{S}|}[n(n-1)q]^t.
  \end{equation}
  For every skeleton \(\sigma\),
  \begin{equation}\label{eq:conditional-reachability-lower}
    d_{\mc{S}}^\sigma(X)\ge1-2q^{1-|\mc{S}|+|\sigma|}.
  \end{equation}
  For full-column sets, replace \(2q^{1-|\mc{S}|}\) in both bounds by
  \(1/|\operatorname{supp}\pi_{\mc{S}}|\).
\end{proposition}
\begin{proof}
  A prescribed value on \(\mc{S}\) has at most
  \(q^{n^2-|\mc{S}|}\) matrix completions. Since
  \(|\mathrm{SL}_n(\F_q)|\ge q^{n^2-1}/2\), every atom of
  \(\pi_{\mc{S}}\) has mass at most \(2q^{1-|\mc{S}|}\).
  The walk has at most \([n(n-1)q]^t\) possible outputs, or
  \(q^t\) after fixing its skeleton. Test membership in this support.
  For full columns, the stationary law is uniform, giving the exact
  atom mass in the final assertion.
\end{proof}

\subsection{A single entry}
\begin{proposition}\label{prop:coupon-lower}
  For every nonempty \(\mc{S}\subseteq[n]\times[n]\) and integer \(t\ge0\),
  \begin{equation}\label{eq:column-support-tv}
    d_{\mc{S}}(t)\ge
    1-\frac{q^{n-1}-1}{q^n-1}-\frac{e^{t/n}-1}{n-1}.
  \end{equation}
  Moreover,
  \begin{equation}\label{eq:accuracy-lower}
    d_{\mc{S}}(t)\ge\frac12\left(1-\frac1n\right)^t.
  \end{equation}
\end{proposition}
\begin{proof}
  Choose \((i,j)\in\mc{S}\) and a row \(h\ne i\). Start at a matrix
  in \(\mathrm{SL}_n(\F_q)\) whose \(j\)-th column is \(e_h\). Observe that the number of rows \(R_t\) reached by transvection paths from \(h\) has
  \(\E[R_{t+1}|R_t]\leq R_t + R_t(n-R_t)/(n(n-1))\le R_t+ R_t/n\), so
  \(\E[R_t]\le(1+1/n)^t\le e^{t/n}\). Symmetry gives
  \[
    \Pr[\text{a path reaches }i\text{ by }t]
    =\frac{\E[R_t]-1}{n-1}\le\frac{e^{t/n}-1}{n-1}.
  \]
  The stationary column distribution is uniform over nonzero vectors, so its
  \(i\)-th coordinate is zero with probability
  \((q^{n-1}-1)/(q^n-1)\le1/2\).
  Testing this entry proves~\eqref{eq:column-support-tv}.

  For the error dependence, put
  \(f(X)=\mathbf1_{\{X_{ij}=0\}}-\pi(X_{ij}=0)\).
  Averaging over the coefficient of a fixed row pair averages over a subgroup action, hence is an orthogonal
  projection on \(L^2(\pi)\). The transition operator \(P\) is therefore
  positive semidefinite. Updates not targeting row \(i\) fix \(f\), so
  \(\langle f,Pf\rangle_\pi\ge(1-1/n)\|f\|_2^2\).
  Jensen's inequality for the spectral measure of \(f\) gives
  \[
    \langle f,P^tf\rangle_\pi
    \ge(1-1/n)^t\|f\|_2^2.
  \]
  Since \(\E_\pi[|f|]=2\|f\|_2^2\), some initial state satisfies
  \(|P^tf(X)|\ge(1-1/n)^t/2\). This single-entry test
  proves~\eqref{eq:accuracy-lower}.
\end{proof}

\subsection{Mixing times and pre-cutoff}
For the next two corollaries, let \(n\geq 256\), \(1\leq k\leq n\), and \(0<\varepsilon\leq \frac{1}{4}\).
\begin{corollary}[Two-term lower bound]\label{cor:all-error-lower}
  \[
    t_{\mathrm{mix}}^{(k)}(\varepsilon)\ge
    \max\left\{\frac{nk-5/2}{1+2\log n/\log q},\;
    \frac{n-1}{2}\log\frac{n}{8\varepsilon}\right\}.
  \]
\end{corollary}
\begin{proof}
  Apply Proposition~\ref{prop:reachability-lower} with
  \(|\mc{S}|=nk\), using \(\log(8/3)<(3/2)\log2\).
  Proposition~\ref{prop:coupon-lower} gives the two lower bounds
  \(n\log((n+3)/4)\) and \((n-1)\log(1/(2\varepsilon))\).
  Their maximum is at least half their sum, which is at least the
  second term displayed above.
\end{proof}

\begin{corollary}[Entry-marginal mixing times]\label{cor:entry-mixing-time}
  For every nonempty
  \(\mc{S}\subseteq[n]\times[n]\),
  every integer \(t\ge0\) with \(d_{\mc{S}}(t)\le\varepsilon\) satisfies
  \[
    t\ge\max\left\{\frac{|\mc{S}|-5/2}{1+2\log n/\log q},\;
    \frac{n-1}{2}\log\frac{n}{8\varepsilon}\right\}.
  \]
  If \(q\ge Q_{\rm all}(n)\), the least such time is
  \(\Theta(|\mc{S}|+n\log(n/\varepsilon))\), with absolute implied constants.
\end{corollary}
\begin{proof}
  The proof of Corollary~\ref{cor:all-error-lower} applies with \(nk\)
  replaced by \(|\mc{S}|\). Under the field hypothesis, combining this
  lower bound with Theorem~\ref{thm:main} gives the asserted order.
\end{proof}

For nonempty entry sets \(\mc{S}_n\subseteq[n]\times[n]\) over
\(\F_{q_n}\), write
\[
  d_n(t)=d_{\mc{S}_n}(t),\qquad
  t_n(\varepsilon)=\min\{t\ge0:d_n(t)\le\varepsilon\},
\]
using the worst-case distance in~\eqref{eq:entry-distance}.
Although the observed entries need not form a Markov chain, \(d_n(t)\)
is nonincreasing by the Markov property of the full walk and convexity
of total variation. For \(\mc{S}_n=[n]\times[k_n]\), this recovers
\(t_n(\varepsilon)=t_{\mathrm{mix}}^{(k_n)}(\varepsilon)\).
The sequence of entry marginals has \emph{pre-cutoff} if
\begin{equation}\label{eq:precutoff-definition}
  \sup_{0<\varepsilon<1/2}\limsup_{n\to\infty}
  \frac{t_n(\varepsilon)}{t_n(1-\varepsilon)}<\infty.
\end{equation}
\begin{corollary}[Pre-cutoff for entry marginals]\label{cor:precutoff}
  Put \(\tau_n=|\mc{S}_n|+n\log n\). If
  \(\log q_n/\log n\to\infty\), then, for every fixed \(0<c<1/2\),
  \[
    d_n(\floor{c\tau_n})\longrightarrow1.
  \]
  The field bound \(q_n\ge Q_{\rm all}(n)\) gives this lower limit and, for a
  sufficiently large absolute constant \(C\),
  \[
    d_n(\ceil{C\tau_n})\longrightarrow0.
  \]
  Hence the sequence has pre-cutoff under this field bound.
\end{corollary}
\begin{proof}
  Fix \(0<c<1/2\). If \(|\mc{S}_n|\ge n\log n\), then
  \(\floor{c\tau_n}\le2c|\mc{S}_n|\), and
  Proposition~\ref{prop:reachability-lower} gives distance tending to one
  because \(\log q_n/\log n\to\infty\). Otherwise
  \(\floor{c\tau_n}\le2cn\log n\), and
  Proposition~\ref{prop:coupon-lower} gives the same conclusion:
  the stationary zero probability and the forward-path hitting bound
  both tend to zero.

  Theorem~\ref{thm:main} gives the upper bound at error \(1/n\).
  Our field bound also implies \(\log q_n/\log n\to\infty\), so the
  lower limit applies.
  Monotonicity then bounds the limsup in~\eqref{eq:precutoff-definition}
  by \(C/c\), uniformly over \(0<\varepsilon<1/2\).
\end{proof}

\section{Trapdoored matrices from transvection walks}
\label{sec:tdms}
\subsection{Constructions for three distributions}
Following~\cite{braverman2025practical,vaikuntanathan2025improving}, an
efficiently samplable distribution of pairs \((M_n,C_n)\) is
\(T(n)\)-efficient and trapdoored relative to \((\mathcal U_n)\) if
\(C_n\) computes \(v\mapsto M_n v\) in at most \(T(n)\) field operations
and \(M_n\) is computationally indistinguishable from \(\mathcal U_n\).

\begin{theorem}\label{thm:transvection-trapdoors}
  Assume Conjecture~\ref{conj:trapdooredness} holds. For every sequence of prime powers
  \(2\le q_n\le 2^{n^{O(1)}}\), there are universal trapdoored-matrix distributions for the uniform distributions on:
  \begin{enumerate}[label={\upshape(\roman*)}]
    \item \(\mathrm{SL}_n(\F_{q_n})\), by a short product of transvections \(X_T\).
    \item \(\GL_n(\F_{q_n})\), by the composition of a short product of transvections \(X_T\) and a scaling of the first row by a uniformly random value \(h\sim \F_{q_n}\setminus \{0\}\).
    \item \(\F_{q_n}^{n\times n}\), by the sum of the previous construction and a uniformly random diagonal matrix.
  \end{enumerate}
  All three ensembles are
  \(O(n\log^2 n)\)-efficient, and their indistinguishability is uniform over the choice of
  \((q_n)\). In cases~(i) and~(ii), there is an identically efficient circuit for multiplication by the inverse.
\end{theorem}

The proof appears in Appendix~\ref{app:trapdoor-proofs}.
Under the field condition of Theorem~\ref{thm:main},
part~(iii) follows directly from~(ii), since a uniform square matrix is
singular with probability at most \(1/(q_n-1)\), negligible in \(n\).

\subsection{Circuit depth}
The direct circuit for a \(T\)-step walk uses two arithmetic gates per
transvection and has depth \(O(T)\). To reduce depth, we precompute a parallelized circuit for the walk by dividing the walk
into blocks of at most \(m\) transvections and precomputing each block's
product. We then apply these block products sequentially to the input vector.

\begin{proposition}\label{prop:sparse-transvection-blocks}
  Let \(n\ge2\) and let \(q\) be a prime power. Let \((B_s)_{s\geq 0}\) be the transvection walk on \(\mathrm{SL}_n(\F_q)\) started at \(I_n\), and put
  \[
    N_s=\operatorname{nnz}(B_s-I_n),
    \qquad
    R_s=\max_{i\in[n]}\operatorname{nnz}\bigl(B_s[i]\bigr).
  \]
  Then
  \begin{equation}\label{eq:expected-block-sparsity}
    \E[N_s]
    \le n(e^{s/n}-1).
  \end{equation}
  Moreover, for \(2\le r\le n\),
  \begin{equation}\label{eq:block-row-support-tail}
    \Pr(R_s\ge r)
    \le n\left[1-\left(1-\frac1n\right)^s\right]^{r-1}
    \le n\left(\frac sn\right)^{r-1}.
  \end{equation}

  Consequently, for integers \(T\ge1\) and \(m\ge1\), grouping a \(T\)-step walk into \(\ceil{T/m}\) consecutive blocks of at most \(m\) steps gives a circuit with expected
  gate count at most
  \[
    \frac{2nT}{m}(e^{m/n}-1).
  \]
  If \(m\le n\), its expected depth is
  \begin{equation}\label{eq:expected-block-depth}
    O\left(
      \ceil{\frac Tm}
      \log\left(
        2+\frac{\log n}{\max\{1,\log(n/m)\}}
      \right)
    \right).
  \end{equation}
\end{proposition}

The proof appears in Appendix~\ref{app:trapdoor-proofs}.

For \(T=\Theta(n\log^2 n)\) and \(m=\lfloor n/\log\log n \rfloor\), this gives circuits for each of the three ensembles (and their inverses in parts~(i) and~(ii)) with expected size \(2T(1+O(\frac{1}{\log \log n}))\) and expected depth \(O(\log^2n (\log \log n)^2)\).

\subsection{Cryptanalysis of short transvection walks}
\label{sec:short-walk-cryptanalysis}
The following test searches for dependent sets of \(\ceil{k/2}\) rows in a full-column-rank
\(n\times k\) matrix \(X\), which are rare in randomly sampled matrices.

\begin{algorithm}[H]
  \caption{Sampled row-dependence test}\label{alg:short-row-dependence}
  \begin{algorithmic}[1]
    \Require full-column-rank \(X\in\F_q^{n\times k}\), with \(1\le k\le n/2\), and an integer \(a\ge1\)
    \State \(r\gets\ceil{k/2}\)
    \For{\(i=1,\ldots,a\)}
    \State Independently sample \(S\) uniformly from \(\binom{[n]}r\)
    \If{\(\rank X[S]<r\)}
    \State \Return \textbf{True}
    \EndIf
    \EndFor
    \State \Return \textbf{False}
  \end{algorithmic}
\end{algorithm}

This algorithm distinguishes efficiently for
\(T\le n\log n\).

\begin{proposition}\label{prop:short-walk-cryptanalysis}
  Let \(k=\floor{n/3}\), \(q\ge n^2\), \(a=\ceil{n/k}\), and start the walk at \(X_0=
    \begin{pmatrix}I_k\\0
  \end{pmatrix}\).
  For all sufficiently large \(n\), Algorithm~\ref{alg:short-row-dependence}
  has distinguishing advantage at least
  \begin{equation}\label{eq:sampled-dependence-lower}
    e^{-T/(n-1)-1}
    -e^{-\Omega(n)}
  \end{equation}
  for every integer \(T\ge0\).
\end{proposition}
The proof appears in Appendix~\ref{app:simultaneous-routing}.

However, for \(q\geq 2^{\Omega(n)}\) and \(T\geq C n\log^2 n\), the attack's behavior changes dramatically: except for low-probability bad skeleton events, the advantage of the attack is exponentially small for one sample and remains negligible for \(2^{o(n)}\) samples, even if query sets are strategically chosen as functions of the skeleton. This happens because, conditional on a good skeleton, \textit{every} subset of \(r\) rows (for every \(r\)) is routed to by \textit{every other} set of \(r\) rows (in the sense of Section~\ref{sec:routing-burnin}), and therefore each strict minor is exponentially close in TV distance to uniform on \(\F_q\).

\begin{proposition}\label{prop:sampled-dependence-upper}
  There are absolute constants \(c,C>0\) such that, for every \(n\ge4\), prime power \(q\), \(1\le r\le k\le n\), and
  full-column-rank \(X_0\in\F_q^{n\times k}\), the following holds.
  Choose \(a\geq 1\) \(r\)-subsets of \([n]\) as arbitrary functions of the skeleton. For every integer \(\tau\) with \(Cn\log^2 n\le\tau\le T\),
  the probability that some sample indexes dependent rows of
  \(X_T\) is at most
  \begin{equation}\label{eq:sampled-dependence-upper}
    e^{-c\tau/n}+a\frac\tau q.
  \end{equation}
\end{proposition}

\begin{proof}
  Split the last \(\tau\) steps into intervals of lengths
  \(\floor{\tau/2}\) and \(\ceil{\tau/2}\).
  By Lemma~\ref{lem:simultaneous-routing}
  (Appendix~\ref{app:simultaneous-routing}), both intervals route every
  pair of equal-sized row sets except on one bad skeleton event of
  probability at most \(e^{-c\tau/n}\), after adjusting \(c,C\).
  Fix the whole skeleton on this event's complement and condition on
  the first \(T-\tau\) coefficients.

  For any minor of order \(1\le s\le k\) with \(s<n\), its columns
  have rank \(s\) at time \(T-\tau\). Take a subset of \(s\) rows forming a basis and route them to an \(s\)-set different from the target
  rows in the first interval, then to the targets in the second.
  The concatenated paths use at least one transmission edge.
  The coefficient argument in Lemma~\ref{lem:routing-determinant}
  therefore makes this minor a nonconstant multiaffine polynomial
  of degree at most \(\tau\).

  Write it as \(zg+h\), where \(z\) is one transmission edge coefficient,
  \(g,h\) do not involve \(z\), and \(g\not\equiv0\).
  Conditional on the other coefficients and \(g\ne0\), its value
  is uniform on \(\F_q\). Since \(\deg g\le\tau-1\), Schwartz--Zippel
  bounds \(\Pr[g=0]\) by \((\tau-1)/q\).
  Averaging over the prefix coefficients, we conclude that for every
  good skeleton, each nonempty minor of order less than \(n\)
  has conditional law within \((\tau-1)/q\) in TV of
  \(\Unif(\F_q)\).

  Each such minor is therefore zero with probability at most \(\tau/q\). For \(r<n\), dependence of the sampled rows forces their minor on the first \(r\) columns to vanish; for \(r=n\), dependence is impossible because \(X_T\) is invertible. Union-bound over the \(a\) queries and add the bad skeleton probability.

  A similar combination of superconcentrators (see Lemma~\ref{lem:simultaneous-routing}) and Schwartz--Zippel for showing rank nondegeneracy was applied in another context by Cheung, Kwok, and Lau~\cite[Appendix~B]{CheungKwokLau2013}.
\end{proof}

At \(\tau=\Theta(n\log^2 n)\), the skeleton term is
\(e^{-\Omega(\log^2 n)}\), independently of \(a\), and acceptance
is negligible for \(a=2^{o(n)}\) when \(q\ge2^{\Omega(n)}\). Whether other nonlocal tests distinguish short walks remains open.

\section{Acknowledgements}
We thank River Newman for discussions of algebraic geometry.

\section{Use of LLM tools}
We used a large language model (GPT Astra 6.0) in the following ways. The first bullet contains the meaningful research contributions of the LLM.
\begin{itemize}
  \item After we came up with the rational slice construction in Section~\ref{sec:rectangular} and asked the LLM to use it to prove mixing, it identified and applied the techniques from previous works to convert a slice into a mixing bound (resulting in Section~\ref{sec:rational-sections}). It also gave a preliminary version of the \(k=n\) slice construction (resulting in Section~\ref{sec:four-segment-alternative}) on request, and provided a proof of Appendix Lemma~\ref{lem:simultaneous-disjoint-routing} on request. With further prodding, the LLM noted that its proof was similar in approach to the previous work referenced in that lemma, and also pointed out that our use of that lemma in Proposition~\ref{prop:sampled-dependence-upper} is similar to the work referenced in that proposition.
  \item It optimized parameters for several constructions in Section~\ref{sec:assembly} and simplified our small-field burn-in argument in
    Appendix~\ref{app:small-field-burnin}.
  \item It drafted standard/straightforward arguments, including the extension to all error levels
    in Section~\ref{sec:assembly}, the construction reduction
    in Appendix~\ref{app:trapdoor-proofs}, and the preliminary lemmas in
    Appendix~\ref{app:standard-estimates}.
  \item It helped align the exact theorem/lemma statements and notation across sections and produced the diagrams under direction. It also helped with many minor edits.
\end{itemize}
The authors remain responsible for the correctness of the work.

\begingroup
\small
\setlength{\bibsep}{2pt}
\bibliographystyle{alpha}
\bibliography{shared/references}
\endgroup

\clearpage
\appendix

\section{Field-uniform rank burn-in}\label{app:small-field-burnin}

For small fields, we make a sampled row set full-rank by taking slightly
more than \(k\) rows. The following estimate is analogous to the burn-in
bound of Pillai and Smith~\cite[Lemma~3.3]{PS}.
\begin{proposition}[small-field rank burn-in]\label{prop:moderate-field-burnin}
  Let \(q\) be a prime power, let \(1\le k<n\), and suppose
  \(0<\beta\le1/10\) and \(\log(1/\beta)\le(n-k)/288\).
  Start the walk at any full-column-rank \(X_0\in\F_q^{n\times k}\),
  and independently choose a uniform row set \(S\subseteq[n]\) of size
  \begin{equation}\label{eq:ell-beta}
    \ell=k+\ceil{4\log_q(1/\beta)}.
  \end{equation}
  Then
  \[
    \Pr_{X_0,S}[\rank X_t[S]<k]\le\beta
    \qquad\text{for every integer }\quad
    t\ge17\frac{n^2}{n-k}\log\frac{nq}{\beta}.
  \]
\end{proposition}
The proof is at the end of this appendix.

Combining this bound with Lemma~\ref{lem:routing-rank} removes the field-size
dependence from the burn-in time. The extra sampled rows suppress rank
deficiency even when the field is small.

\begin{theorem}[field-uniform random-subset burn-in]\label{thm:field-uniform-subset-burnin}
  Let \(q\) be a prime power, let \(1\le k<n\), and suppose
  \(0<\beta\le1/10\) and \(\log(1/\beta)\le(n-k)/288\).
  Start the walk at any full-column-rank \(X_0\in\F_q^{n\times k}\),
  and independently choose a uniform row set \(S\subseteq[n]\) of size
  \(k+\ceil{4\log_q(1/\beta)}\). Then
  \begin{equation}\label{eq:uniform-subset-burnin}
    \Pr_{X_0,S}[\rank X_t[S]<k]\le\beta
    \qquad\text{for every integer }\quad
    t\ge60\frac{n^2}{n-k}\log\frac n\beta.
  \end{equation}
\end{theorem}

\begin{proof}
  Set
  \[
    T=\ceil{9\frac{n^2}{n-k}\log\frac n\beta}.
  \]
  With \(\zeta=\min\{\beta/2,(16n)^{-1}\}\), we have
  \(\log(n/\zeta)\le2\log(n/\beta)\), so the shorter bound in
  Lemma~\ref{lem:routing-rank} applies. The hypotheses also give \(n-k\ge32\).
  If \(q>2T/\beta\), Lemma~\ref{lem:routing-rank} gives failure probability at
  most \(\beta\) at every time \(t\ge T\). Otherwise, since
  \(T\le10\frac{n^2}{n-k}\log(n/\beta)\),
  \[
    \log\frac{nq}{\beta}
    \le\log\left(\frac{20n^3}{(n-k)\beta^2}\log\frac n\beta\right)
    \le\frac72\log\frac n\beta.
  \]
  The last inequality uses \(n-k\ge32\) and \(\log x\le x/2\).
  Proposition~\ref{prop:moderate-field-burnin} therefore gives the claimed
  coefficient \((7/2)\cdot17<60\). The stated time bound also implies
  \(t\ge T\), covering the routing case.
\end{proof}

To prove Proposition~\ref{prop:moderate-field-burnin}, we bound a potential
that dominates the rank-failure probability. Its evolution is described by
a one-dimensional chain that drifts toward \(n\), where the potential vanishes.

\subsection{A rank-deficiency potential and its exact dual chain}

For \(0\le \ell\le n\), let \(S\) be a uniformly random \(\ell\)-subset of
\([n]\). The expected number of one-dimensional subspaces of \(\ker X[S]\) is
\begin{equation}\label{eq:def-rank-potential}
  \mathcal F_\ell(X)
  =
  \E_S\left[
    \frac{q^{\,k-\rank X[S]}-1}{q-1}
  \right].
\end{equation}
Since a rank-deficient sample has nullity at least one,
\begin{equation}\label{eq:rank-failure-by-potential}
  \Pr_S[\rank X[S]<k]\le \mathcal F_\ell(X).
\end{equation}

Let \(\widehat L_s\) be the birth--death chain on \(\{0,\ldots,n\}\) with
absorbing endpoints and, for \(1\le j<n\), transition probabilities
\begin{equation}\label{eq:dual-discrete-transitions}
  \begin{aligned}
    \Pr_j[\widehat L_1=j+1]&=\frac{j(n-j)}{n(n-1)},\\
    \Pr_j[\widehat L_1=j-1]&=\frac1q\frac{j(n-j)}{n(n-1)}.
  \end{aligned}
\end{equation}
The chain otherwise stays at \(j\). Its state indexes the sample size.
Conditional on moving, \(\widehat L\) moves upward with probability
\(q/(q+1)\) and downward with probability \(1/(q+1)\).

\begin{lemma}[rank-moment duality]\label{lem:rank-moment-duality}
  For every full-column-rank \(X_0\in\F_q^{n\times k}\), every integer \(s\ge0\), and every
  \(0\le\ell\le n\),
  \begin{equation}\label{eq:rank-moment-duality}
    \E_{X_0}\left[\mathcal F_\ell(X_s)\right]
    =
    \E_\ell\left[\mathcal F_{\widehat L_s}(X_0)\right].
  \end{equation}
\end{lemma}

\begin{proof}
  For \(A\subseteq[n]\), let \(H_X(A)\) be the number of projective
  functionals annihilating every row of \(X[A]\). Thus
  \[
    H_X(A)=\frac{q^{k-\rank X[A]}-1}{q-1},
    \qquad \mathcal F_\ell(X)=\E_S[H_X(S)].
  \]
  Suppose \(1\le\ell<n\). Only a transvection \(I\mapsto J\) with
  \(J\in S\) and \(I\notin S\) can change this quantity. For such a
  transvection, denote the updated matrix by \(X'\). Counting the solutions of
  \(\xi(X[J])+a\xi(X[I])=0\) for each functional annihilating
  \(X[S\setminus\{J\}]\) gives
  \[
    \E_a[H_{X'}(S)]
    =H_X(S\cup\{I\})
    +\frac1q\bigl[H_X(S\setminus\{J\})
    -H_X((S\setminus\{J\})\cup\{I\})\bigr].
  \]
  Averaging over a uniform \(S\) and a uniform transvection
  \(I\mapsto J\) with \(J\in S\), \(I\notin S\), the three sets on
  the right are uniform of sizes \(\ell+1,\ell-1,\ell\), respectively.
  The crossing probability is \(\ell(n-\ell)/(n(n-1))\), so
  \begin{equation}\label{eq:rank-potential-generator}
    P\mathcal F_\ell
    =\mathcal F_\ell+\frac{\ell(n-\ell)}{n(n-1)}
    \left[\mathcal F_{\ell+1}-\mathcal F_\ell
    +\frac1q(\mathcal F_{\ell-1}-\mathcal F_\ell)\right].
  \end{equation}
  At \(\ell=0,n\), the potential is constant. Hence, at every level,
  \[
    (P\mathcal F_\ell)(X)
    =\E_\ell[\mathcal F_{\widehat L_1}(X)].
  \]
  Iterating proves the result.
\end{proof}

\subsection{Initial rank-potential and discrete birth--death estimates}

Set
\[
  w=1+(q-1)\frac kn.
\]

\begin{lemma}\label{lem:initial-potential-bound}
  For every \(0\le v\le n\),
  \begin{equation}\label{eq:initial-potential-bound}
    \mathcal F_{n-v}(X_0)
    \le
    \frac{w^v-1}{q-1}.
  \end{equation}
\end{lemma}

\begin{proof}
  Choose \(\mathcal B_0\subseteq[n]\) so that the rows of \(X_0\) indexed by
  \(\mathcal B_0\) form a basis. Let \(E\) be a uniform \(v\)-subset of \([n]\).
  The rows indexed by
  \(\mathcal B_0\setminus E\) are linearly independent, so
  \[
    k-\rank X_0\bigl[[n]\setminus E\bigr]\le |\mathcal B_0\cap E|.
  \]
  Randomly order the \(v\) sampled indices. Expanding over these positions gives
  \[
    \E_E\!\left[q^{|\mathcal B_0\cap E|}\right]
    =\sum_{r=0}^{v}\binom vr(q-1)^r
    \frac{\binom kr}{\binom nr}
    \le\sum_{r=0}^{v}\binom vr
    \left(\frac{(q-1)k}{n}\right)^r
    =w^v.
  \]
  Here \(\binom kr/\binom nr\le(k/n)^r\): for \(r\le k\),
  each factor \((k-j)/(n-j)\) is at most \(k/n\), and for
  \(r>k\) the numerator is zero. Substituting this
  bound into~\eqref{eq:def-rank-potential} proves the inequality.
\end{proof}

The next lemma bounds how quickly the dual reaches a high level and how
its distance from \(n\) then contracts.

\begin{lemma}[killed generating-function contraction]
  \label{lem:discrete-pgf-contraction} 
  Let \(H\ge2\) be an integer, let \(0<\lambda\le1\), and let \(Z^+,Z^-\) be birth--death
  chains on \(\{0,\ldots,H\}\) with absorbing endpoints. From \(1\le j<H\),
  \(Z^+\) moves up with probability \(jc_j\), down with probability
  \(jc_j/q\), and otherwise stays at \(j\), where \(c_j\ge\lambda\).
  The chain \(Z^-\) has these upward and downward probabilities interchanged.
  Killing at \(H\) assigns zero contribution to paths that reach \(H\).
  Set
  \[
    a(u)=\frac{1+qu}{1+u},\qquad
    \rho=\frac1{1+\lambda(1-q^{-1})}.
  \]
  For every \(u\ge0\), \(0\le j<H\), and integer \(s\ge0\),
  \begin{equation}\label{eq:killed-pgf-contraction}
    \begin{aligned}
      \E_j\!\left[\left(\frac{a(u)}q\right)^{Z_s^+}
      \mathbf1_{\{Z_s^+<H\}}\right]
      &\le\left(\frac{a(\rho^s u)}q\right)^j,\\
      \E_j\!\left[(a(u)^{Z_s^-}-1)\mathbf1_{\{Z_s^-<H\}}\right]
      &\le a(\rho^s u)^j-1.
    \end{aligned}
  \end{equation}
  In particular,
  \[
    \Pr_j[Z_s^+<H]\le3\left(\frac1q+H\rho^s\right)^j.
  \]
\end{lemma}

\begin{proof}
  Since \((1-\rho)/\rho=\lambda(1-q^{-1})\),
  \begin{equation}\label{eq:pgf-base-contraction}
    \begin{aligned}
      a(u)-a(\rho u)
      &=\frac{(q-1)(1-\rho)u}{(1+u)(1+\rho u)}\\
      &\le\lambda(a(u)-1)\left(1-\frac{a(u)}q\right).
    \end{aligned}
  \end{equation}
  For \(1\le j<H\), put \(a=a(u)\). The transition rules,
  \(c_j\ge\lambda\), the tangent-line inequality for \(x\mapsto x^j\),
  and~\eqref{eq:pgf-base-contraction} give
  \[
    \begin{aligned}
      q^j\E_j[(a/q)^{Z_1^+}]
      &=\E_j[a^{Z_1^-}]\\
      &=a^j-jc_ja^{j-1}(a-1)(1-a/q)\\
      &\le[a-\lambda(a-1)(1-a/q)]^j
      \le a(\rho u)^j.
    \end{aligned}
  \]
  Divide by \(q^j\) for \(Z^+\), and subtract one for \(Z^-\).
  Killing at \(H\) decreases both observables; at zero the inequalities
  are equalities. Iteration proves~\eqref{eq:killed-pgf-contraction}.
  Finally, \(a(H)/q\ge H/(H+1)\), so
  \((a(H)/q)^{H-1}>e^{-1}>1/3\). Use the first bound with \(u=H\)
  and \(a(u)/q\le q^{-1}+u\) to obtain the hitting-time estimate.
\end{proof}

\subsection{Dual absorption and deficit contraction}

Under the hypotheses of Proposition~\ref{prop:moderate-field-burnin}, we
prove the stronger estimate
\begin{equation}\label{eq:moderate-field-conclusion}
  \E_{X_0}\left[\mathcal F_\ell(X_t)\right]\le\beta.
\end{equation}

\begin{proof}[Proof of Proposition~\ref{prop:moderate-field-burnin}]
  Let \(\widehat L\) be the discrete dual chain from
  Lemma~\ref{lem:rank-moment-duality}, started at \(\ell\), and set
  \[
    d_0=\floor{\frac{n-\ell}{2}},
    \qquad
    j_\star=n-d_0.
  \]
  We first bound the time for the dual to reach \(j_\star\) without
  hitting zero. We then bound the exponential moment of the deficit
  \(n-\widehat L\), stopping if the dual falls back to \(d_0\). Duality converts
  these estimates into a bound on the rank-deficiency potential.

  Since \(\log(1/\beta)>1\) and \(\log q>1/2\), the hypothesis gives
  \(\ell-k\le9\log(1/\beta)\le(n-k)/32\), and hence
  \[
    n-k\ge32,
    \qquad
    n-\ell\ge\frac{31}{32}(n-k),
    \qquad
    k\le\ell<j_\star.
  \]
  Set
  \[
    R=\ceil{8\frac{n^2}{n-\ell}\log\frac{nq}{\beta}},
    \qquad
    E=\exp\!\left(-\frac{(n-\ell)R}{4n^2}\right).
  \]
  Since \(E\le(\beta/(nq))^2\) and \(\ell\le n\),
  \begin{equation}\label{eq:moderate-parameter-choices}
    qn\ell E\le\frac{\beta^2}{q}\le\frac\beta2.
  \end{equation}

  \medskip
  \noindent\emph{Reaching the top band.}
  Apply Lemma~\ref{lem:discrete-pgf-contraction} to the dual stopped at
  \(H=j_\star\), with
  \[
    \lambda=\frac{d_0+1}{n(n-1)}
    \ge\frac{n-\ell}{2n(n-1)}.
  \]
  Since \(\lambda\le1/(n-1)\), its contraction factor satisfies
  \[
    -\log\rho\ge\log(1+\lambda/2)
    \ge\frac{\lambda}{2+\lambda}\ge\frac{n-\ell}{4n^2},
  \]
  so \(\rho^R\le E\).
  The lemma therefore bounds the probability of not reaching \(j_\star\)
  by time \(R\) by
  \[
    3\left(\frac1q+j_\star E\right)^\ell.
  \]
  This failure event includes absorption at zero. Since
  \(\mathcal F_0(X_0)\le q^k\) and \(qj_\star\ell E\le1\), our choice
  of \(\ell-k\) yields
  \begin{equation}\label{eq:first-phase-errors-general}
    \mathcal F_0(X_0)
    \Pr_\ell\!\left[\max_{0\le s\le R}\widehat L_s<j_\star\right]
    \le3e\,q^{k-\ell}
    \le3e\,\beta^4
    \le\frac\beta4.
  \end{equation}

  \medskip
  \noindent\emph{Contracting the deficit.}
  On hitting \(j_\star\), restart the clock. Write
  \(\widehat U_s=n-\widehat L_s\) for the discrete deficit chain, started from
  \(\widehat U_0=d_0\), and write \(\widehat\tau_j\) for its hitting time of
  \(j\). Until it hits \(j_\star\), this chain has the same
  per-particle factors \(c_j=(n-j)/(n(n-1))\) as the first-stage chain, with
  upward and downward moves interchanged. Thus the same \(\lambda,\rho\) apply.
  The process \(\bigl(q^{\widehat U_s}-1\bigr)_{s\ge0}\) is a martingale, so
  optional stopping at \(\{0,j_\star\}\) gives
  \begin{equation}\label{eq:deficit-escape-general}
    \begin{aligned}
      \mathcal F_0(X_0)
      \Pr_{d_0}[\widehat\tau_{j_\star}<\widehat\tau_0]
      &\le q^k\frac{q^{d_0}-1}{q^{j_\star}-1}\\
      &\le q^{k+2d_0-n}
      \le q^{k-\ell}\le\beta^4\le\frac\beta8.
    \end{aligned}
  \end{equation}

  To bound the remaining potential, set \(u_0=k/(n-k)\), so that
  \(w=a(u_0)\) in Lemma~\ref{lem:discrete-pgf-contraction}. We have
  \(d_0u_0\le n/2\) and \(d_0(q-1)Eu_0\le nqE/2\le1\). The second bound of
  the lemma, \(\rho^s\le E\) for \(s\ge R\), and the inequalities
  \(a(u)\le e^{(q-1)u}\) and \(e^x-1\le2x\) for \(0\le x\le1\) therefore give
  \begin{equation}\label{eq:deficit-moment-final}
    \begin{aligned}
      \sup_{s\ge R}\frac1{q-1}
      \E_{d_0}\left[
        (w^{\widehat U_s}-1)
        \mathbf 1_{\{\widehat\tau_{j_\star}>s\}}
      \right]
      &\le\frac{a(Eu_0)^{d_0}-1}{q-1}\\
      &\le 2d_0Eu_0\le nE\le\frac\beta8.
    \end{aligned}
  \end{equation}

  \medskip
  \noindent\emph{Combining the stages.}
  A path that reaches \(j_\star\) by time \(R\) leaves at least \(R\) steps for
  the deficit chain before time \(2R\). Apply the strong Markov property at this
  hitting time. If the deficit subsequently reaches \(j_\star\), bound the potential
  by \(\mathcal F_0(X_0)\); otherwise apply
  Lemma~\ref{lem:initial-potential-bound} and
  \eqref{eq:deficit-moment-final}. Failure to reach \(j_\star\),
  subsequent deficit escape, and paths with neither failure contribute at
  most \(\beta/4\), \(\beta/8\), and \(\beta/8\), respectively, by
  \eqref{eq:first-phase-errors-general}, \eqref{eq:deficit-escape-general}, and
  \eqref{eq:deficit-moment-final}.  Hence
  \[
    \E_\ell\left[\mathcal F_{\widehat L_{2R}}(X_0)\right]
    \le\frac\beta2.
  \]
  Since \(n-\ell\ge31(n-k)/32\) and \(n\ge33\),
  \[
    2R\le\frac{512}{31}\frac{n^2}{n-k}\log\frac{nq}{\beta}+2
    \le17\frac{n^2}{n-k}\log\frac{nq}{\beta}.
  \]
  Duality~\eqref{eq:rank-moment-duality} proves the estimate at time \(2R\),
  and Lemma~\ref{lem:fixed-time-extension} extends it to every time in the
  stated range. Averaging~\eqref{eq:rank-failure-by-potential} over \(X_t\)
  gives the proposition's rank-failure bound.
\end{proof}

\section{Proofs for the trapdoored-matrix constructions}
\label{app:trapdoor-proofs}

\begin{proof}[Proof of Theorem~\ref{thm:transvection-trapdoors}]
  Fix \(n\), write \(q=q_n\) and \(T=f(n)\), and sample a
  length-\(T\) walk history \(\tau=((I_s\mapsto J_s,a_s))_{s=1}^T\)
  with skeleton \(\sigma\). Its product is
  \[
    M_\tau
    =(I_n+a_TE_{J_T,I_T})\cdots(I_n+a_1E_{J_1,I_1})
    \in\mathrm{SL}_n(\F_q).
  \]
  The history \(\tau\) specifies a circuit for \(M_\tau v\): starting from
  \(y=v\), perform
  \[
    y_{J_s}\leftarrow y_{J_s}+a_sy_{I_s},
    \qquad s=1,\ldots,T.
  \]
  This uses at most \(2T\) field operations, in addition to \(O(n)\)
  input and output work. Conjecture~\ref{conj:trapdooredness} gives
  pseudorandomness with \(\sigma\) public, proving~(i).

  Reversing the history and negating its coefficients computes
  \(M_\tau^{-1}v\), with the same operation count, because
  \((I_n+aE_{ji})^{-1}=I_n-aE_{ji}\) for \(i\ne j\).

  For~(ii), independently sample \(h\sim\Unif(\F_q^\times)\), set
  \[
    H_h=\operatorname{diag}(h,1,\ldots,1),
    \qquad
    G_{\tau,h}=H_hM_\tau,
  \]
  and append one scalar multiplication to the circuit above. If \(S\) is uniform on
  \(\mathrm{SL}_n(\F_q)\), then \(H_hS\) is uniform on \(\GL_n(\F_q)\): every
  \(G\in\GL_n(\F_q)\) has the unique representation
  \[
    G=H_{\det G}S,
    \qquad S=H_{\det G}^{-1}G\in\mathrm{SL}_n(\F_q).
  \]
  This randomized postprocessing preserves indistinguishability
  jointly with \(\sigma\).

  To compute \(G_{\tau,h}^{-1}v=M_\tau^{-1}H_h^{-1}v\), first scale
  the first input coordinate by \(h^{-1}\), then use the inverse
  circuit. Inversion preserves uniform measure and is efficiently
  computable on both groups, so the inverse ensembles are also
  computationally indistinguishable from uniform.

  For~(iii), independently sample \(z\sim\Unif(\F_q^n)\), write
  \(D_z=\operatorname{diag}(z_1,\ldots,z_n)\), and use
  \[
    G_{\tau,h}+D_z.
  \]
  Its trapdoor circuit computes \(G_{\tau,h}v\) and adds \(D_zv\), using
  \(O(n+T)\) field operations. Lemma~\ref{lem:diagonal-smoothing} below shows
  that \(G+D_z\) is negligibly close to uniform on \(\F_q^{n\times n}\) when
  \(G\sim\Unif(\GL_n(\F_q))\) is independent of \(z\). By~(ii),
  \((\sigma,G_{\tau,h})\) and \((\sigma,G)\) are computationally
  indistinguishable, with \(G\) independent of \(\sigma\).
  Adding \(D_z\) preserves this property; the triangle inequality
  proves~(iii).

  In each case, the public skeleton determines the circuit; the coefficients
  and any added scalars are secret inputs. Thus the families are universal.
  The reductions above and the statistical error in
  Lemma~\ref{lem:diagonal-smoothing} are uniform in \(q\). Thus all three
  ensembles inherit the uniformity in Conjecture~\ref{conj:trapdooredness}.
\end{proof}

\begin{lemma}[Diagonal smoothing]\label{lem:diagonal-smoothing}
  Let \(G\sim\Unif(\GL_n(\F_q))\) and
  \(z\sim\Unif(\F_q^n)\) be independent. Then
  \begin{equation}\label{eq:diagonal-smoothing}
    \left\|
    \mathcal L(G+D_z)-\Unif(\F_q^{n\times n})
    \right\|_{\TV}
    \le 4\left(\frac{\sqrt3}{2}\right)^n.
  \end{equation}
\end{lemma}

\begin{proof}
  We bound the Fourier mass that remains after adding the random diagonal.
  Let \(\mathsf{ZD}_n\) be the subspace of zero-diagonal matrices in
  \(\F_q^{n\times n}\), fix a nontrivial additive character \(\psi\) of
  \(\F_q\), and put
  \(\Psi(B)=\E[\psi(\operatorname{tr}(B^{\top} G))]\).
  We use \(\chi^2(\mu\|\nu)=\sum_x\mu(x)^2/\nu(x)-1\) when \(\nu\) has full support.
  Averaging over \(D_z\) kills the Fourier coefficients outside
  \(\mathsf{ZD}_n\), so Plancherel gives
  \begin{equation}\label{eq:diagonal-smoothing-chisq}
    \chi^2\!\left(\mathcal L(G+D_z)\middle\|\Unif(\F_q^{n\times n})\right)
    =\sum_{\substack{B\in\mathsf{ZD}_n\\B\ne0}}|\Psi(B)|^2.
  \end{equation}
  Put \(p=|\GL_n(\F_q)|/q^{n^2}\). The product formula gives
  \[
    p=\prod_{j=1}^n(1-q^{-j})
    \ge\tfrac12\left(1-\sum_{j=2}^n2^{-j}\right)\ge\tfrac14,
  \]
  and Plancherel gives \(\sum_{B\in\F_q^{n\times n}}|\Psi(B)|^2=p^{-1}\).

  For \(B\ne0\) and independent uniform \(x,y\in\F_q^n\),
  one has \(\Pr[By=0]\le q^{-1}\), hence
  \[
    \Pr[x^{\top} By=0]
    =q^{-1}+(1-q^{-1})\Pr[By=0]
    \le2q^{-1}-q^{-2}\le\tfrac34.
  \]
  If \(P,Q\) are independent uniform matrices in \(\F_q^{n\times n}\),
  their column pairs are independent, so
  \(\Pr[P^{\top} BQ\in\mathsf{ZD}_n]\le(3/4)^n\).
  Conditioning on \(P,Q\in\GL_n(\F_q)\) costs at most \(p^{-2}\).

  Now take \(P,Q\) independent and uniform on \(\GL_n(\F_q)\).
  Invariance of \(G\) under left and right multiplication gives
  \(\Psi(P^{\top} BQ)=\Psi(B)\). Changing variables \(B\mapsto P^{\top} BQ\)
  in~\eqref{eq:diagonal-smoothing-chisq} and averaging yields
  \[
    \begin{aligned}
      \sum_{\substack{B\in\mathsf{ZD}_n\\B\ne0}}|\Psi(B)|^2
      &=\sum_{B\ne0}|\Psi(B)|^2
      \Pr[P^{\top} BQ\in\mathsf{ZD}_n]\\
      &\le p^{-2}(3/4)^n\sum_{B\ne0}|\Psi(B)|^2\\
      &\le p^{-3}(3/4)^n\le64(3/4)^n.
    \end{aligned}
  \]
  The total-variation distance is at most half the square root of this
  bound, proving~\eqref{eq:diagonal-smoothing}.
\end{proof}

\begin{proof}[Proof of Proposition~\ref{prop:sparse-transvection-blocks}]
  If the next transvection is \(I\mapsto J\), the new
  row \(J\) of \(B_s-I_n\) is supported on the union of
  its old support, the support of row \(I\) of \(B_s-I_n\), and coordinate \(I\).
  Therefore
  \[
    N_{s+1}\le N_s+\operatorname{nnz}\bigl((B_s-I_n)[I]\bigr)+1.
  \]
  Conditional on \(B_s\), the source \(I\) is uniform on \([n]\), so
  \[
    \E[N_{s+1}\mid B_s]
    \le \left(1+\frac1n\right)N_s+1.
  \]
  Since \(N_0=0\), iteration and \((1+1/n)^s\le e^{s/n}\)
  prove~\eqref{eq:expected-block-sparsity}.

  To bound the support of one output row, trace its possible source
  coordinates backward through the skeleton.
  Start with its row label. For each exposed transvection \(i\mapsto j\),
  add \(i\) if \(j\) belongs to the current set, irrespective of the coefficient.
  This set contains the output row's support and is independent of
  the next exposed transvection. Its size \(A_t\) starts at \(1\)
  and, at size \(a\), grows with probability
  \[
    \frac{a(n-a)}{n(n-1)}\le\frac an.
  \]
  Put \(\rho_t=1-(1-1/n)^t\). We claim that
  \(\Pr(A_t\ge r)\le\rho_t^{r-1}\) for \(2\le r\le n\).
  The cases \(t=0,1\) are immediate. For \(t\ge1\), the birth probability
  at size \(r-1\) is at most \((r-1)/n\), so the tail recursion and induction give
  \[
    \begin{aligned}
      \Pr(A_{t+1}\ge r)
      &\le\left(1-\frac{r-1}{n}\right)\Pr(A_t\ge r)
      +\frac{r-1}{n}\Pr(A_t\ge r-1)\\
      &\le\rho_t^{r-1}
      +\frac{r-1}{n}(1-\rho_t)\rho_t^{r-2}\\
      &\le\left(\rho_t+\frac{1-\rho_t}{n}\right)^{r-1}
      =\rho_{t+1}^{r-1}.
    \end{aligned}
  \]
  The last inequality is the binomial theorem.
  A union bound over the output rows, followed by \(\rho_s\le s/n\), proves
  \eqref{eq:block-row-support-tail}.

  Compute \(B_sv=v+(B_s-I_n)v\) using at most
  \(2\operatorname{nnz}(B_s-I_n)\) gates: multiply once for each nonzero entry of
  \(B_s-I_n\), then add the resulting terms to the corresponding coordinates of
  \(v\). Combining the diagonal correction with the original coordinate does
  not increase this count and leaves at most \(R_s\) summands in each output
  coordinate. Balanced addition trees therefore give depth
  \(O(\log(R_s+1))\). For a block of length \(s\le m\), convexity and
  \eqref{eq:expected-block-sparsity} bound the expected gate count by
  \[
    2n(e^{s/n}-1)\le\frac{2ns}{m}(e^{m/n}-1).
  \]
  The block lengths sum to \(T\), giving the size bound.
  Sequential application adds their depths.

  We finish by bounding the expected depth of each block.
  For \(1\le s\le n\), one has
  \(\rho_s\le\min\{s/n,3/4\}\), since
  \((1-1/n)^n\ge1/4\). Split the tail sum for \(R_s\) at
  \(1+\ceil{\log n/\log(1/\rho_s)}\). The remaining geometric series
  has sum at most \(1/(1-\rho_s)\le4\), so
  \[
    \E[R_s]
    \le 5+\frac{\log n}{\log(1/\rho_s)}
    =O\left(1+
    \frac{\log n}{\max\{1,\log(n/s)\}}\right).
  \]
  Jensen's inequality and \(s\le m\le n\) therefore give
  \[
    \E[\log(R_s+1)]
    =O\left(
      \log\left(
        2+\frac{\log n}{\max\{1,\log(n/m)\}}
      \right)
    \right).
  \]
  Summing over the blocks proves~\eqref{eq:expected-block-depth}.
\end{proof}

\section{Proofs of the standard preliminary lemmas}
\label{app:standard-estimates}

\begin{proof}[Proof of Lemma~\ref{lem:fixed-time-extension}]
  For (i), take integers \(0\le s\le t\), condition on \(X_{t-s}\), and
  apply the supremum bound for the remaining \(s\) steps.
  Part (ii) follows from contraction of total variation under \(P^{t-s}\)
  and stationarity of \(\pi\).
\end{proof}

\begin{proof}[Proof of Lemma~\ref{lem:geometric-waiting}]
  Conditional on the history before stage \(i\), its waiting time \(W_i\)
  is stochastically dominated by a geometric variable with success
  probability \(p_i\).
  Thus, for \(0\le\lambda<p_i\),
  \[
    \E\left[e^{\lambda W_i}\mid\text{history before stage }i\right]
    \le\frac{p_i}{e^{-\lambda}-1+p_i}
    \le\frac1{1-\lambda/p_i},
  \]
  using \(e^{-\lambda}\ge1-\lambda\). For \(0\le z<1\), the power series gives
  \(-\log(1-z)\le z+z^2/(2(1-z))\). Applying this with \(z=\lambda/p_i\)
  and iterating conditional expectations gives, for \(0\le\lambda<p_*\),
  \[
    \log\E[e^{\lambda(T-\mu)}]
    \le\sum_{i=1}^m\frac{\lambda^2}{2p_i^2(1-\lambda/p_i)}
    \le\frac{\lambda^2\mu}{2p_*(1-\lambda/p_*)}.
  \]
  Put \(L=\log(1/\varepsilon_0)\). Chernoff's bound with
  \[
    \lambda=\frac{p_*\sqrt{2L}}{\sqrt{\mu p_*}+\sqrt{2L}}\in(0,p_*)
  \]
  gives the claimed probability bound \(e^{-L}=\varepsilon_0\).
\end{proof}

\begin{proof}[Proof of Lemma~\ref{lem:full-rank-uniform}]
  If \(X\) is rank deficient, some projective nonzero functional on
  \(\F_q^k\) annihilates all its rows. A union bound gives the first estimate.
  Conditioning a uniform \(n\times k\) matrix on full rank changes its law in TV by
  exactly the rank-failure probability.
\end{proof}

\section{Proofs of bounds for Algorithm~\ref{alg:short-row-dependence}}\label{app:simultaneous-routing}

\begin{proof}[Proof of Proposition~\ref{prop:short-walk-cryptanalysis}]
  Put \(r=\ceil{k/2}\) and fix an initially zero row.
  With probability \((1-1/n)^T\), it is never targeted and remains zero.
  Independently, at least one query contains this row with probability
  \[
    1-\left(1-\frac rn\right)^a
    \ge1-e^{-ar/n}\ge1-e^{-1/2}.
  \]
  Thus the walk's acceptance probability is at least
  \[
    (1-e^{-1/2})(1-1/n)^T\ge e^{-T/(n-1)-1},
  \]
  using \(1-e^{-1/2}\ge e^{-1}\) and
  \(\log(1-1/n)\ge-1/(n-1)\).

  Lemma~\ref{lem:full-rank-uniform}, applied to the transpose of a uniform
  \(r\times k\) matrix, bounds its probability of dependent rows by \(q^{r-k}\).
  Conditioning the whole matrix on full column rank costs at most a
  factor of two. A union bound therefore gives stationary acceptance
  probability at most \(2a q^{r-k}\le e^{-\Omega(n)}\).
  Subtracting proves~\eqref{eq:sampled-dependence-lower}.
\end{proof}

For Proposition~\ref{prop:sampled-dependence-upper}, we use one skeleton
event that supplies disjoint paths to every target set.

\begin{lemma}[simultaneous disjoint-path routing]\label{lem:simultaneous-routing}
  \label{lem:simultaneous-disjoint-routing}
  There are absolute constants \(c,C>0\) such that, for \(n\ge4\) and
  every integer \(T\ge Cn\log^2 n\), a random length-\(T\) skeleton
  has the following property with probability at least \(1-e^{-cT/n}\):
  for every \(1\le r\le n\) and
  \(\mathcal B,\mathcal T\subseteq[n]\) with
  \(|\mathcal B|=|\mathcal T|=r\), its time-expanded graph contains
  \(r\) vertex-disjoint directed paths from
  \(\mathcal B\times\{0\}\) to \(\mathcal T\times\{T\}\).
\end{lemma}

Such a graph is known as a \emph{superconcentrator}~\cite{Valiant1975,Pippenger1977}.

\begin{proof}
  For \(r\le n/2\), we show that no set of fewer than \(r\) vertices
  separates the sources from the targets in the time-expanded graph.
  Put \(L=\ceil{128n\log n}\) and \(m=\floor{T/(2L)}\).
  Split the first \(2mL\) steps into two halves of \(m\) blocks
  of length \(L\); persistence extends the resulting paths to time \(T\).
  If each half has enough good blocks, forward reachability from the
  sources and backward reachability from the targets each grow to
  \(3n/4\) rows despite the deletions. An intact row in their overlap
  connects the two. A complement argument handles larger \(r\).

  \paragraph{Expansion in a block.}
  We now follow a technique similar in spirit to that of Smith and Zhang~\cite[Appendix~D]{SmithZhang2013}, though applied on a different graph: divide the graph into layers, show expansion in each layer, and argue that dropping rows affects this by a limited amount.

  Let \(N^+(A)\) consist of \(A\) and its outgoing neighbors in the
  block, and define \(N^-(A)\) using incoming neighbors.
  Call a block good if, for every \(A\subseteq[n]\),
  \[
    \min\{|N^+(A)|,\ |N^-(A)|\}\ge \min\{8|A|,7n/8\}.
  \]
  If this condition fails for an \(a\)-set \(A\), let \(U\) be a
  deficient neighborhood. Then \(|U|<\min\{8a,7n/8\}\), and the block
  misses all arrows between \(A\) and \([n]\setminus U\) in that direction.
  Since \(A\subseteq U\) and \(|[n]\setminus U|>n/8\), there are at least
  \(an/8\) such possible arrows. For fixed \(A,U\), the probability of
  missing them throughout the block is at most \(e^{-La/(8n)}\).
  There are at most \(n^a\) choices of \(A\) and \(n^{8a}\) choices of \(U\).
  A union bound over \(a\ge1\) and both directions gives
  \[
    \Pr[\text{a block is bad}]
    \le2\sum_{a=1}^n n^{9a}e^{-La/(8n)}
    \le n^{-6}.
  \]

  \paragraph{Growth after deleting vertices.}
  Fix \(r\le n/2\) and delete fewer than \(r\) vertices.
  Start from any \(r\) rows. At each block boundary, let \(s\) count
  reachable rows and let \(b\) be \(r\) minus all deletions so far,
  including time zero. Then \(s\ge b\ge1\).
  If the next block has \(d\) further deletions, its new values
  \(s'\) and \(b'=b-d\) satisfy, by persistence,
  \[
    s'\ge s-d,\qquad \frac{s'}{b'}\ge\frac sb.
  \]
  At least \(s-d\) reachable rows persist throughout the block.
  Their outgoing neighbors are reachable at its end except on at most \(d\)
  rows. Thus a good block gives
  \[
    s'\ge\min\{8(s-d),7n/8\}-d.
  \]
  If \(s<n/4\) and \(d\le s/2\), then \(s'/b'\ge2s/b\).
  If \(s<n/4\) and \(d>s/2\), then \(b'<b/2\).
  If \(s\ge n/4\) and \(d\le n/8\), then \(s'\ge3n/4\).
  Before reaching \(3n/4\) rows, the first two cases occur at most
  \(\ceil{\log_2 n}\) times each, since \(s/b\le n\) and \(1\le b\le r\).
  The remaining good blocks each use more than \(n/8\) deletions,
  so there are at most three.
  Thus \(M=2\ceil{\log_2 n}+4\) good blocks force \(s\ge3n/4\).

  \paragraph{Routing all targets.}
  Suppose each half contains at least \(M\) good blocks.
  Apply the growth argument forward from \(\mathcal B\) in
  the first half and backward from \(\mathcal T\) in the second.
  At two ordered times, the resulting sets each have at least
  \(3n/4\) rows, so share at least \(n/2\). Fewer than \(r\le n/2\)
  deletions leave one common row intact between those times.
  Its persistence edges join the two paths, so no such set of deletions
  separates the sources and targets. Menger's theorem gives \(r\)
  disjoint paths, simultaneously for all \(\mathcal B,\mathcal T\)
  and \(r\le n/2\).

  \paragraph{Larger row sets.}
  We now use a deterministic variant of the classic particle-hole duality to prove the result for large row sets. Define the \emph{twist} of a skeleton by replacing each \(i\mapsto j\)
  with \(j\mapsto i\) at the same time.
  A family of \(r\) vertex-disjoint paths from \(\mathcal B\) to
  \(\mathcal T\) corresponds to \(n-r\) vertex-disjoint paths between
  their complements in the twist.

  Start the complementary paths on \([n]\setminus\mathcal B\).
  Whenever an original path moves \(i\mapsto j\), row \(j\) must be
  unoccupied just before the move: any path already there would persist
  and cause a collision.
  The complementary paths occupy \(j\) but not \(i\), so send the path
  at \(j\) to \(i\). At all other positions and times, use persistence.
  The occupied sets remain complementary at every time, so the new paths
  stay disjoint and end at \([n]\setminus\mathcal T\).
  Twisting twice restores the original skeleton, giving the converse.

  For \(n/2<r<n\), apply the proved case to the complements in the twist.
  Twisting swaps \(N^+\) and \(N^-\), so it preserves the good blocks
  and the event under consideration. For \(r=n\), use all persistence paths.

  \paragraph{Completing the proof.}
  For \(C\) large enough, \(m\ge2M\), and independence bounds the
  probability of fewer than \(M\) good blocks in either half by
  \[
    2\cdot2^m(n^{-6})^{m/2}
    =2^{m+1}n^{-3m}\le e^{-cT/n},
  \]
  for an absolute \(c>0\), since \(m=\Theta(T/(n\log n))\).
\end{proof}

\end{document}